\documentclass[aps,10pt,prd,citeautoscript,longbibliography,superscriptaddress,twocolumn,nofootinbib]{revtex4-1}
\usepackage[T1]{fontenc}
\usepackage{color}
\usepackage{graphicx}
\usepackage{amsmath,amssymb,amsfonts,bm,dsfont,units}
\usepackage{hyperref}
\usepackage{float}

\begin{document}

\title{
Vortex-enabled Andreev processes in quantum Hall-superconductor hybrids
}

\author{Yuchen Tang}
\affiliation{Department of Physics, University of California, Berkeley, California 94720, USA}
\affiliation{Department of Physics and Institute for Quantum Information and Matter, California Institute of Technology, Pasadena, CA 91125, USA}

\author{Christina Knapp}
\affiliation{Microsoft Station Q, Santa Barbara, California 93106-6105 USA}
\affiliation{Department of Physics and Institute for Quantum Information and Matter, California Institute of Technology, Pasadena, CA 91125, USA}
\affiliation{Walter Burke Institute for Theoretical Physics, California Institute of Technology, Pasadena, CA 91125, USA}

\author{Jason Alicea}
\affiliation{Department of Physics and Institute for Quantum Information and Matter, California Institute of Technology, Pasadena, CA 91125, USA}
\affiliation{Walter Burke Institute for Theoretical Physics, California Institute of Technology, Pasadena, CA 91125, USA}

\date{\today}

\begin{abstract}
	Quantum Hall-superconductor heterostructures provide possible platforms for intrinsically fault-tolerant quantum computing.  Motivated by several recent experiments that successfully integrated these phases, we investigate transport through a proximitized integer quantum Hall edge---paying particular attention to the impact of vortices in the superconductor.  By examining the downstream conductance, we identify regimes in which sub-gap vortex levels mediate Andreev processes that would otherwise be frozen out in a vortex-free setup.  Moreover, we show that at finite temperature, and in the limit of a large number of vortices, the downstream conductance can average to zero, indicating that the superconductor effectively behaves like a normal contact.  Our results highlight the importance of considering vortices when using transport measurements to study superconducting correlations in quantum Hall-superconductor hybrids.
\end{abstract}

\maketitle

\section{Introduction}

Quantum Hall (QH)-superconductor hybrids provide fertile ground for pursuing non-Abelian defects that can be exploited for intrinsically fault-tolerant quantum computing~\cite{Qi10, Lindner12, Cheng12, Clarke13, Vaezi13, Vaezi14, Mong14, Clarke14,Alicea16}.  As a fascinating prerequisite,  numerous experiments have successfully demonstrated proximity-induced superconductivity in quantum Hall edge states---both in the integer~\cite{Takayanagi98,Komatsu12,Rickhaus12,Zhong15,Amet16,Lee17,Park17,Guiducci18,Draelos18,Seredinski19,Zhi19,Zhao20,Gul20,Hatefipour21} and fractional~\cite{Gul20} regime.  These experiments raise fundamental questions concerning the interplay between chiral electron transport and Cooper pairing that has been the subject of several recent theory works~\cite{Ostaay11, Clarke14, Zocher16, Huang17, Gamayun17, Chaudhary20, Manesco21, Nikolaenko21, Kurilovich22,Schiller22}.

In the absence of superconductivity, electrons injected into a quantum Hall edge state propagate with negligible probability of backscattering, thus underlying exquisite conductance quantization.  Proximity-induced Cooper pairing enriches the story by enabling a new process: on traversing a superconducting region, edge electrons can in principle convert into co-moving holes via a chiral counterpart of Andreev reflection~\cite{Gamayun17}.  When the probability for such an Andreev process exceeds 1/2, the conductance quite strikingly becomes \emph{negative}---a clear demonstration of superconducting correlations induced in a quantum Hall edge.  Interestingly, negative conductance has been observed experimentally in Refs.~\onlinecite{Lee17, Park17, Zhao20, Gul20, Hatefipour21}. 

Chirality can nevertheless frustrate the effects of Cooper pairing.  For reference, in a clean, time-reversal-invariant system, symmetry guarantees that an electron with momentum ${\bf k}$ has a partner with momentum $-{\bf k}$ at the same energy with which it can resonantly form a Cooper pair.  For a quantum Hall edge, by contrast, the energy for states with opposite momenta differ except with fine tuning.  At least in the clean limit, kinematic constraints therefore generically suppress Andreev processes \cite{Gamayun17, Kurilovich22}.  This suppression prompted Ref.~\onlinecite{Kurilovich22} to invoke disorder as a means of resurrecting Andreev processes observed in experiment.  

Here we incorporate an alternative mechanism that can  revive Andreev processes even in an otherwise clean system: hybridization between edge electrons and vortices in the proximitizing superconductor.  Vortices are generally expected to appear in QH-superconductor hybrids given the strong magnetic fields required to reach the QH regime.  Moreover, they bind a series of sub-gap Caroli-de Gennes-Matricon states with typical splittings that are small compared to accessible temperatures, and that thus play a role even at the lowest energy scales probed in experiment.  
Previous experimental~\cite{Lee17,Gul20,Hatefipour21} and theoretical~\cite{Kurilovich22, Schiller22} works have considered vortices to \emph{reduce} the visibility of  superconducting correlations in a QH edge, for instance by allowing chiral electrons to escape the edge by inter-vortex tunneling events.  

\begin{figure}[t]
    \centering
    \includegraphics[width=\columnwidth]{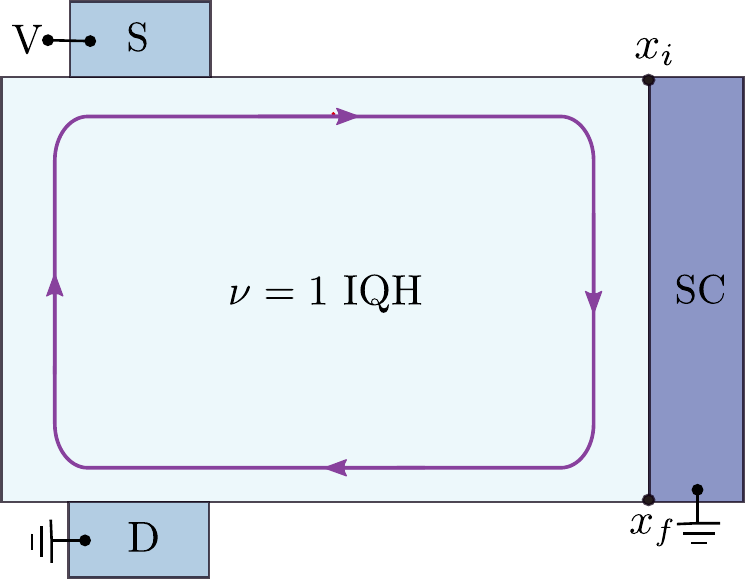}
    \caption{Setup: 
    A $\nu=1$ integer quantum Hall (IQH) state borders a grounded superconductor (SC).
    For edge coordinates $x_i<x<x_f$ the chiral edge mode (purple) inherits proximity-induced Cooper pairing from the adjacent superconductor.
    We are interested in the conductance describing current flow between the source (S) and drain (D) as a function of bias voltage $V$ in the presence of vortices in the superconductor.
    }
    \label{fig:experimental_setup}
\end{figure}

We consider an alternative regime in which vortices are sufficiently well-separated that inter-vortex hopping is negligible, focusing on a proximitized $\nu = 1$ integer quantum Hall edge for simplicity; see the circuit from Fig.~\ref{fig:experimental_setup}.  In this case an edge electron that tunnels onto a nearby vortex must (eventually) return to the edge\footnote{In principle electrons might also be able to escape by propagating along a given vortex line into some low-lying states, e.g., in the substrate; we assume that  such paths are unavailable.}---but crucially can undergo a particle-hole rotation mediated by resonances with the sub-gap vortex levels.  We show that vortices can correspondingly provide a mechanism to \emph{enhance} Andreev processes.  In fact, hybridization with just a \emph{single} vortex in principle allows the thermally averaged conductance to become negative at low bias voltages, even when Andreev processes are absent entirely in the vortex-free limit.  We further show that as the number of vortices that couple to the quantum Hall edge increases, the thermally averaged low-bias conductance approaches zero.  That is, resonances from many vortices tend to randomize the net particle-hole rotation experienced at the edge, so that an incident edge electron exits the superconducting region as either an electron or hole with equal probability.  In this case the superconductor effectively behaves like a normal contact.  

The remainder of this paper is organized as follows. 
In Section \ref{sec:vortex-free}, we review the vortex-free problem using a scattering matrix approach that generalizes to the case with vortices. We incorporate a single vortex in Section~\ref{sec:single-vortex}, elucidating the role of the vortex bound states on the conductance. After solving for the single vortex scattering matrix, we deduce the multi-vortex solution in Section~\ref{sec:multi-vortex}. In both sections, we first present the general solution to the problem before examining a simple limit to gain intuition for the physics and finite-temperature effects. We conclude in Section~\ref{sec:discussion} by discussing the implications of our results.
Details of the calculations are relegated to the appendices.

\section{Review: vortex-free conductance}\label{sec:vortex-free}

We review the transport properties of a proximitized $\nu=1$ IQH edge in the absence of vortices in the superconductor, following the elegant treatment of Ref.~\onlinecite{Gamayun17}.
The setup, illustrated in Fig.~\ref{fig:experimental_setup}, features a single $\nu = 1$ edge mode with induced Cooper pairing along the interface with the superconductor occurring between positions $x_i$ and $x_f$.
We model the system with an effective Hamiltonian
\begin{align}
    H &= H_0 + H_\Delta 
    \\ H_0 &= \int_x \Tilde{\psi}^{\dagger} [-iv\partial_x-\tilde \mu(x)] \Tilde{\psi}. \label{eq:vortex_free_H0}
    \\ H_{\Delta} &=  \frac{1}{2} \int_x \Delta(x)[ie^{i \phi(x)} \Tilde{\psi} \partial_x \Tilde{\psi} + h.c.]. \label{eq:vortex_free_Hdelta}
\end{align}
(Throughout this paper we set $\hbar = e = k_B = 1$.)
Here, $\tilde{\psi}(x)$ is a fermion operator that removes an electron from position $x$ along the edge.
The first term, $H_0$, describes the kinetic energy along the edge with associated velocity $v$ and chemical potential $\tilde{\mu}(x)$.
Position dependence in $\tilde \mu(x)$ encodes possible charge transfer from the superconductor to the edge mode in the proximitized region.  
We fix $\tilde{\mu}(x)=0$ away from the superconductor but, to account for such charge transfer, allow for a non-zero $\tilde \mu(x)$ adjacent to the superconductor. 

The second term, $H_\Delta$, encodes pairing processes generated by the parent superconductor, with $\Delta(x)$ and $\phi(x)$ respectively denoting the (real) pairing amplitude and phase; note that $\Delta(x)$ is nonzero only in the proximitized region.  
It is convenient to hereafter remove the superconducting phase from Eq.~\eqref{eq:vortex_free_Hdelta} by defining $\tilde \psi(x) = \psi(x) e^{-i \phi(x)/2}$. 
In these variables, phase winding has been recast as a chemical potential renormalization such that the new effective chemical potential in the proximitized region is $\mu(x) = \tilde \mu(x) + \frac{v}{2} \partial_x \phi(x)$.  
For simplicity, we will always assume spatially independent $\Delta(x) \equiv \Delta_{\rm sc}$ and $\mu(x) \equiv \mu_{\rm sc}$ adjacent to the superconductor.

Useful insight can be gleaned by examining limits of $\mu_{\rm sc}$.  
First, when $\mu_{\rm sc} = 0$, one can profitably view the IQH edge state as two co-propagating Majorana fermions $\gamma_{1,2}$ by writing $\psi = \gamma_1 + i \gamma_2$.  
In this representation, the full $\mu_{\rm sc} = 0$ Hamiltonian reads
\begin{align}\label{eq:Hamiltonian_gamma}
    H = \int_x\{-i\gamma_1[v-\Delta(x)]\partial_x\gamma_1
    &-i\gamma_2[v+\Delta(x)]\partial_x\gamma_2\}.
\end{align}
Away from the superconductor, where again $\Delta(x) = 0$, both Majorana fermions propagate with the same velocity $v$ as required by local charge conservation.  
In the proximitized region, the induced pairing instead yields unequal velocities $v-\Delta_{\rm sc}$ and $v+\Delta_{\rm sc}$ for $\gamma_1$ and $\gamma_2$, respectively.  
For an incident electron with energy $E$, the Majorana fermion $\gamma_2$ accordingly acquires a phase of
\begin{equation}
      \delta \phi(E) = -2\pi E/\tilde V
      \label{deltaphi}
\end{equation}
relative to $\gamma_1$, 
where
\begin{equation}\label{eq:til-V}
    \tilde V = \frac{\pi}{x_f-x_i}\left(\frac{v^2-\Delta_{\rm sc}^2}{\Delta_{\rm sc}} \right)
\end{equation}
is an important energy scale that sets the periodicity of $\delta \phi(E)$.
An edge with $v = 10^4$\,m/s,  $\Delta_{\rm sc} = v/10$, and $x_f-x_i = 1\mu$m is characterized by $\tilde V \approx 0.4$K.  Acquisition of the relative phase $\delta \phi(E)$ morphs the incoming electron into an outgoing superposition of electron and hole with amplitudes $A_{\rm e}^{\rm out} = \left(1+e^{i\delta \phi(E)}\right)/2$ and $A_{\rm h}^{\rm out} = \left(1-e^{i\delta \phi(E)}\right)/2$, respectively.  
This rotation in particle-hole space underlies Andreev conversion processes for the proximitized QH edge.  
In particular, the zero-temperature conductance at bias voltage $V$ follows as 
\begin{equation}
  G(V) = g_0(|A_{\rm e}^{\rm out}|^2-|A_{\rm h}^{\rm out}|^2) = g_0 \cos \left[\delta \phi(V)\right],
  \label{G_periodic}
\end{equation}
where $g_0$ represents the conductance quantum $e^2/h$ that becomes $1/2\pi$ in our units.  
The conductance accordingly oscillates between $+g_0$ (incident electrons transmit perfectly as electrons) and $-g_0$ (incident electrons transmit perfectly as holes) as $V$ varies. 

\begin{figure}[t]
    \centering
    \includegraphics[width=1\columnwidth]{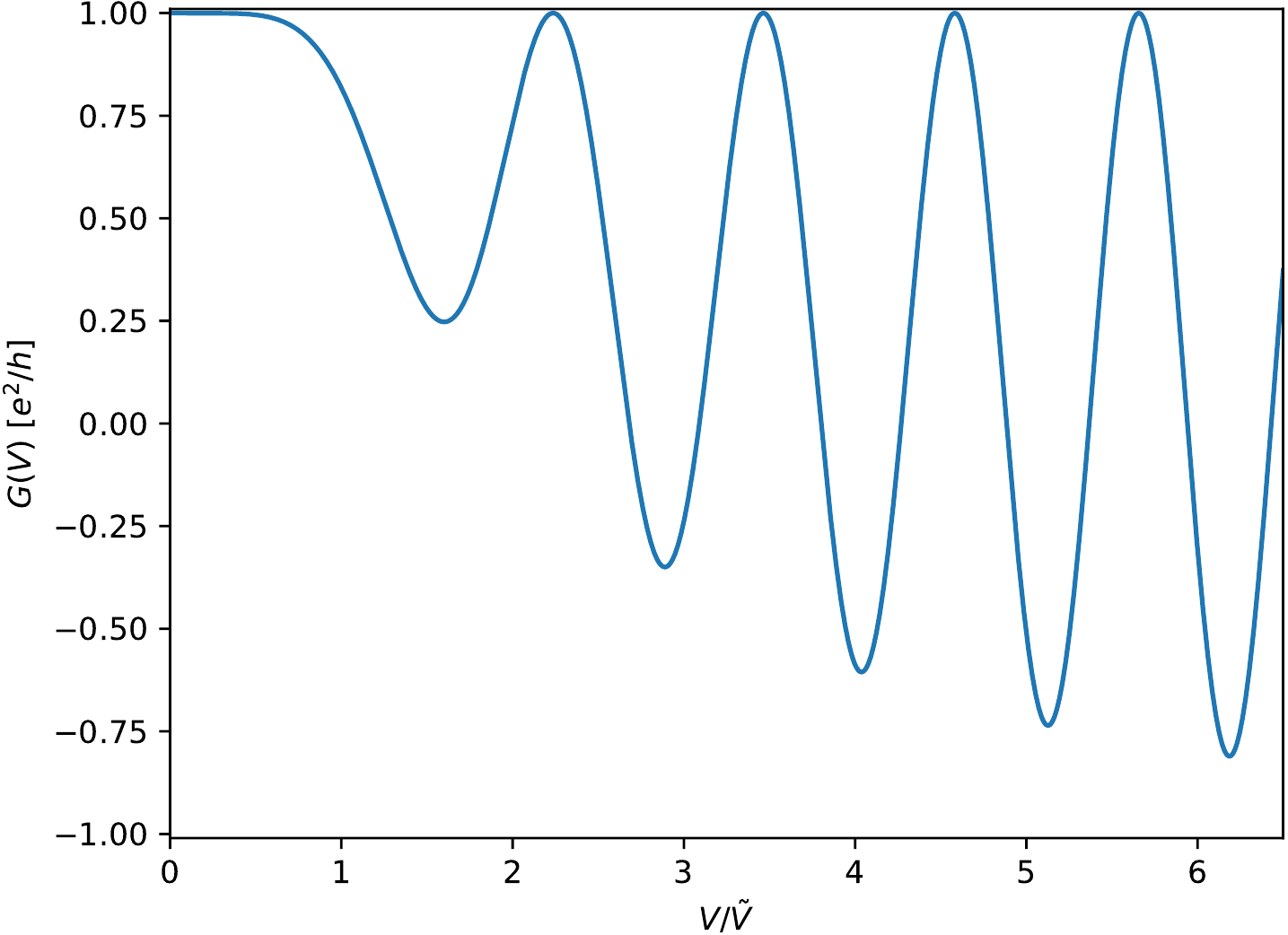}
    \caption{Vortex-free conductance versus $V/\tilde{V}$, with $\tilde{V}$ given by Eq.~\eqref{eq:til-V} and $V_0 = 2\ \tilde V$ [Eq.~\eqref{eq:f(E)}]. Oscillations are suppressed for small bias voltages $V/\Tilde{V}\lesssim 1$, as here the pairing amplitude cannot overcome the energy difference between opposite-momentum states created by the effective chemical potential $\mu_{\rm sc}$ (which is non-zero due to $V_0 \neq 0$). 
    In contrast, for $V/\Tilde{V} \gg 1$, Cooper pairs form efficiently and the conductance oscillates approximately sinusoidally.
    }
    \label{fig:no_vortex}
\end{figure}

With $\mu_{\rm sc}$ nonzero, $\gamma_1$ and $\gamma_2$ couple and thus no longer form the natural basis.
In this regime, kinematic constraints tend to obstruct Andreev processes.  
The pairing term in Eq.~\eqref{eq:vortex_free_Hdelta} favors Cooper pairing $\psi(k)$ and $\psi(-k)$ with amplitude $\propto \Delta_{\rm sc} k$, where $k$ is a momentum. 
Due to the edge state's chirality however, $\mu_{\rm sc} \neq 0$ pushes this pairing process off resonance, rendering Cooper pairing ineffective at energy scales for which $\mu_{\rm sc} \gtrsim \Delta_{\rm sc} k$.  
Incident electrons are then largely unaffected by the superconductor and exit as electrons with near-unity probability.  
The zero-temperature conductance at ${\mu_{\rm sc} \gtrsim (\Delta_{\rm sc}/v)V}$ is then simply $G(V) \approx g_0$.  

To address the crossover between these extremes, Appendix~\ref{appendix:diagonalize} diagonalizes the piece-wise constant Hamiltonian.  
With the wavefunctions in hand, one can extract the conductance using standard scattering matrix formalism.  
The scattering matrix,
\begin{equation}\label{eq:S_hole_electron}
       S = \begin{pmatrix}
        S_{\rm ee} & S_{\rm eh}\\
        S_{\rm he} & S_{\rm hh} 
        \end{pmatrix} ,
\end{equation}
relates incoming and outgoing electron and hole amplitudes according to
\begin{equation}\label{eq:relate_outgoing_ingoing}
        \begin{pmatrix}
        A_{\rm e}^{\rm out} \\
        A_{\rm h}^{\rm out} 
        \end{pmatrix} = S 
        \begin{pmatrix}
        A_{\rm e}^{\rm in} \\
        A_{\rm h}^{\rm in}
        \end{pmatrix}.
\end{equation}
In this framework, the zero-temperature conductance is
\begin{equation}\label{eq:conductance_formula}
    G(V) = g_0 (|S_{\rm ee}|^2 - |S_{\rm eh}|^2),
\end{equation}
where on the right side the scattering matrix elements are evaluated at energy $E = V$.

The bulk of the problem is then to derive the scattering matrix $S$.
Denoting the scattering matrix without vortices by $S_0$ to distinguish it from the general scattering matrix considered later, we find that 
\begin{equation}\label{eq:scattering_matrix_orthogonal_no_vortex}
    S_0(x_f-x_i) = e^{i\omega} \mathcal{O}^T D(x_f-x_i) \mathcal{O},
\end{equation}
where $\mathcal{O}$ is an orthogonal matrix rotating into the energy basis of the proximitized region and the diagonal matrix 
\begin{equation}
    D(x_f-x_i) = 
        \begin{pmatrix}
        e^{ik_+(x_f-x_i)} & 0 \\
        0 & e^{ik_-(x_f-x_i)} \label{eq:transmission_matrix}
        \end{pmatrix}
\end{equation}
describes the wavefunction propagation within the superconducting region $x_i<x<x_f$. 
The overall phase factor $e^{i\omega}$ is unimportant for the transport properties considered here and will be dropped hereafter.  
Momenta $k_\pm$ appearing in Eq.~\eqref{eq:transmission_matrix} are given by
\begin{equation}
    k_{\pm} = \frac{E[v\pm \Delta_{\rm sc} f(E)]}{v^2-\Delta_{\rm sc}^2},
    \label{kpm}
\end{equation}
where we have defined 
\begin{align}\label{eq:f(E)}
    f(E) &= \sqrt{1+\left(\frac{V_0}{E}\right)^2}, & V_0 &= \frac{\sqrt{v^2-\Delta_{\rm sc}^2}}{\Delta_{\rm sc}}\mu_{\rm sc}. 
\end{align}
In the limit  $\mu_\text{sc}=0$, the orthogonal matrix reduces to
\begin{equation}
    \mathcal{O} = \frac{1}{\sqrt{2}}
    \begin{pmatrix}
     1 & 1\\
     -1 & 1
    \end{pmatrix}
\end{equation}
indicating that the decoupled quasiparticles are Majorana fermions in agreement with Eq.~\eqref{eq:Hamiltonian_gamma}.
The derivation of the vortex-free scattering matrix and the explicit expression of $\mathcal{O}$ at $\mu_{\rm sc} \neq 0$ can be found in Appendix~\ref{appendix:vortex_free_S}.

The vortex-free, zero-temperature conductance obtained from the scattering matrix is 
\begin{equation}\label{eq:vortex_free_conductance}
    G(V) = g_0 \left( \left[1-f(V)^{-2}\right] + f(V)^{-2} \cos[\delta\phi(V) f(V)] \right).
\end{equation}
Figure~\ref{fig:no_vortex} plots $G(V)$ versus $V/\tilde V$ assuming $V_0 = 2\ \tilde V$.  Consistent with the intuition laid out earlier, at $V \lesssim V_0$ we obtain $G(V) \approx g_0$, while at larger $V$ pronounced oscillations develop that conform approximately to Eq.~\eqref{G_periodic} at $V \gg V_0$.

\section{Single-vortex problem}\label{sec:single-vortex}

We are now prepared to discuss the case in which the QH edge state hybridizes with a single vortex in the parent superconductor located near edge coordinate $x_1$.  For this setup, we modify the Hamiltonian to 
\begin{equation}
    H = H_0 + H_\Delta + H_v + H_{\psi-v}.
    \label{singlevortexH}
\end{equation}
The first two terms are the same as in Eqs.~\eqref{eq:vortex_free_H0} and \eqref{eq:vortex_free_Hdelta} (but recall that we traded in $\tilde \psi$ for $\psi$ fermions).  
The third term,
\begin{equation}
    H_v = \sum_{n = 0 }^{n_{\rm max}} \epsilon \left(n+\frac{1}{2}\right) a_n^{\dagger}a_n, \label{eq:Hv}
\end{equation}
describes the Caroli-de Gennes-Matricon sub-gap states bound to the vortex~\cite{Caroli64}, where $a_n$ is a fermion annihilation operator associated with the $n^{th}$ sub-gap level.  Equation~\eqref{eq:Hv} assumes a non-degenerate, harmonic-oscillator-like spectrum with energy spacing $\epsilon$.  Note the absence of a zero-energy vortex level as appropriate for the case of a gapped, non-topological parent superconductor considered here.  The level spacing scales as $\epsilon \sim \Delta_{\rm parent}^2/E_F$ with $\Delta_{\rm parent}$ and $E_F$ respectively denoting the parent superconductor's pairing gap and Fermi energy.  For conventional systems with $\Delta_{\rm parent}/E_F \ll 1 $, a very small spacing $\epsilon \sim 1$\,mK is expected, in which case the upper limit $n_{\rm max}$ on the sum in Eq.~\eqref{eq:Hv} can be in the thousands.  
The last term, 
\begin{equation}\label{eq:Hint}
    H_{\psi-v} =  \sum_{n = 0}^{n_{\rm max}} \left[t a_n \psi(x_1) + t' a_n^{\dagger}\psi(x_1) + h.c.\right], 
\end{equation}
describes tunneling between the edge and vortex states.  
We assume that the edge couples to the vortex levels only at position $x_1$, with amplitudes $t$ and $t'$ that are taken to be $n$-independent for simplicity.  Without loss of generality we fix $t \in \mathbb{R}$, though $t'$ can then generally be complex.  Our treatment allows for hybridization with the vortex but once again does not include dissipation.

Similar to the vortex-free case in Eq.~\eqref{eq:relate_outgoing_ingoing}, the outgoing wavefunction is expressed in terms of the incoming wavefunction using a scattering matrix. 
The single-vortex scattering matrix can be expressed as
\begin{equation}
    S = S_0(x_f-x_1) M_v S_0(x_1-x_i).
    \label{eq:single_vortex_scattering_matrix}
\end{equation}
Here $S_0$ is the vortex-free scattering matrix from Eq.~\eqref{eq:scattering_matrix_orthogonal_no_vortex} and 
$M_v$ is a new unitary matrix that accounts for the effects from the vortex levels.
The derivation of this scattering matrix and the full expression for $M_v$ appear in Appendix~\ref{appendix:general}.

\subsection{Toy limit}
\label{toylimit}

To gain physical insight, it is helpful to examine the special case $\mu_{\rm sc} = 0$ and $t = t'$ at zero temperature before presenting results for the general case. 
In this limit, the Majorana fermions obtained by writing $\psi = \gamma_1 + i\gamma_2$ decouple, and the vortex states only interact with $\gamma_2$: 
\begin{equation}
    H_{\psi-v} =  2it\sum_n (a_n + a_n^{\dagger}) \gamma_2(x_1). \label{eq:single _vortex_simplified}
\end{equation}
For an incident electron with energy $E$, hybridization with the vortex causes $\gamma_2$ to pick up an \emph{additional} relative phase of
\begin{equation}
    e^{i\theta(E)} = \frac{v+\Delta_{\rm sc} - it^2\sum_{n = 0}^{n_{\rm max}}\frac{2E}{E^2-\epsilon^2\left(n+\frac{1}{2}\right)^2}}{v+\Delta_{\rm sc} + it^2\sum_{n = 0}^{n_{\rm max}}\frac{2E}{E^2-\epsilon^2\left(n+\frac{1}{2}\right)^2}}
    \label{eq:thetaE}
\end{equation}
compared to $\gamma_1$; see App.~\ref{appendix:general}.  The single-vortex conductance correspondingly becomes
\begin{equation}\label{eq:simplified_conductance}
    G(V) = g_0 \cos \left[\delta\phi(V) + \theta(V)\right]
\end{equation}
with $\delta \phi(V)$ defined in Eq.~\eqref{deltaphi}.

Whenever the incident energy is resonant with a vortex level, i.e., at $E = \epsilon(n+1/2)$, Eq.~\eqref{eq:thetaE} yields $\theta(E) = \pi$---producing a sign change in the conductance compared to the vortex-free case.  The width of these resonances is set by the dimensionless ratio
\begin{equation}
    \alpha = \frac{t^2}{\epsilon v}
\end{equation}
in the $\Delta_{\rm sc} \ll v$ limit.  Moreover, the spacing $\epsilon$ between resonances will in practice be vastly smaller than the energy scale $\tilde V$ [Eq.~\eqref{eq:til-V}] that sets the period of conductance oscillations in the absence of a vortex.  Hybridization with the vortex levels thus generates wild oscillations that can dramatically alter the structure of the vortex-free conductance.  These finely spaced oscillations highlight the importance of finite-temperature effects, to which we now turn.

\subsection{Finite temperature}

Given the zero-temperature conductance $G(E,T=0)$ considered so far, 
the finite-temperature conductance follows as~\cite{Dolcini09}
\begin{equation}\label{eq:G-finite-T}
    G(V,T) = \int dE\,  \frac{d}{dV}\left[f_S(E,V)-f_D(E,0)\right] G(E,T=0) .
\end{equation}
Here 
\begin{equation}
    f_S(E,V)-f_D(E,0) = \frac{1}{1+e^{\frac{E-V}{T}}}-\frac{1}{1+e^{\frac{E}{T}}}
\end{equation}
is the difference in Fermi-Dirac distribution functions at the source and drain.  
At zero temperature, the first derivative of the Fermi-Dirac distribution gives a Dirac-delta function; the peak widens as temperature increases, resulting in a smearing out of the zero-temperature conductance.  Throughout we are interested in the experimentally relevant regime $\epsilon \ll T \ll \tilde V$ where thermal effects efficiently average over many fine vortex-induced oscillations but not the longer-period oscillations characteristic of the vortex-free limit.

\begin{figure}[H]
  \centering
    \includegraphics[width=\columnwidth]{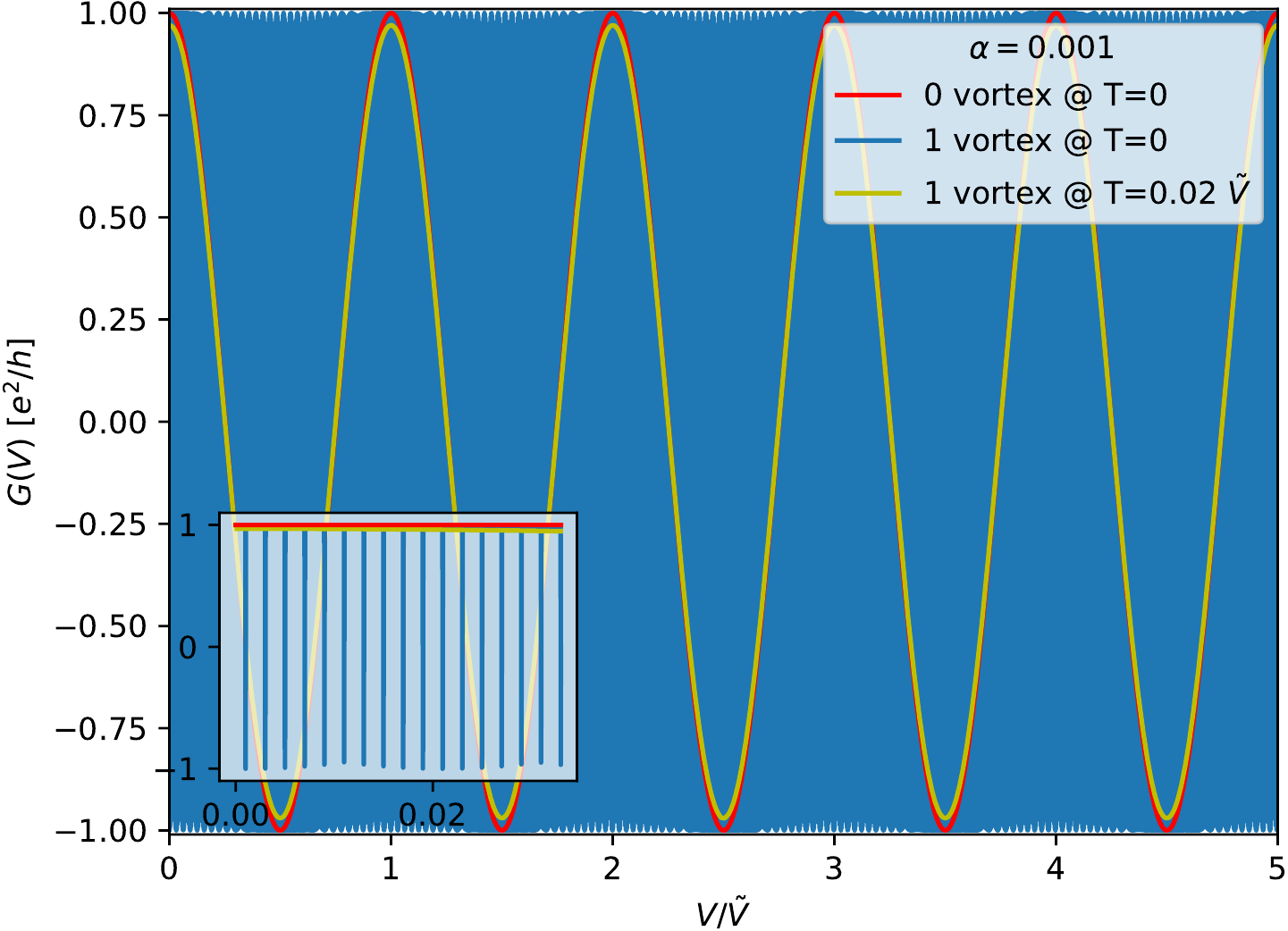}
    \includegraphics[width=\columnwidth]{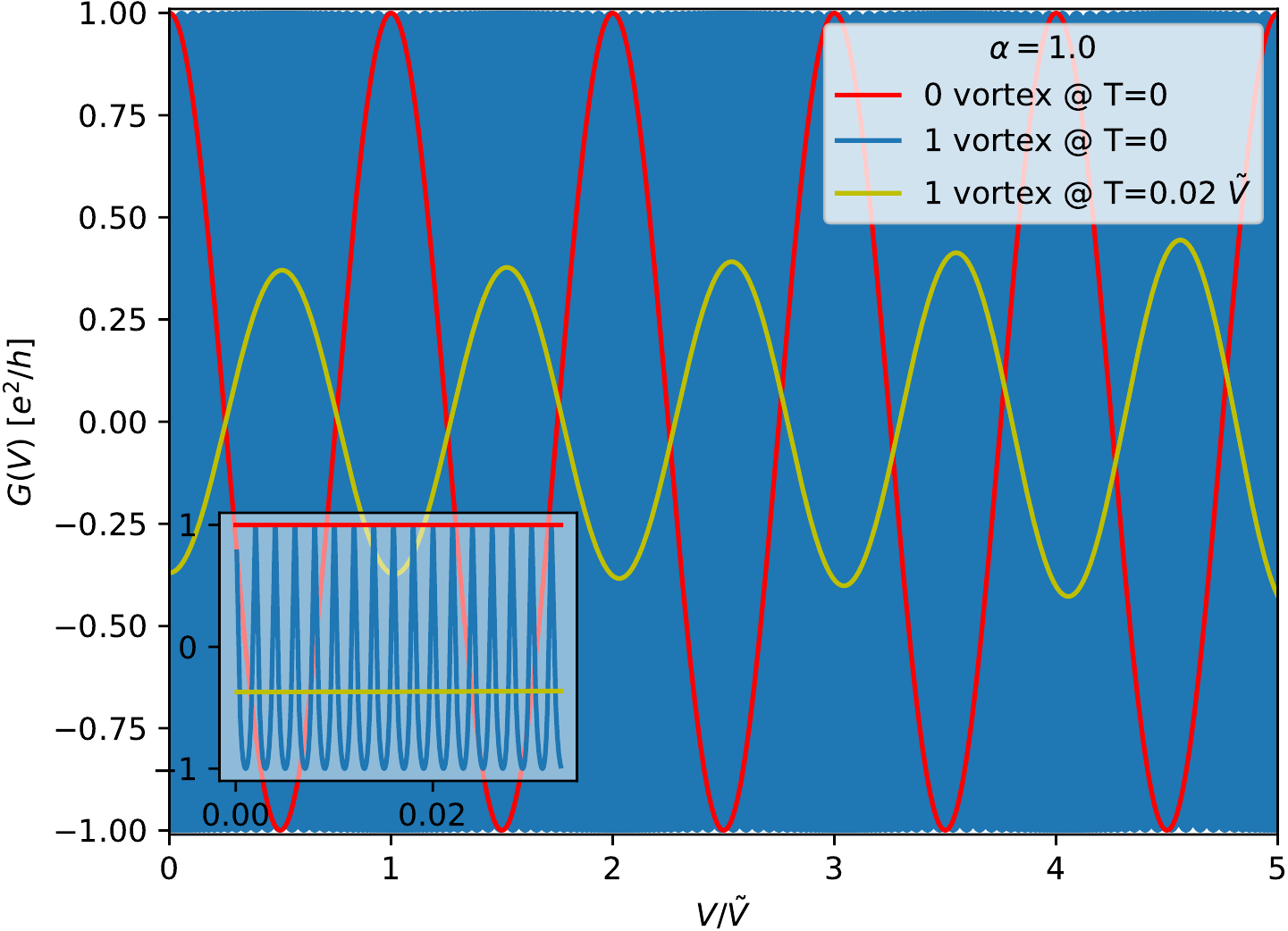}
    \includegraphics[width=\columnwidth]{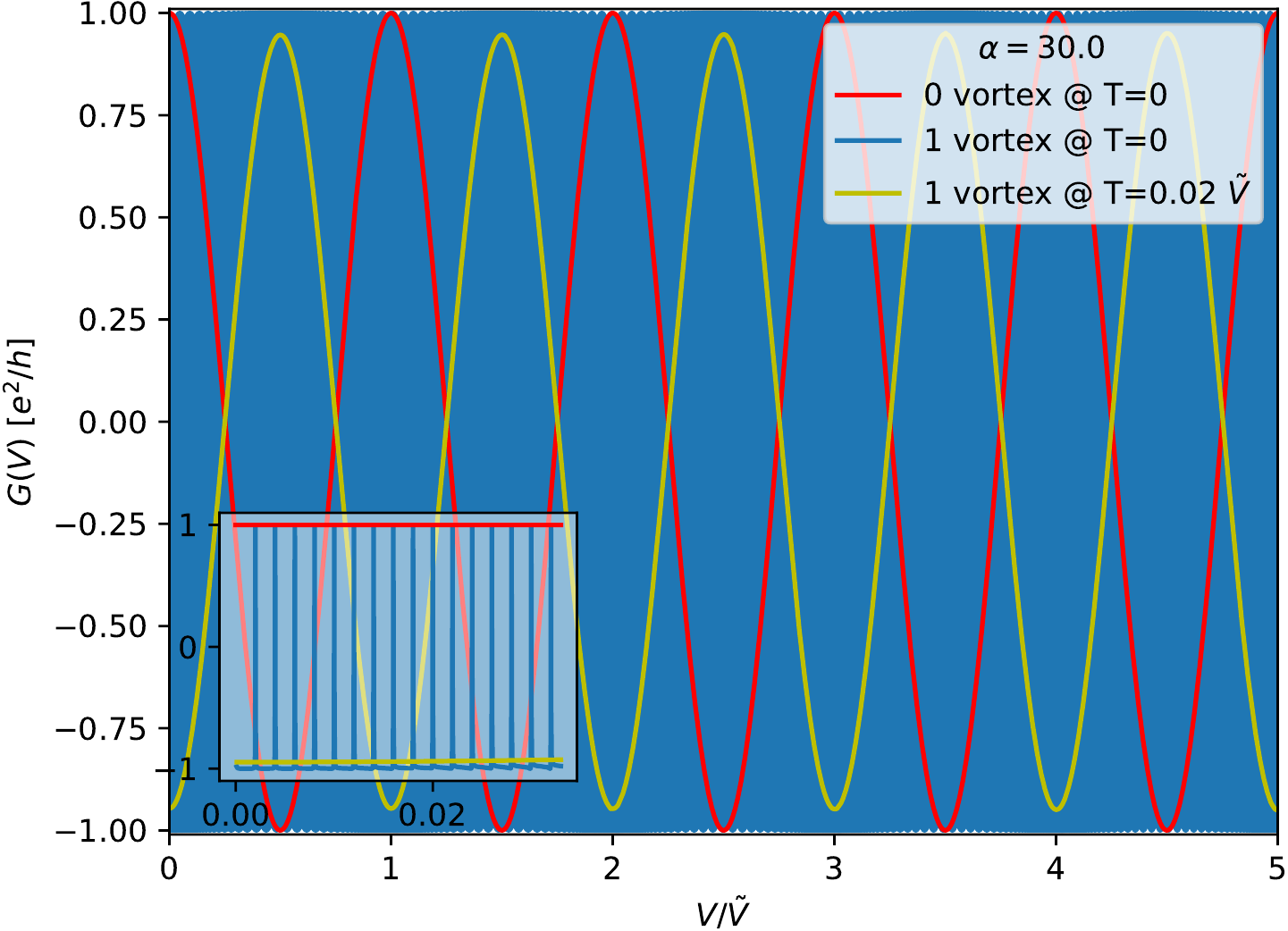} 
\caption{Single-vortex conductance versus bias voltage in the toy limit (Sec.~\ref{toylimit}) for varying resonance widths.  See main text for parameters.  Insets: zoom-in of the conductance near zero bias.  
}
\label{fig:single_vortex_all_toy}
\end{figure}

For the toy limit of the vortex-hybridization problem considered above, the structure of the thermally smeared conductance depends on the resonance width $\alpha$ as illustrated in Fig.~\ref{fig:single_vortex_all_toy}.  The panels correspond to $\alpha = 0.001$ (top), $1$ (middle), and $30$ (bottom) with parameters $v/\Delta_{sc} = 10$,  $\epsilon = 0.002 \tilde V$,  and $n_{\rm max} = 4000$.  (In the toy limit considered here, the vortex position $x_1$ drops out.)  Red curves show the vortex-free conductance versus $V/\tilde V$ as a reference, while the blue and green curves respectively present the single-vortex conductance at $T = 0$ and $T = 0.02\tilde V$ ($\epsilon = 0.1 \ T$ at the latter temperature).  Insets zoom in on the voltage window near zero.  In the narrow-resonance limit $\alpha \ll 1$ (top panel), thermal smearing approximately reproduces the vortex-free conductance since electrons `see' the vortex levels only in tiny incident-energy windows.  In the wide-resonance limit $\alpha \gg 1$ (bottom panel), the Majorana fermion $\gamma_2$ nearly always experiences a vortex-induced $\pi$ phase shift compared to $\gamma_1$; accordingly, the finite-temperature conductance is approximately negated compared to the vortex-free conductance.  
As the system crosses over from the narrow-resonance to the wide-resonance regime, the thermally averaged conductance exhibits oscillations that are initially suppressed in amplitude but eventually revive and become out-of-phase with the vortex-free conductance as illustrated in the middle panel.

\subsection{General case}

\begin{figure}[t]
\centering
\includegraphics[width=\columnwidth]{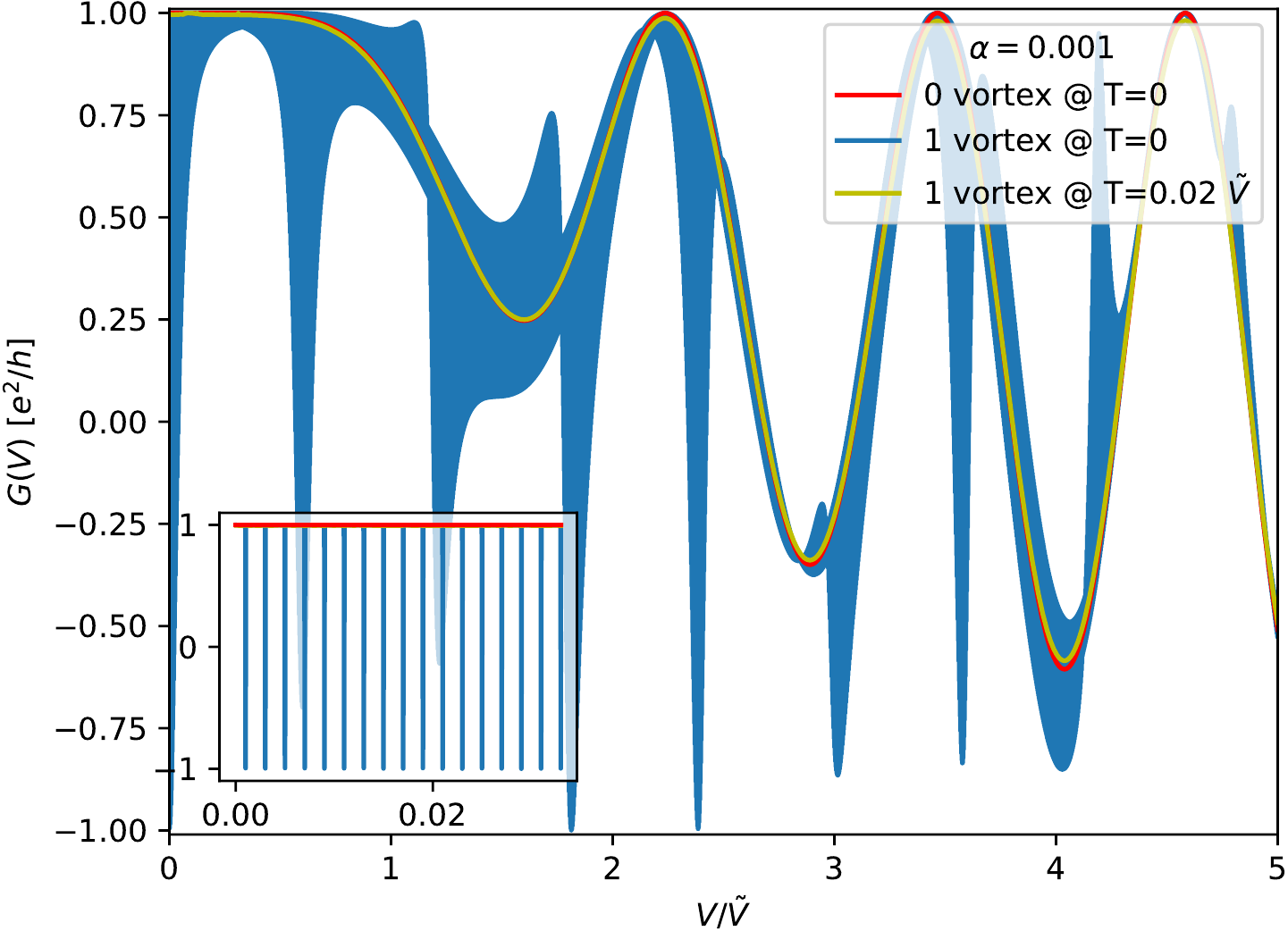}
\includegraphics[width=\columnwidth]{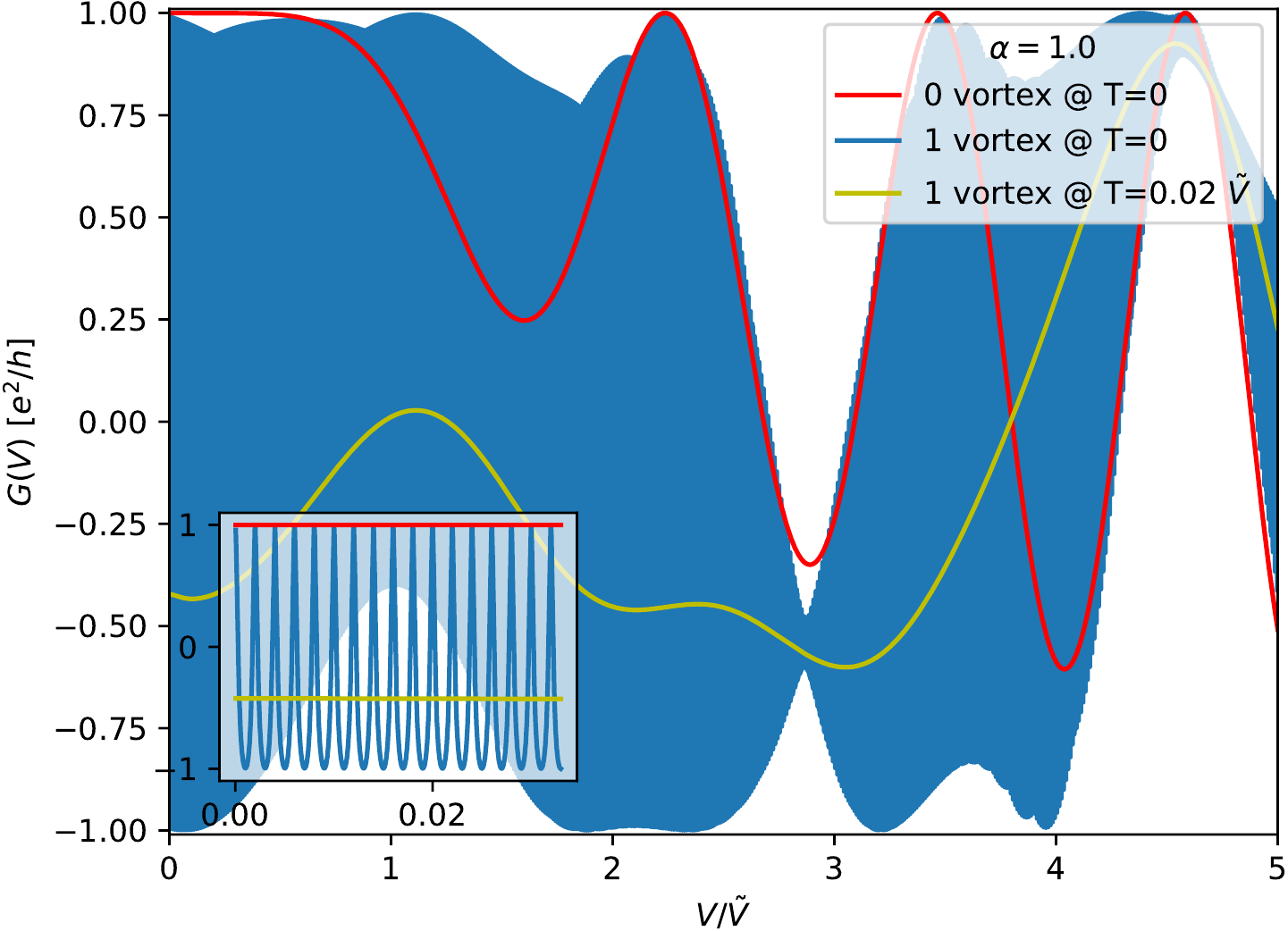}
 \includegraphics[width=\columnwidth]{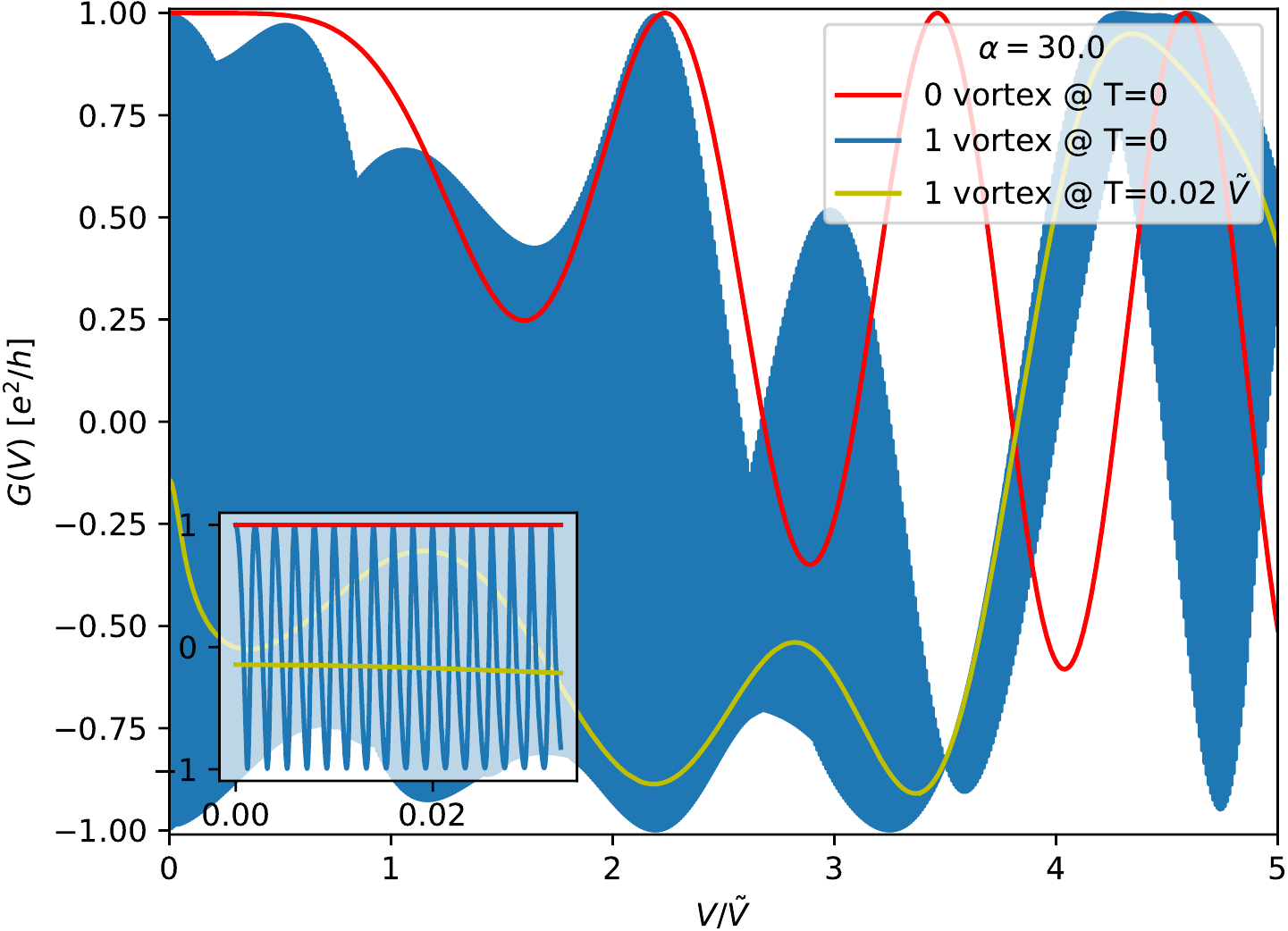} 
\caption{Same as Fig.~\ref{fig:single_vortex_all_toy}, but with parameters $\mu_{\rm sc} = 0.2 \tilde V$ and $t' = 0.95 e^{i 2\pi/3}t$ that place the system away from the toy limit. The vortex is located at $x_1 = (x_f-x_i)/3$.
}
\label{fig:single_vortex_all}
\end{figure}

Figure~\ref{fig:single_vortex_all} plots the single-vortex conductance versus bias voltage $V/\tilde V$ for the more general case with scattering matrix given by Eq.~\eqref{eq:single_vortex_scattering_matrix}.
Parameters are the same as for Fig.~\ref{fig:single_vortex_all_toy} except that now $\mu_{\rm sc} = 0.2 \ \Tilde{V}$ and $t' = 0.95 e^{i\frac{2\pi}{3}}\ t$, placing the system away from the toy limit examined above.  Additionally, we fix the vortex position---which is no longer arbitrary---to $x_1 = (x_f-x_i)/3$.
Just as for the toy limit, the finite-temperature conductance for the narrow-resonance case closely tracks the vortex-free conductance. The intermediate- and wide-resonance cases behave more nontrivially compared to the toy limit, though both display an important characteristic: The finite-temperature conductance becomes negative over an extended voltage window near zero bias due to averaging over vortex-induced resonances, even though the vortex-free conductance is positive and nearly maximized because of kinematic constraints produced by $\mu_{\rm sc} \neq 0$. 

\begin{figure}[t]
\centering
\includegraphics[width=\columnwidth]{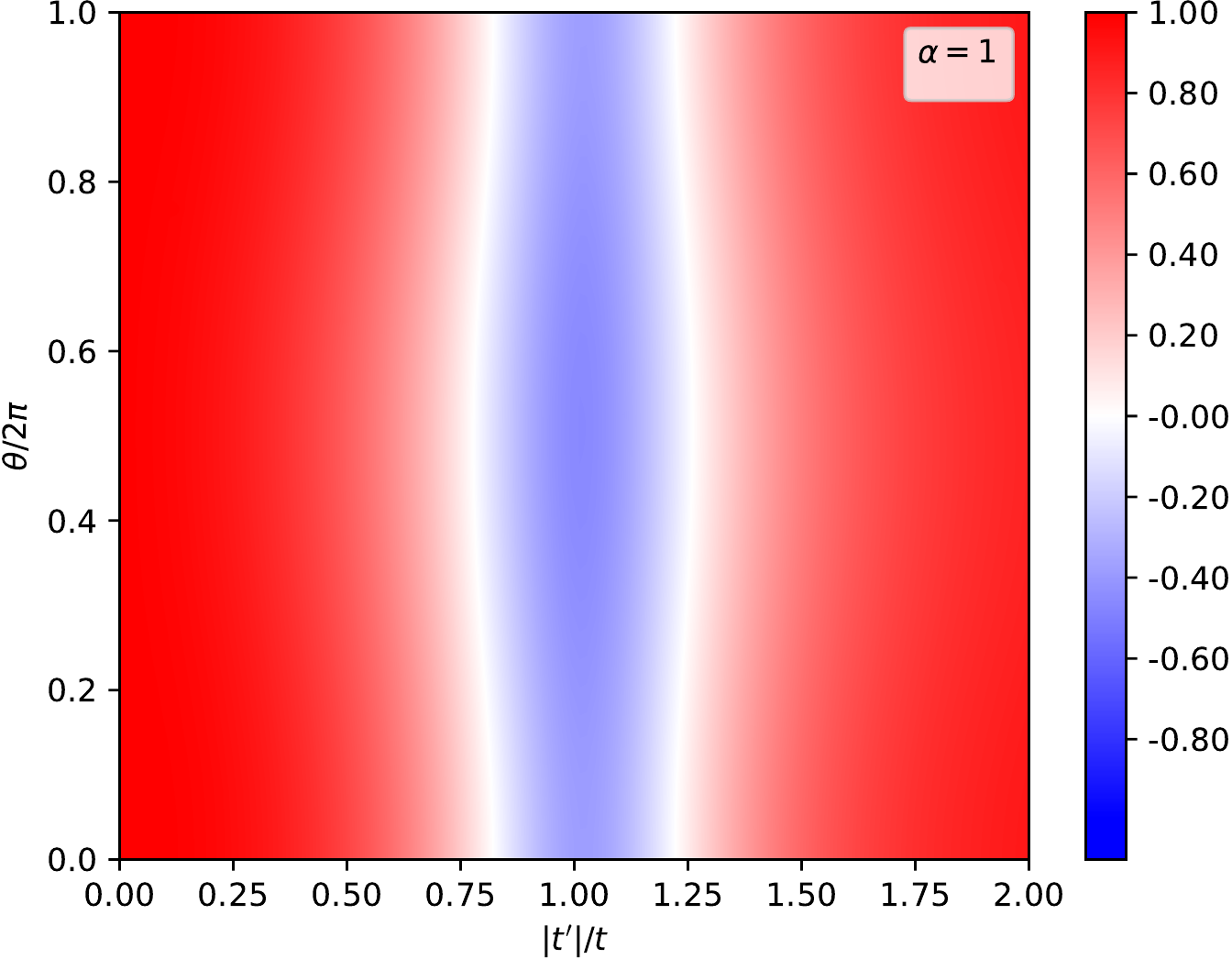}
\includegraphics[width=\columnwidth]{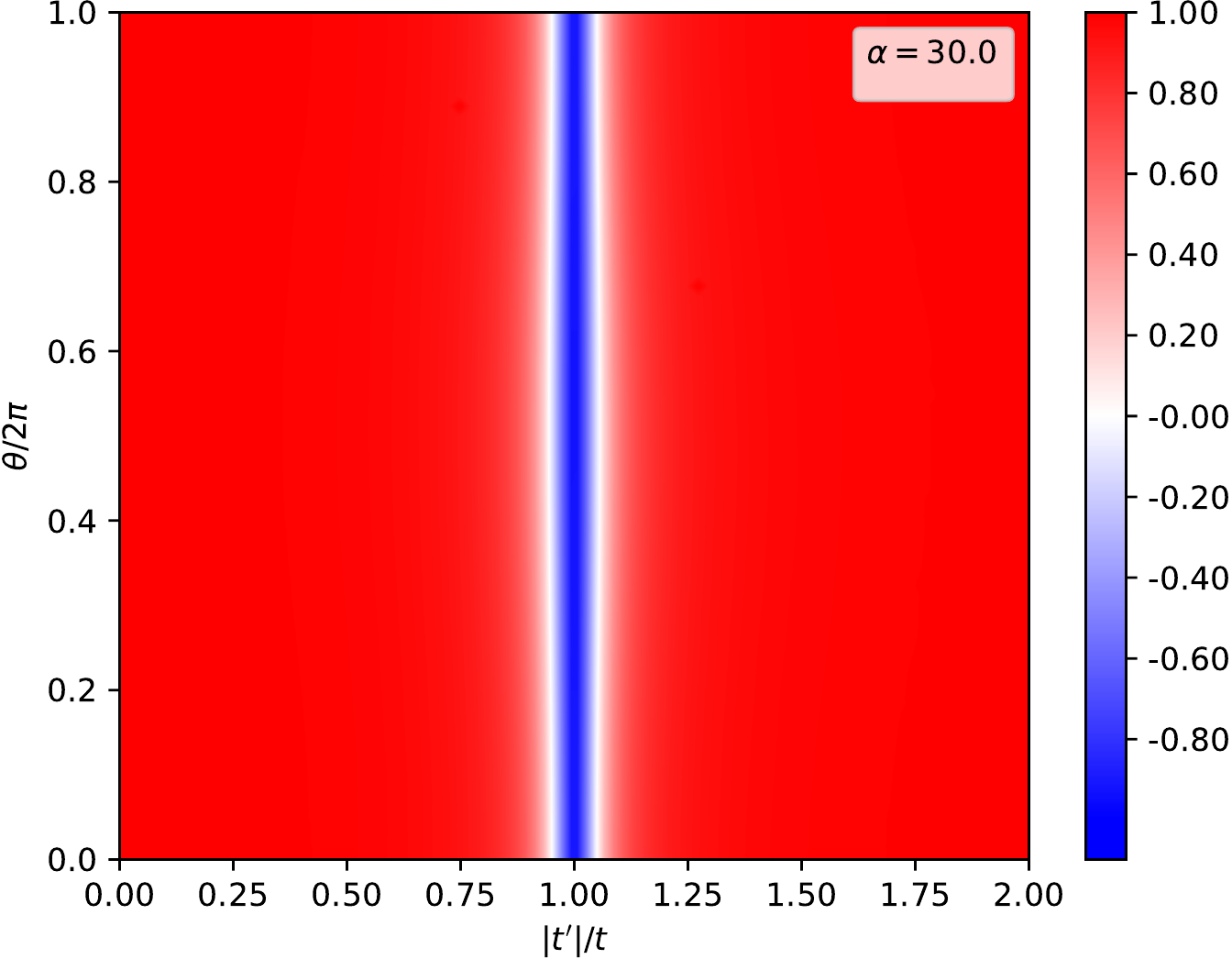}
\caption{
Finite temperature zero-bias conductance with a single vortex versus the magnitude and phase (denoted $\theta$) of $t'$. Upper and lower panels respectively correspond to the intermediate- and wide-resonance cases.  
}
\label{fig:mag_phase}
\end{figure}

To illustrate the dependence on tunneling parameters, Fig.~\ref{fig:mag_phase} shows the finite-temperature zero-bias conductance versus the magnitude and phase of $t'$.  The top and bottom panels correspond to the intermediate- and wide-resonance cases, with all parameters aside from $t'$ the same as for Fig.~\ref{fig:single_vortex_all}. (We do not show results for the narrow-resonance case since there the finite-temperature conductance is essentially insensitive to $t'$.)  Evidently the conductance depends much more sensitively on the magnitude of $t'$ compared to its phase.  In particular, negative zero-bias conductance sets in only for $|t'|/t$ sufficiently close to one, with a window that (at least for these generic parameters) decreases in the wide-resonance regime.

\section{Multi-vortex problem}\label{sec:multi-vortex}

\begin{figure*}[t]
  \centering
    \includegraphics[width=\columnwidth]{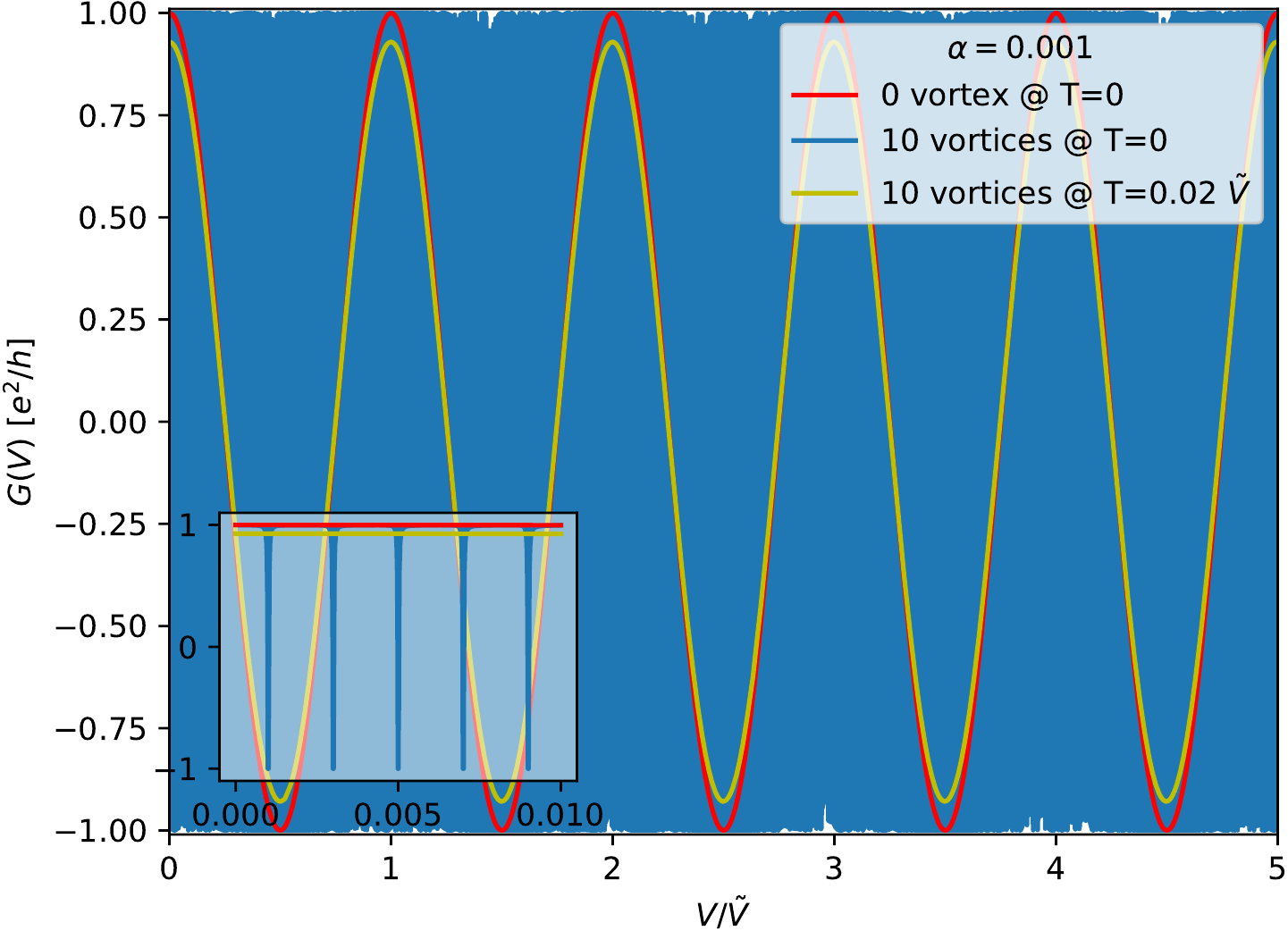}
    \includegraphics[width=\columnwidth]{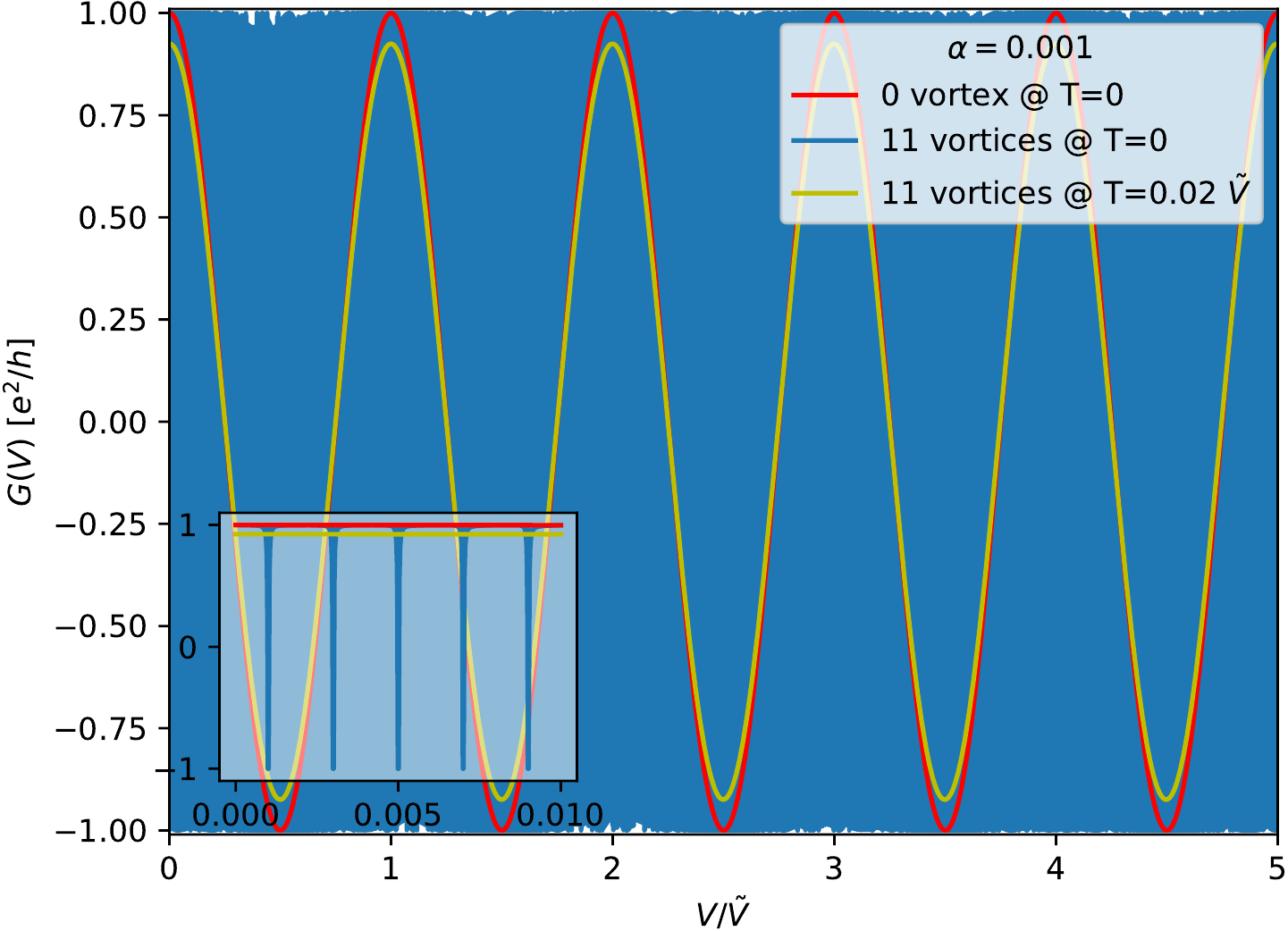}
    \includegraphics[width=\columnwidth]{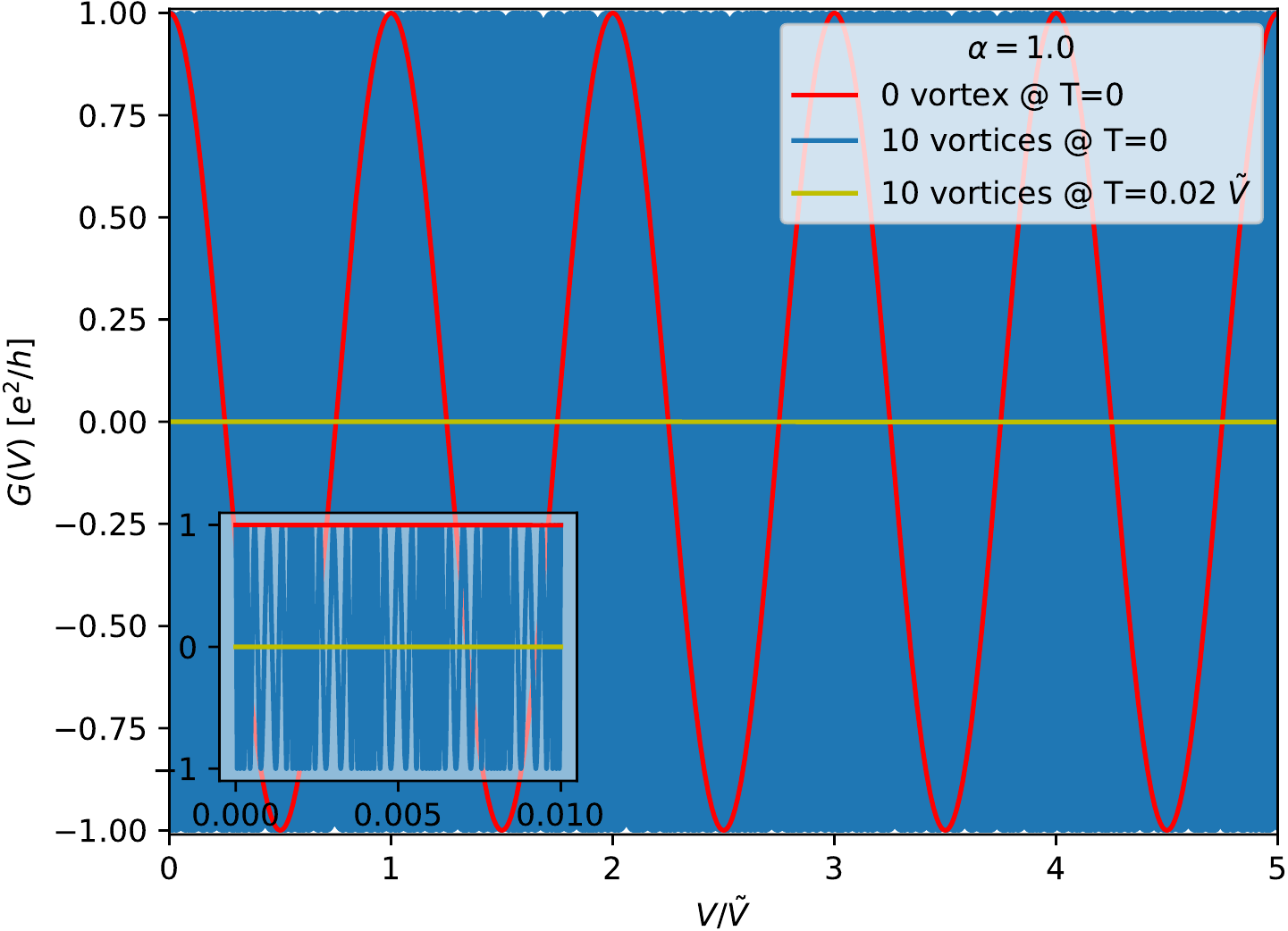}
    \includegraphics[width=\columnwidth]{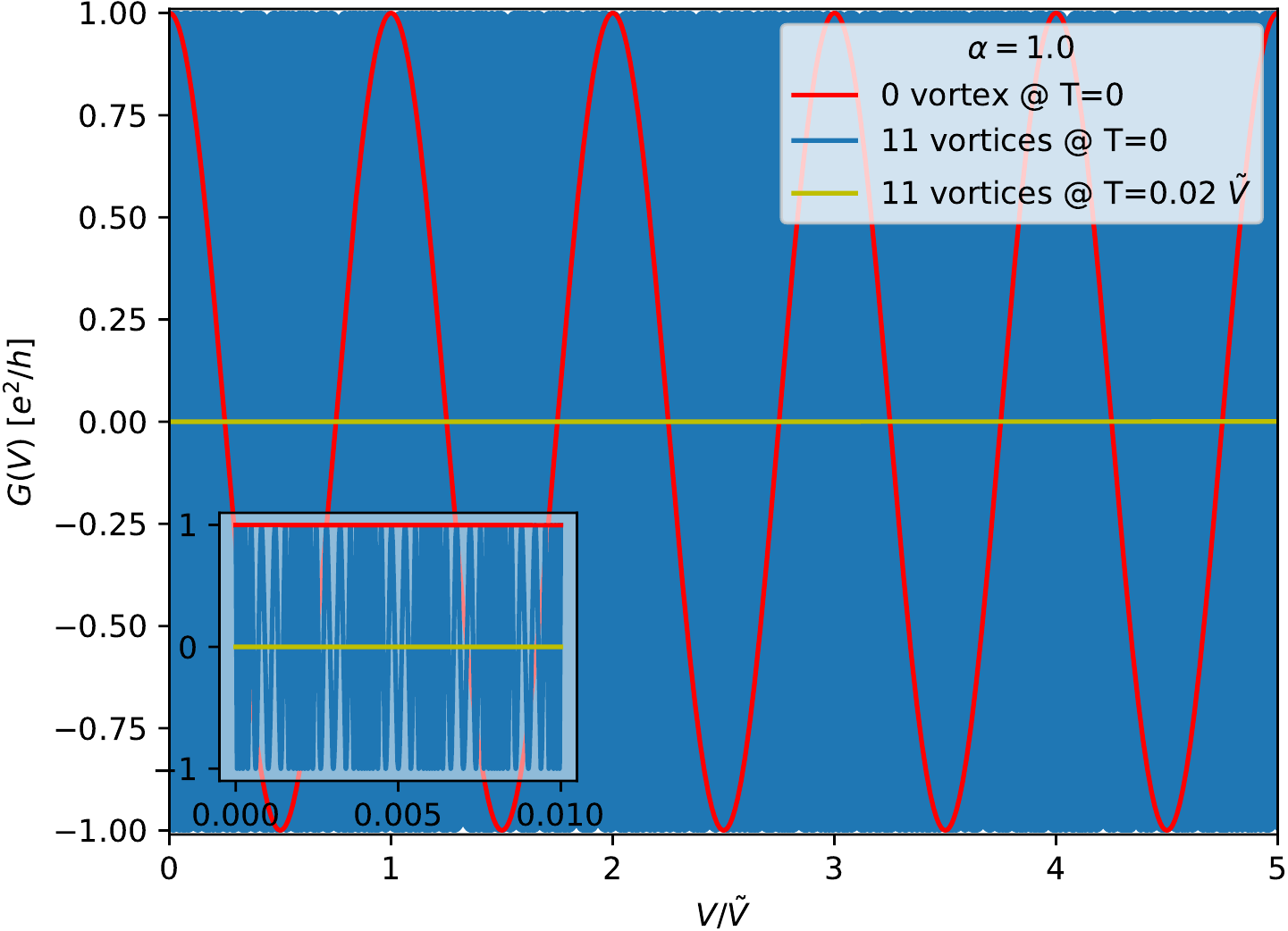}
    \includegraphics[width=\columnwidth]{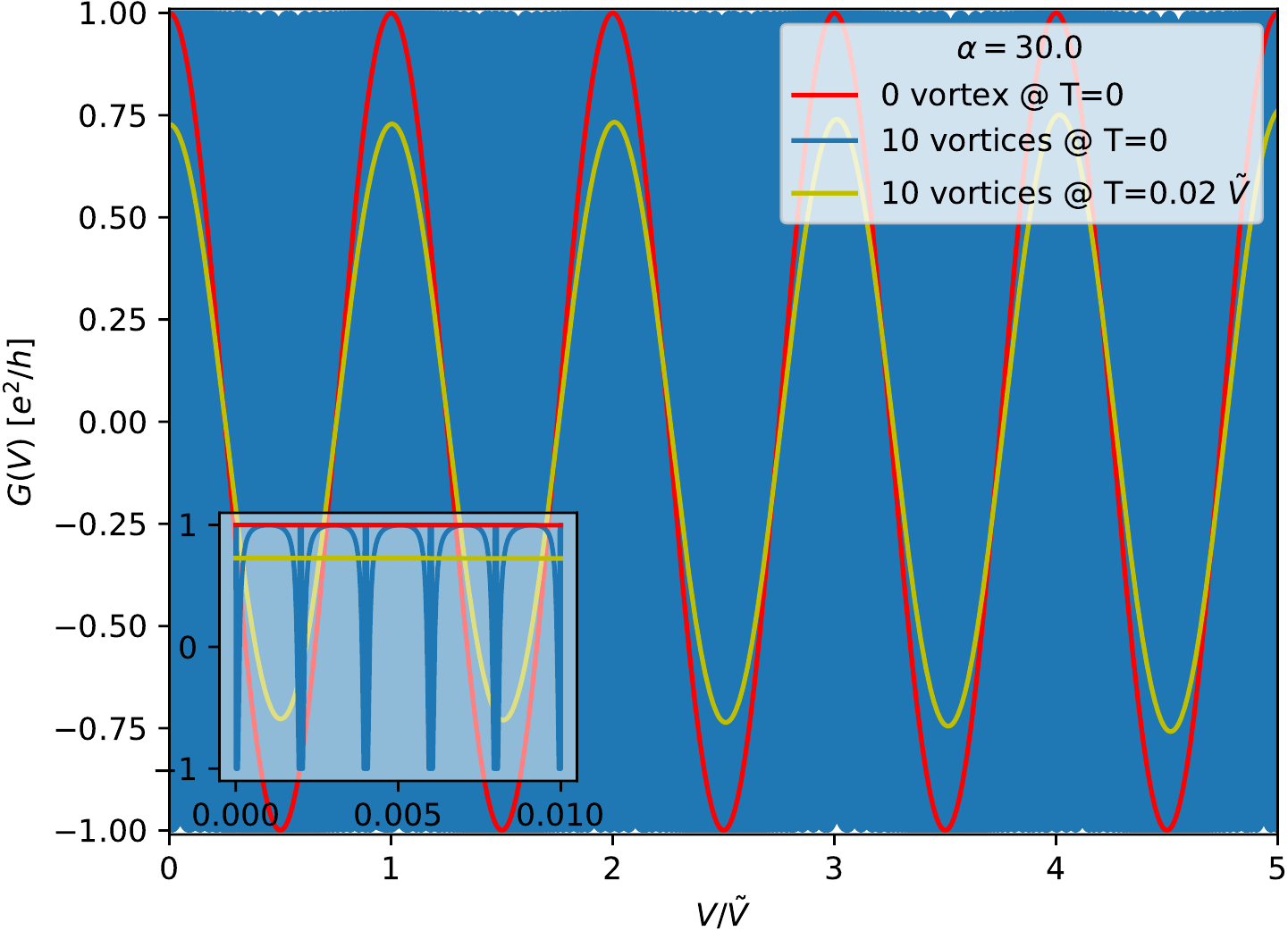}
     \includegraphics[width=\columnwidth]{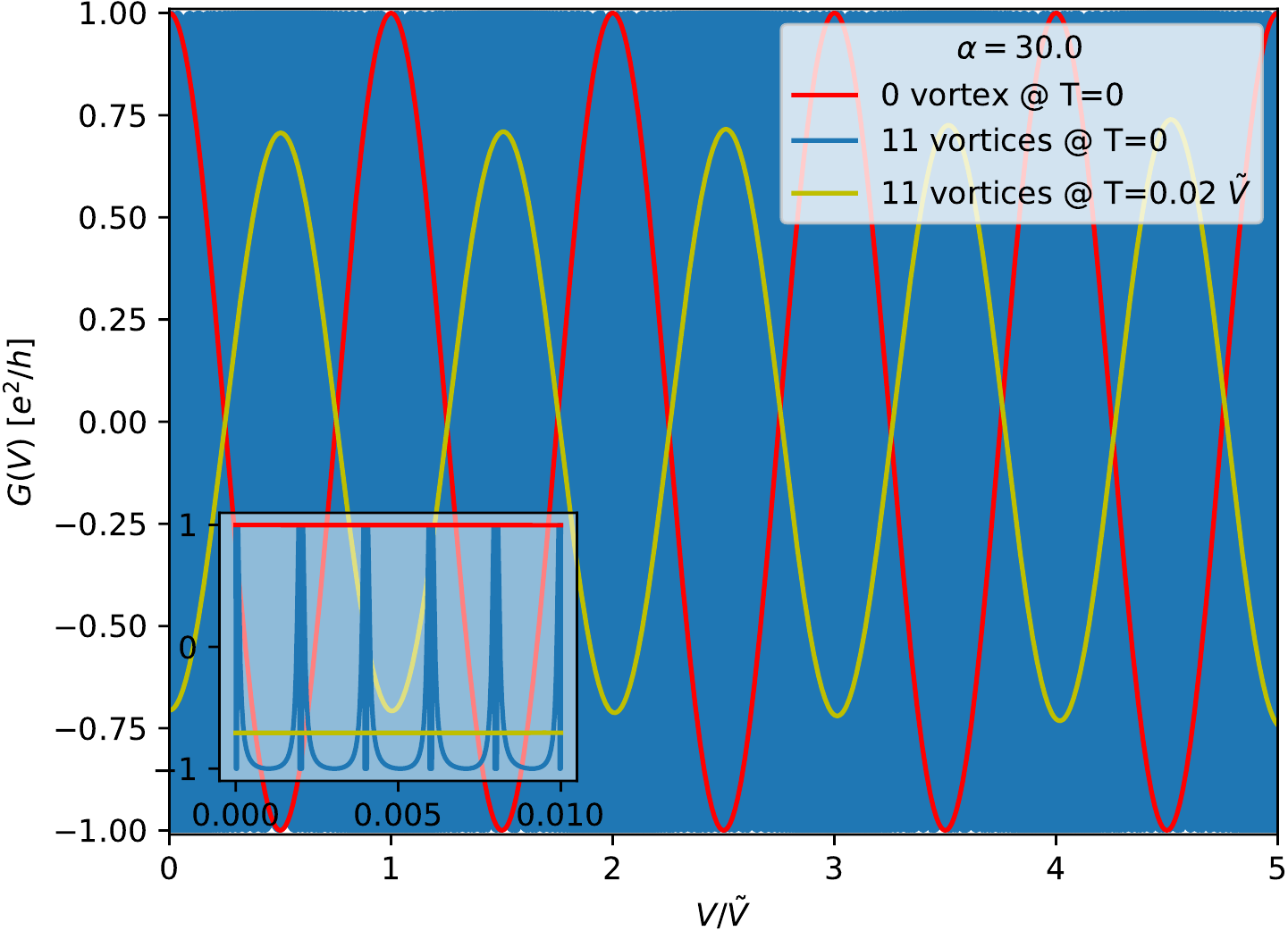} 
\caption{Same as Fig.~\ref{fig:single_vortex_all_toy}, but with $10$ (left column) and $11$ (right column) identical vortices.}
\label{fig:10_11_vortices_all_toy}
\end{figure*}

The single-vortex Hamiltonian studied in the previous section straightforwardly generalizes to the case where chiral edge electrons encounter $N_v>1$ vortices at coordinates $x_1,\ldots,x_{N_v}$ along the proximitized region.  In particular, the multi-vortex Hamiltonian takes the same form as in Eq.~\eqref{singlevortexH} but now with 
\begin{equation}
      H_v = \sum_{n = 0}^{n_{\rm max}} \sum_{j = 1}^{N_v} \epsilon \left(n+\frac{1}{2}\right) a_{n,j}^{\dagger}a_{n,j} \label{eq:Hv_multi} 
\end{equation}
and
\begin{equation}
    H_{\psi-v} =  \sum_{n = 0}^{n_{\rm max}} \sum_{j = 1}^{N_v} [t_{j} a_{n,j} \psi(x_j) + t_{j}' a_{n,j}^{\dagger}\psi(x_j) + h.c.]. \label{eq:Hint_multi}
\end{equation}
Here $a_{n,j}$ is a fermion annihilation operator associated with the $n^{th}$ sub-gap level at vortex $j$; for simplicity we assumed that each vortex has the same level spacing $\epsilon$.  Additionally, $t_j, t_j'$ describe the amplitudes for tunneling between the edge and the vortex at position $x_j$.  To estimate $N_v$, we first note that vortex-edge coupling is expected to decay exponentially with the distance between a vortex and the edge, so that only the line of vortices closest to the QH-superconductor interface need to be considered.
We thus have 
\begin{equation}
    N_v \sim \sqrt{\frac{B}{\Phi_0}} \ell
    \label{Nvestimate}
\end{equation}
with $B$ the average magnetic field penetrating the superconductor, $\Phi_0 = \frac{h}{2e}$ the superconductor flux quantum, and $\ell = x_f-x_i$ the length of the superconducting region. 
For $B\sim 1$\,T and $\ell\sim 1\,\mu$m, Eq.~\eqref{Nvestimate} yields $N_v\sim 20$.

Due to the edge state's chirality, the corresponding multi-vortex scattering matrix may be decomposed in terms of a product of scattering matrices associated with each of the $N_v$ vortices:
\begin{align}\label{multi_vortex_matrix}
    S = \prod_{j=1}^{N_v}  S_0(x_{j+1}- x_j) M_{v,j} S_0(x_j - x_{j-1}).
\end{align}
Here $x_0 = x_i$ and $x_{N_v+1} = x_f$ correspond to the left and right endpoints of the proximitized region, and $M_{v,j}$ is a unitary matrix that incorporates effects of vortex $j$ (which depend on the parameters $t_j,t_j'$).

\subsection{Toy limit}\label{sec:multi-vortex-toy}

Consider the special case $\mu_{\rm sc} = 0$ and $t_j = t_j' = t \in \mathbb{R}$ analogous to the toy limit explored for the single-vortex problem in Sec.~\ref{toylimit}.  Upon writing $\psi = \gamma_1 + i \gamma_2$, we similarly find that here only $\gamma_2$ couples to the $N_v$ vortices.  Correspondingly, \emph{each} vortex generates an additional phase shift for $\gamma_2$ given by Eq.~\eqref{eq:thetaE} (which is independent of the vortex position), yielding a zero-temperature conductance
\begin{equation}\label{eq:GV-multi-toy}
G(V) = g_0 \cos \left[\delta\phi(V) + N_v\theta(V)\right]
\end{equation}
that straightforwardly extends Eq.~\eqref{eq:simplified_conductance}.  
Figure~\ref{fig:10_11_vortices_all_toy} plots Eq.~\eqref{eq:GV-multi-toy} for the same parameters as Fig.~\ref{fig:single_vortex_all_toy} with 10 and 11 vortices.
Similar to the toy-limit conductance for a single vortex, the structure of the thermally smeared conductance depends on the resonance width $\alpha$. The narrow-resonance limit $\alpha\ll 1$ (top panels) again approximately reproduces the vortex-free conductance.  The wide-resonance limit $\alpha\gg 1$ (bottom panels) either approximately reproduces the vortex-free conductance in the case of 10 vortices (left column) or picks up a $\pi$ phase shift compared to the vortex-free conductance in the case of 11 vortices (right column).  This difference indicates an even-odd effect that naturally follows from Eq.~\eqref{eq:GV-multi-toy}: whenever the incident energy is resonant with a vortex level, the argument of the cosine jumps by $N_v \pi$, hence the dependence on the parity of $N_v$.   
In the middle panels---corresponding to the crossover between the narrow- and wide-resonance limits---the thermally smeared conductance is pinned near zero for all voltages, indicating that the superconductor effectively behaves like a normal contact.  Note that thermal smearing is more efficient at suppressing the conductance in all panels of Fig.~\ref{fig:10_11_vortices_all_toy} compared to Fig.~\ref{fig:single_vortex_all_toy}.  

\begin{figure}[t]
  \begin{minipage}[b]{\columnwidth}
    \includegraphics[width=\columnwidth]{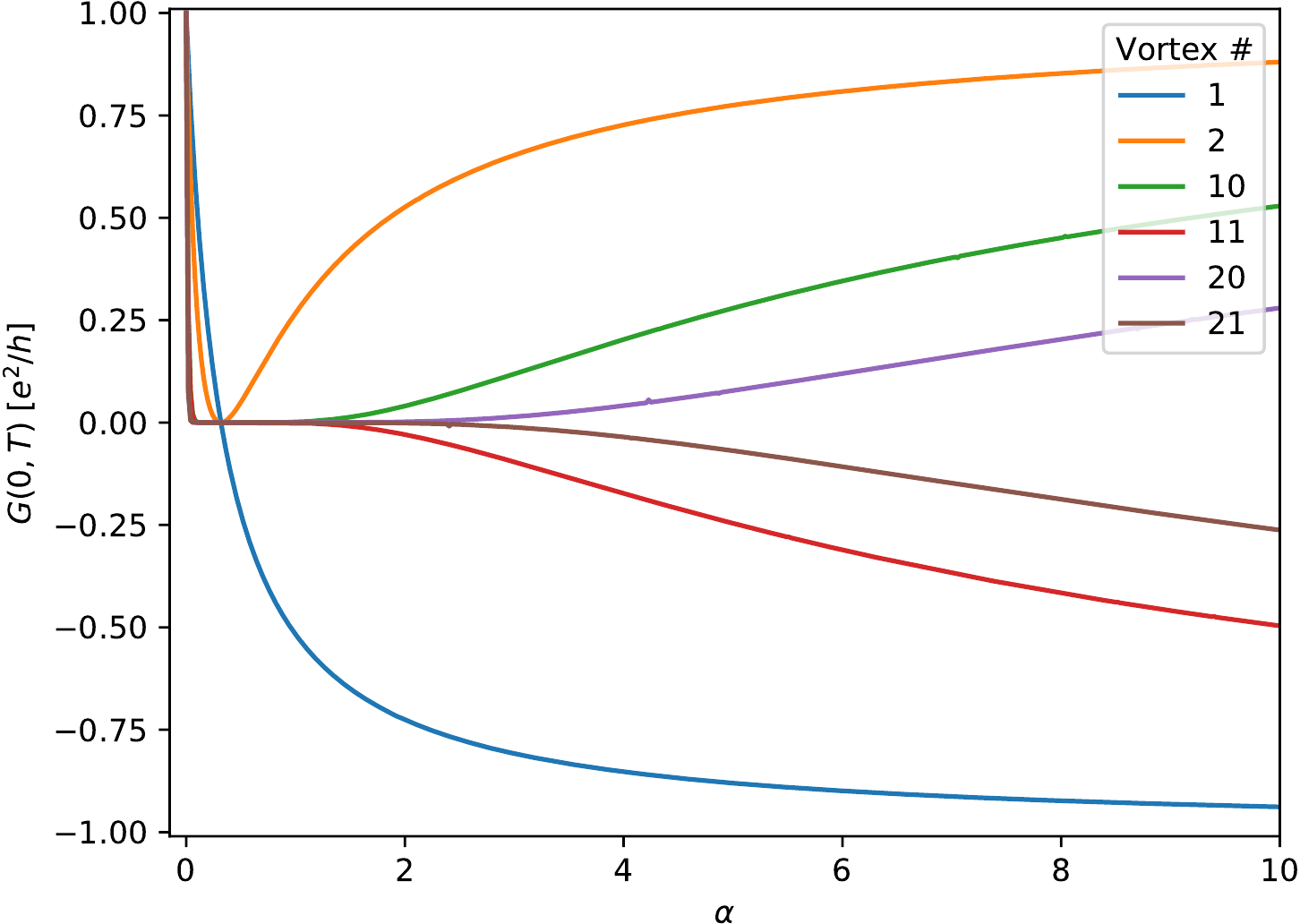}
  \end{minipage}
  \hfill
  \begin{minipage}[b]{\columnwidth}
    \includegraphics[width=\columnwidth]{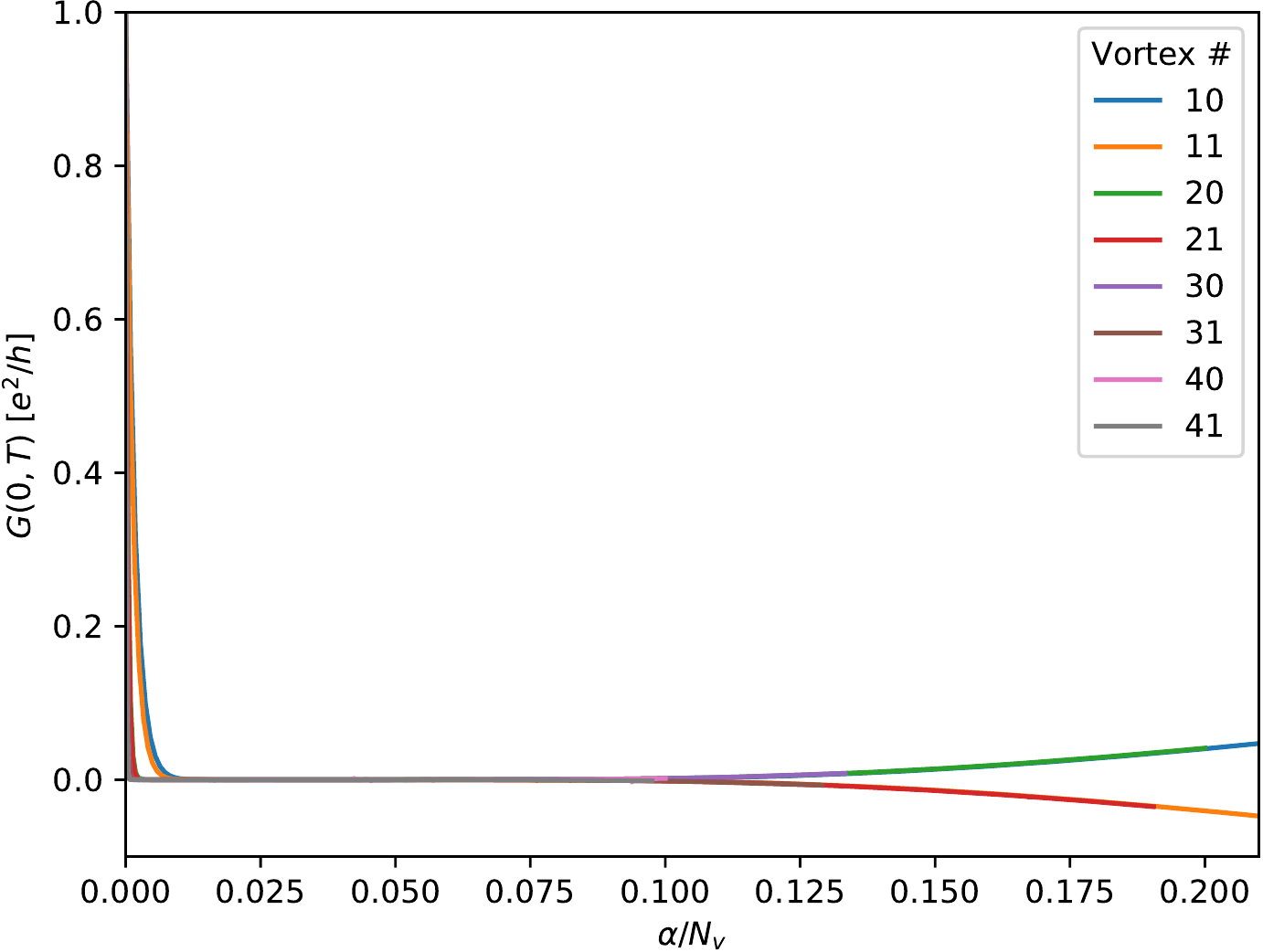}
  \end{minipage}
\caption{Top panel: Finite-temperature zero-bias conductance versus $\alpha$ with multiple vortices in the toy limit.  Data were obtained with $\Delta_{\rm sc} = 0$ and $T/\epsilon = 10$; all other parameters are the same as used for Fig.~\ref{fig:10_11_vortices_all_toy}.  Bottom panel: Finite-temperature zero-bias conductance plotted versus $\alpha/N_v$, illustrating data collapse.
}
\label{fig:conductance_alpha}
\end{figure}

For deeper insight into the interplay between the number of vortices and resonance width $\alpha$, we further examine the thermally smeared conductance at zero-bias.  In the temperature regime of interest, suppression of the zero-bias conductance below $g_0$ arises predominantly from averaging over vortex-induced oscillations (as opposed to proximity-induced pairing encoded through $\Delta_{\rm sc}$).  We therefore set $\Delta_{\rm sc} = 0$ here; dimensional analysis then indicates that the zero-bias conductance takes the form
\begin{equation}
    G(V = 0, T) = g_0 \mathcal{G}(T/\epsilon,\alpha,N_v)
    \label{eq:G-multi-0}
\end{equation}
for some scaling function $\mathcal{G}$ that depends on the remaining dimensionless parameters specified in the arguments.  Figure~\ref{fig:conductance_alpha}, top panel, plots Eq.~\eqref{eq:G-multi-0} versus $\alpha$ with $T/\epsilon = 10$ and various $N_v$.  As $N_v$ increases, three regimes become apparent.  For $\alpha$ near zero, we see $G(0,T) \approx g_0$ as expected.  As $\alpha$ increases the conductance precipitously drops and forms a plateau near zero whose width broadens as $N_v$ increases (notice that the middle panels of Fig.~\ref{fig:10_11_vortices_all_toy} sit within the plateau for $N_v = 10,11$).
At still larger $\alpha$ the system enters the wide-resonance regime, and the conductance tends toward $G(0,T) \rightarrow (-1)^{N_v} g_0$.  The bottom panel plots Eq.~\eqref{eq:G-multi-0} versus $\alpha/N_v$ with $N_v$ ranging from 10 to 41.  Excellent data collapse is observed (except at the smallest $\alpha$ regime) for both the even- and odd-$N_v$ branches.  This collapse demonstrates that the plateau region with $G(0,T) \approx 0$ populates a window of $\alpha$ that grows linearly with the number of vortices.  In other words, with more vortices the superconductor more readily behaves like a normal contact.

\subsection{General case}

\begin{figure*}[t]
\centering
\includegraphics[width=\columnwidth]{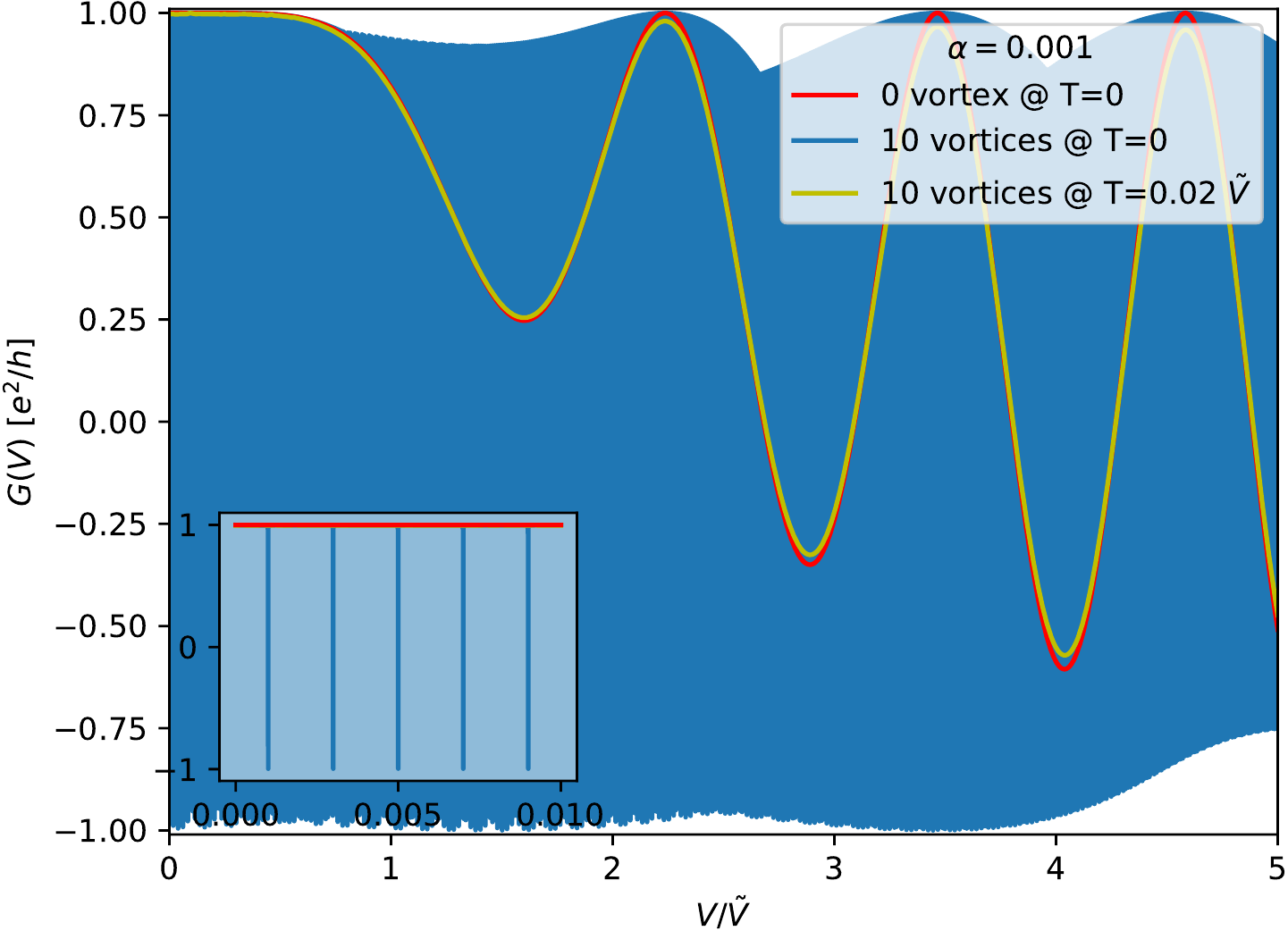}
\includegraphics[width=\columnwidth]{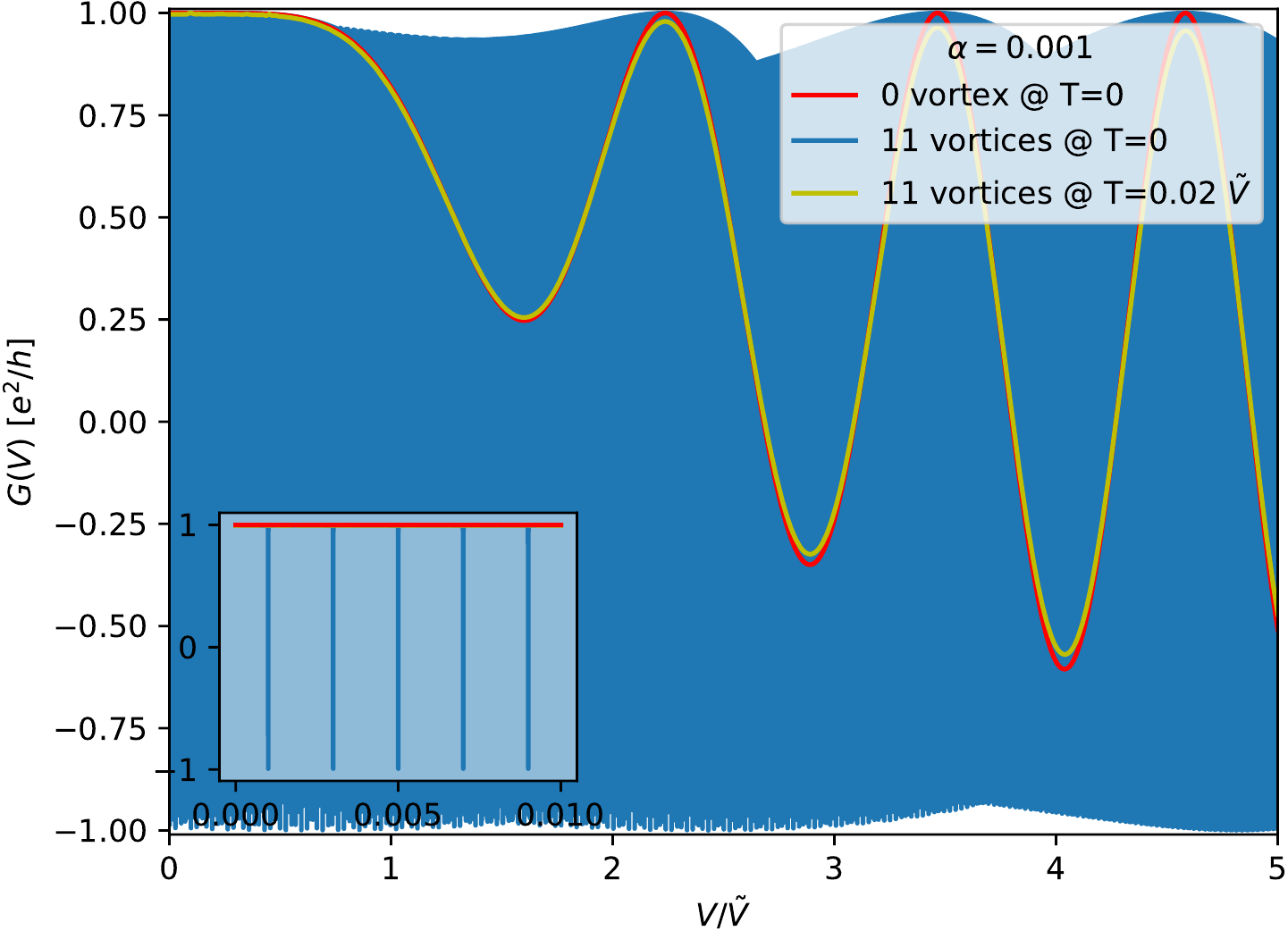}
\includegraphics[width=\columnwidth]{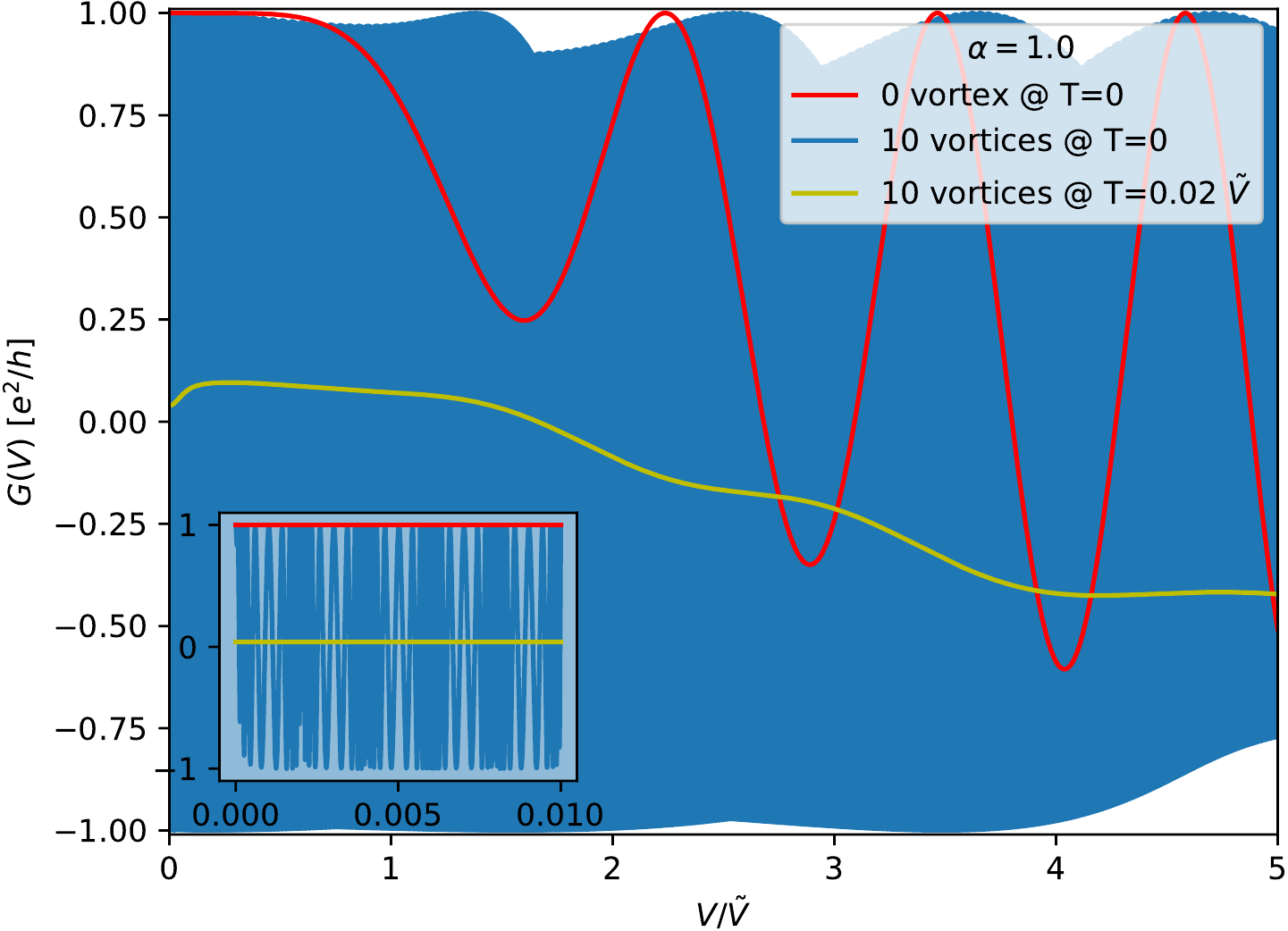}
\includegraphics[width=\columnwidth]{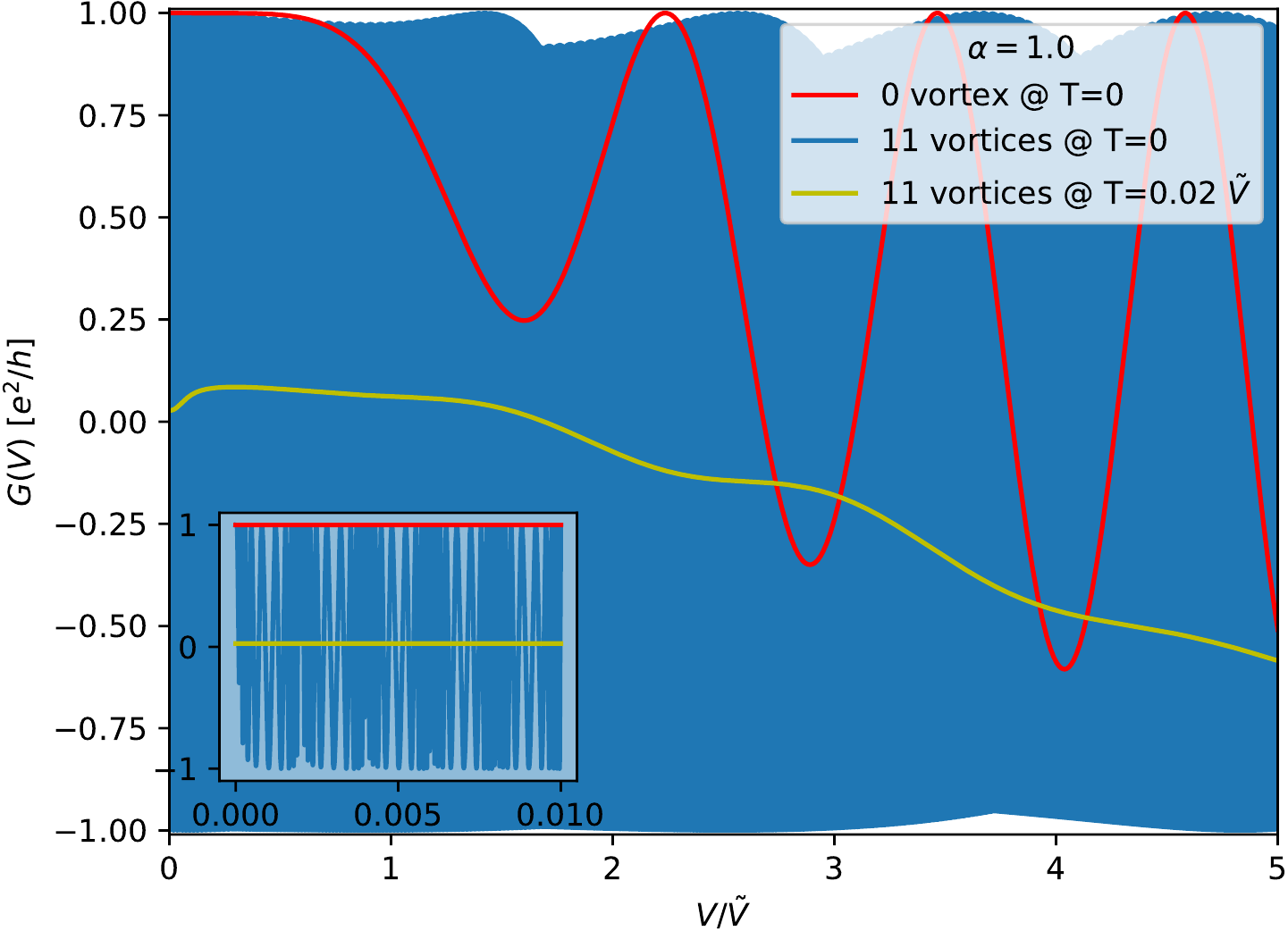}
\includegraphics[width=\columnwidth]{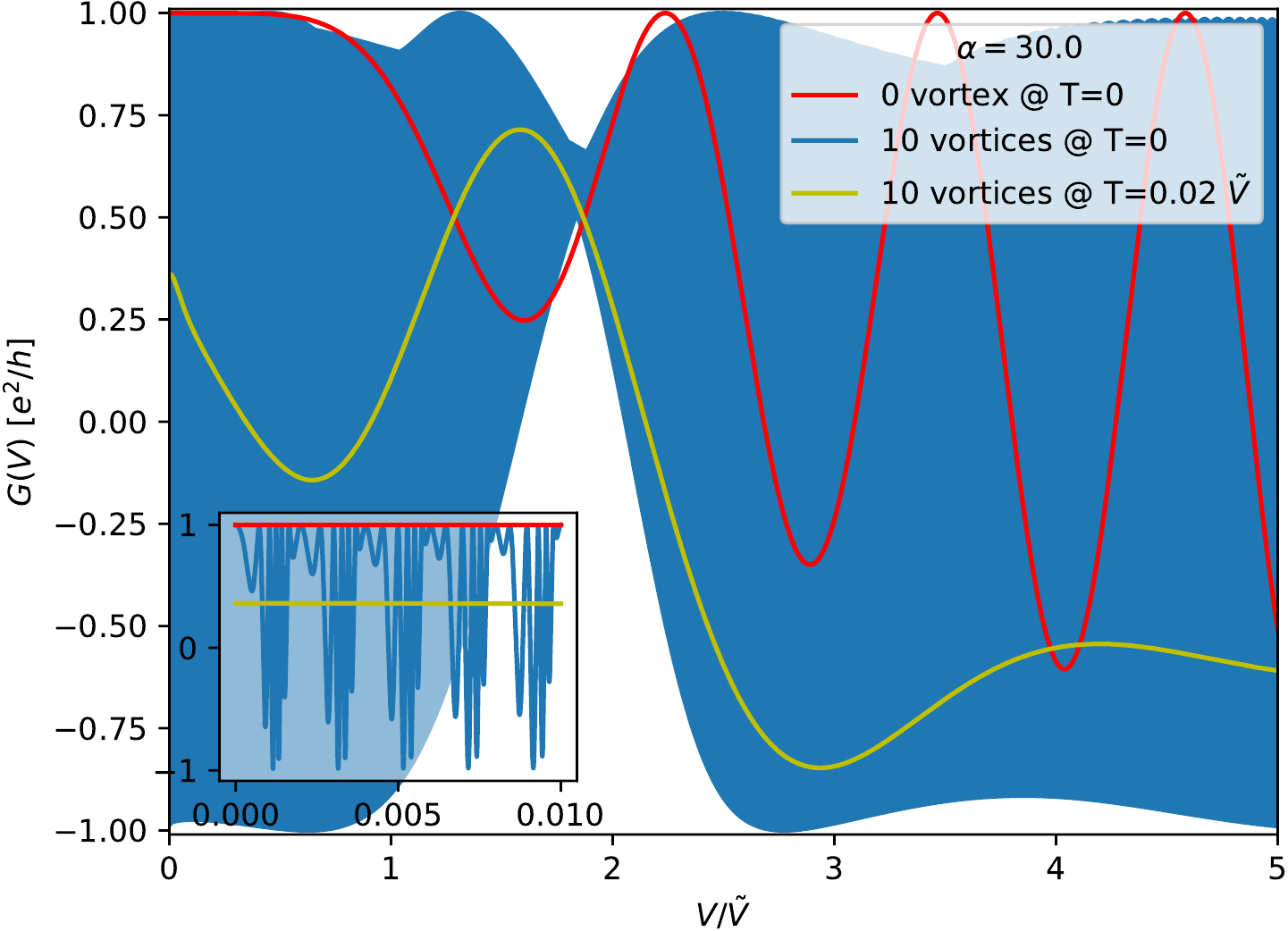} 
\includegraphics[width=\columnwidth]{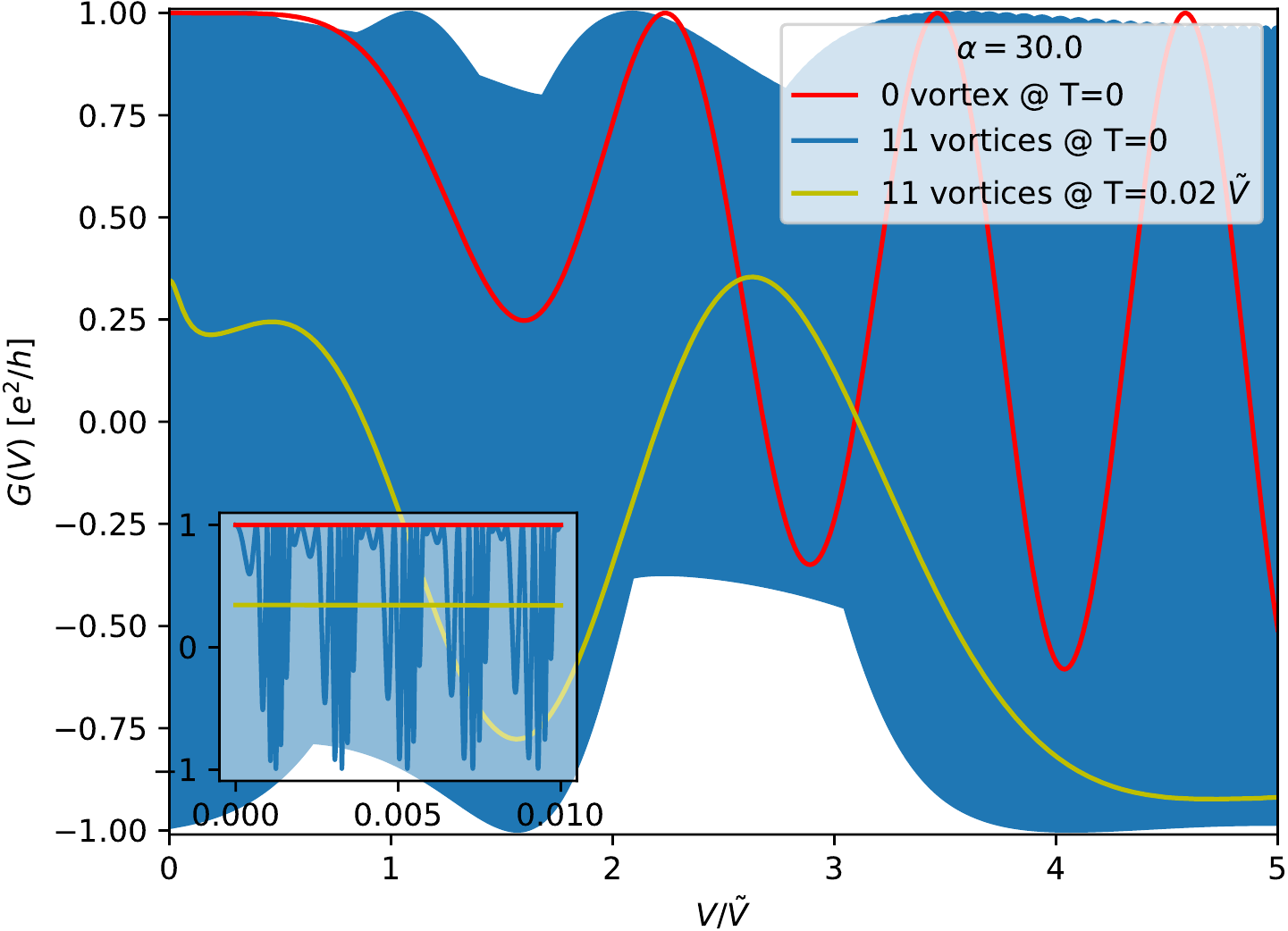} 
\caption{Same as Fig.~\ref{fig:single_vortex_all}, but with $10$ (left column) and $11$ (right column) identical vortices.}
\label{fig:10_11_vortices_all}
\end{figure*}

Figure~\ref{fig:10_11_vortices_all} plots the conductance for systems with 10 and 11 vortices using the same parameters as for Fig.~\ref{fig:single_vortex_all}, i.e., beyond the toy limit examined in the previous subsection.  [For all simulations in this subsection  the vortices are evenly distributed at positions $x_n = \frac{n}{N_v+1}(x_f-x_i)$.]  At least for these parameters, the negative zero-bias finite-temperature conductance regimes identified in the single-vortex case (Fig.~\ref{fig:single_vortex_all}, middle and bottom panels) are no longer present.  Moreover, in the intermediate-resonance case corresponding to the middle panels, the conductance retains significantly more structure compared to the toy limit (Fig.~\ref{fig:10_11_vortices_all_toy}) but which is clearly diminished compared to the single-vortex limit.  This structure points to a nontrivial interplay between the edge chemical potential, proximity-induced pairing strength, resonance width, and number of vortices.  

\begin{figure*}[t]
\centering
    \includegraphics[width=\columnwidth]{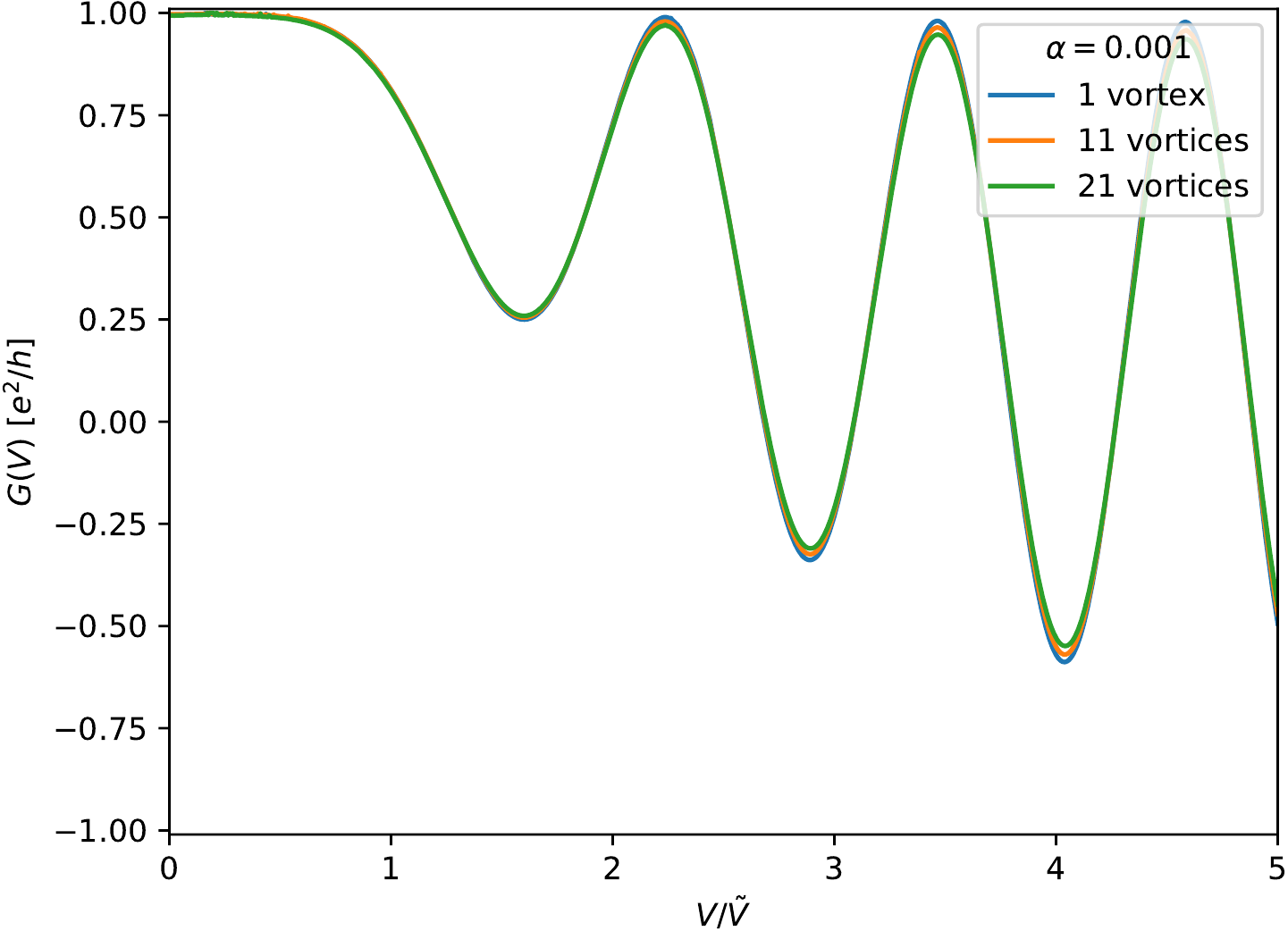}
    \includegraphics[width=\columnwidth]{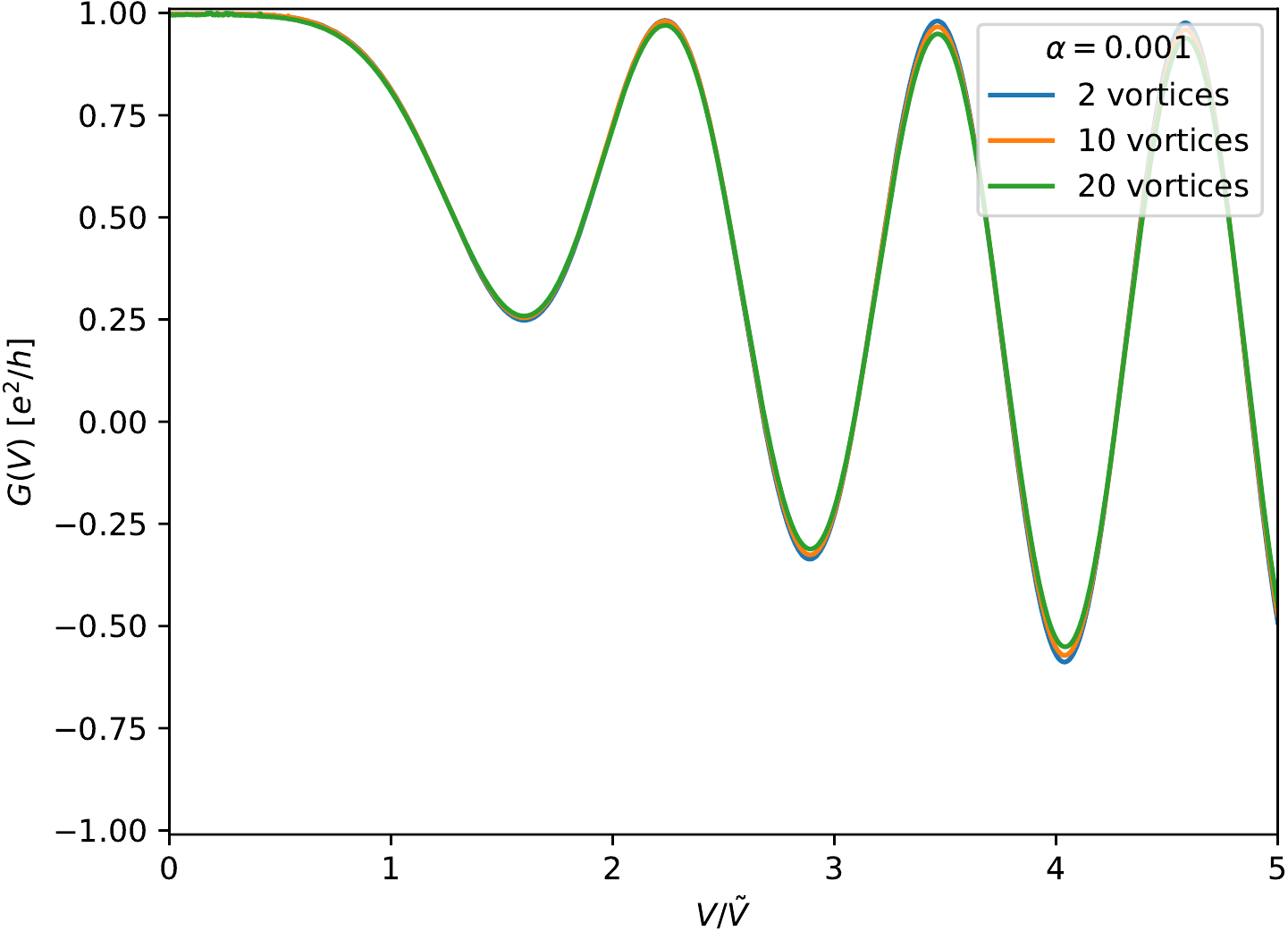}
    \includegraphics[width=\columnwidth]{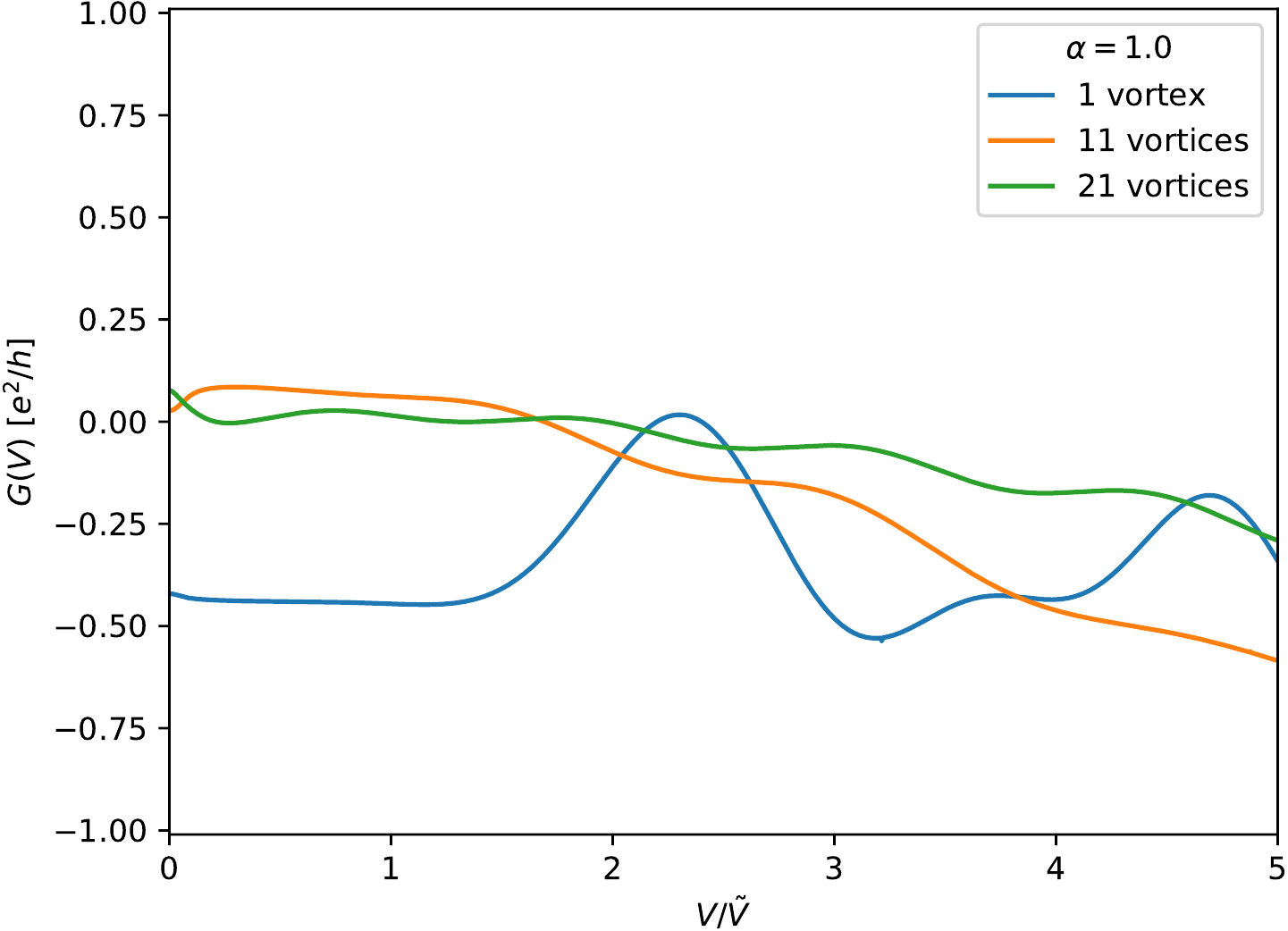}
    \includegraphics[width=\columnwidth]{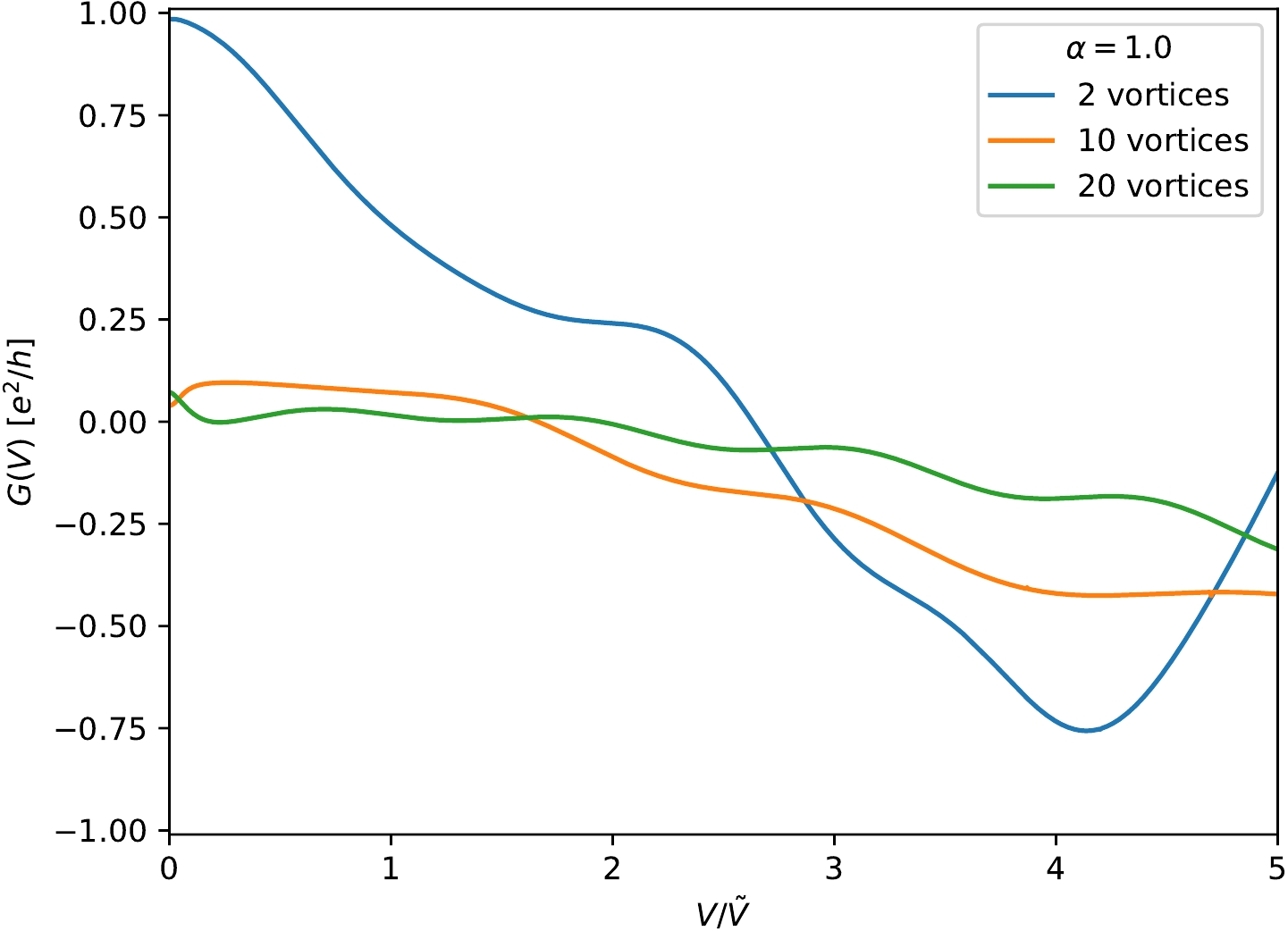}
    \includegraphics[width=\columnwidth]{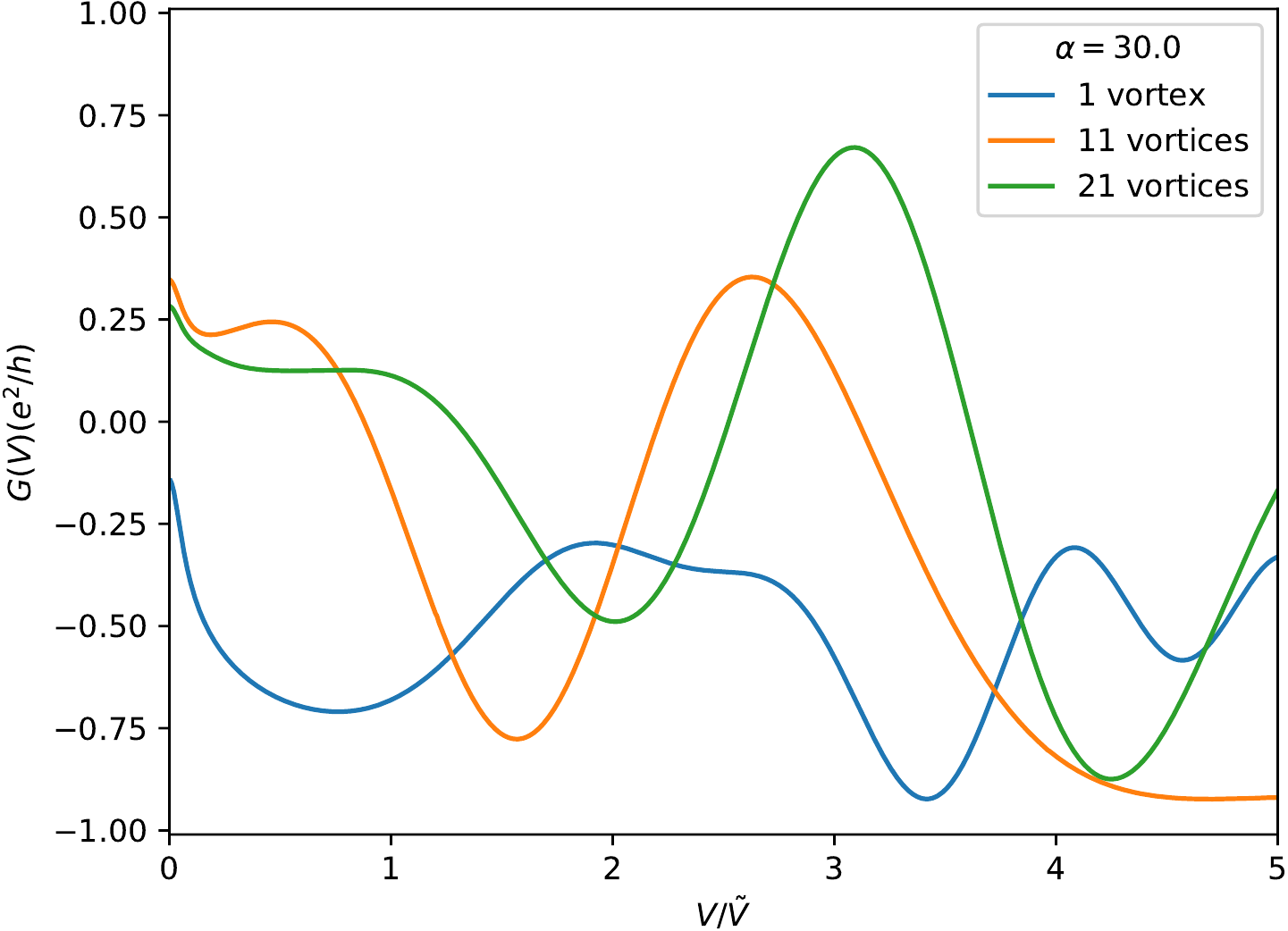}
    \includegraphics[width=\columnwidth]{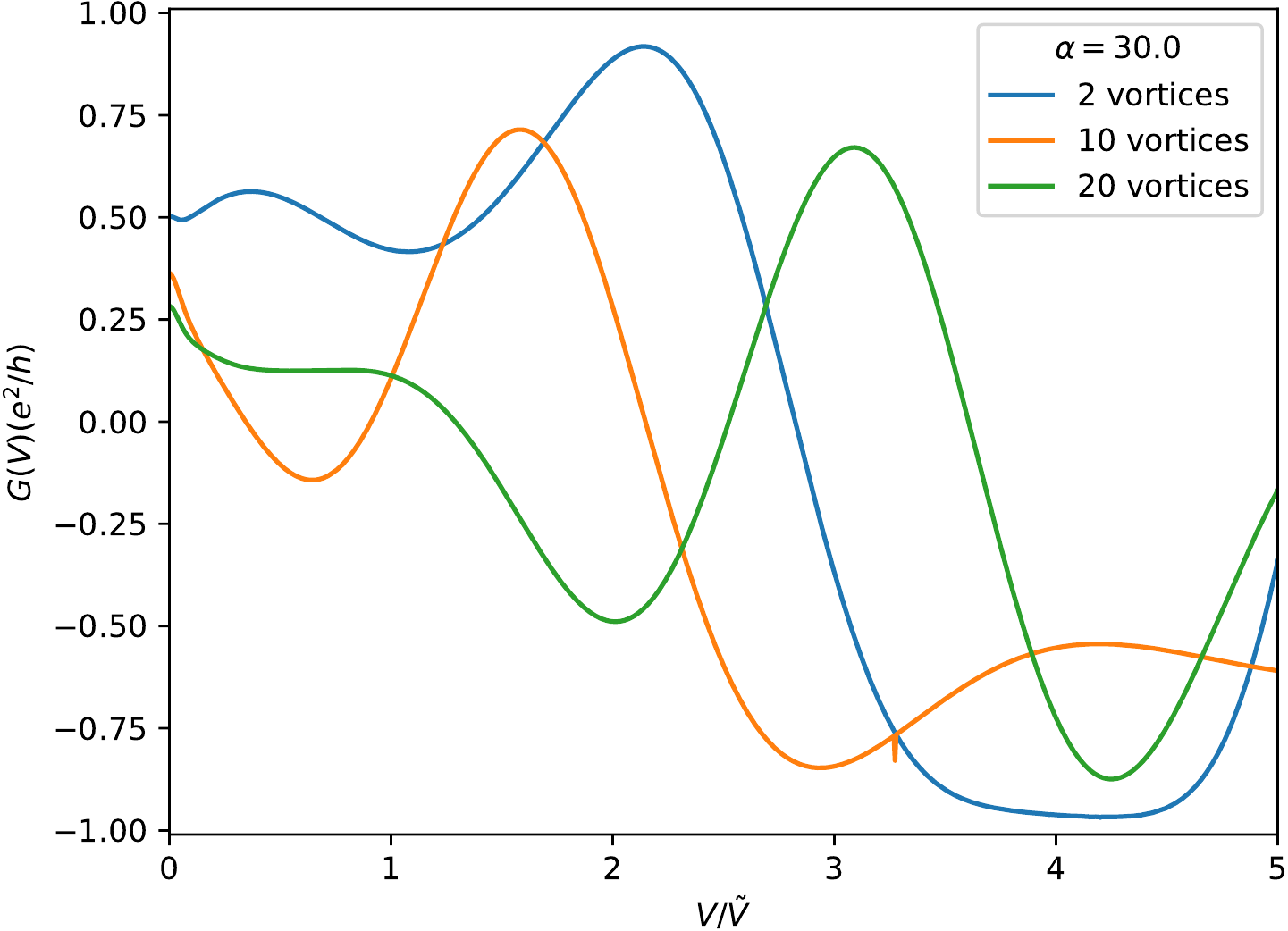}
    \caption{Finite temperature conductance for systems with odd (left column) and even (right column) numbers of vortices.  Parameters are the same as in Fig.~\ref{fig:10_11_vortices_all}.}
    \label{fig:finite_temp_plot}
\end{figure*}

We can, nevertheless, ascertain that the suppression of conductance with the number of vortices holds more generally, at least for intermediate-width resonances.   Figure~\ref{fig:finite_temp_plot} shows the evolution of the finite-temperature conductance as the vortex number increases.  (Note that for $N_v =1$ the vortex position is different compared to Fig.~\ref{fig:single_vortex_all}, explaining the difference in finite-temperature conductance.)  The central panels reveal a clear tendency for the structure in the conductance to wash out and approach zero as $N_v$ increases.  We expect a similar trend also in the narrow and wide-resonance regimes---but setting in at much larger $N_v$ than what we consider here.

\section{Discussion}\label{sec:discussion}

We examined the effect of vortices on the transport properties of a proximitized $\nu=1$ QH edge in the setup of Fig.~\ref{fig:experimental_setup}.
In the vortex-free limit, proximity-induced Cooper pairing enables Andreev processes that can result in negative conductance, a striking demonstration of superconducting correlations.
Previous work~\cite{Gamayun17} found that for a clean, vortex-free system, such Andreev processes are, however, suppressed by kinematic constraints near zero-bias.  
In this paper, we found that coupling to vortices resurrects these processes, resulting in rapid zero-temperature conductance oscillations even at low bias voltages.  
Finite temperature washes out these rapid oscillations, but can still result in negative low-bias conductance in both the single-vortex and multi-vortex limits; see Figs.~\ref{fig:single_vortex_all_toy}, \ref{fig:single_vortex_all}, and \ref{fig:10_11_vortices_all_toy}.

As more vortices couple to the edge, the finite-temperature conductance tends to vanish, reproducing the expected behavior for a QH edge adjacent to a normal contact (instead of a superconductor).  We anticipate that randomness---e.g., in the vortex positions, sub-gap spacing, and tunneling amplitudes---will enhance this tendency towards normal-contact behavior beyond the trends that we captured in our simulations.  Reference~\onlinecite{Lee17} measured a similar effect in which a decrease in the zero-bias conductance with increasing magnetic field was attributed to vortex-mediated dissipation.
Similarly, Refs.~\onlinecite{Kurilovich22,Schiller22} identify electrons leaving the edge through vortices as a mechanism for reducing superconducting correlations in the proximitized edge.  
In contrast, our dissipation-free analysis uncovers a new mechanism through which vortex-enabled Andreev processes can suppress the magnitude of the finite-temperature conductance.  

Natural extensions of this work include considering higher integer, and eventually fractional, filling factors.  It would also be useful to obtain a more microscopic understanding of the edge-vortex hybridization processes---treated here on a phenomenological level.  Finally, adapting the scattering-matrix based analysis to study crossed Andreev reflection in the setup of Ref.~\onlinecite{Gul20} could provide further insight into the role that vortices play in that system.

\acknowledgements
This work was supported by the Army Research Office under
Grant Award W911NF-17-1-0323; the Caltech Institute for Quantum
Information and Matter, an NSF Physics Frontiers Center with support of the Gordon and Betty Moore Foundation through Grant GBMF1250; and the Walter Burke
Institute for Theoretical Physics at Caltech.

\appendix

\section{Solution of the vortex-free Hamiltonian}
\label{appendix:diagonalize}

Here we sketch the derivation of wavefunctions for the vortex-free Hamiltonian that enable computation of the conductance.  After writing $\tilde \psi(x) = \psi(x) e^{-i \phi(x)/2}$,  Eqs.~\eqref{eq:vortex_free_H0} and \eqref{eq:vortex_free_Hdelta} become 
\begin{align}
      H_0 &= \int_x \psi^{\dagger} [-iv\partial_x-\mu(x)] \psi 
      \label{eq:vortex_free_H0b}
      \\
      H_{\Delta} &=  \frac{1}{2} \int_x \Delta(x)(i \psi \partial_x\psi + h.c.)
      \label{eq:vortex_free_Hdeltab}
\end{align}
with
\begin{equation}\label{eq:mu_eff}
    \mu(x) = \tilde \mu(x) + \frac{v}{2} \partial_x \phi(x).
\end{equation}
We assume piecewise-constant $\mu(x)$ and $\Delta(x)$ such that adjacent to the superconductor (i.e., for $x_i<x<x_f$) $\mu(x) = \mu_{\rm sc}$ and $\Delta(x) = \Delta_{\rm sc}$ while elsewhere both quantities vanish.  The Hamiltonian takes a conventional Bogoliubov-de Gennes form when expressed using the two-component operator $\Psi = [{\psi}, {\psi^{\dagger}}]^T$:
\begin{align}
    H &= \frac{1}{2} \int dx \Psi^{\dagger} h \Psi \label{eq:vortex_free_H_bdg}
\\ h &=
    \begin{pmatrix}
    -iv \partial_x - \mu(x) & \frac{i}{2}\{\Delta(x), \partial_x\} \\
    \frac{i}{2}\{\Delta(x),\partial_x\} & -iv \partial_x + \mu(x)
    \end{pmatrix} \label{eq:vortex_free_H_bdg_h}
\end{align}
We seek energy eigenstates satisfying $h \chi_E(x) = E \chi_E(x)$, where $\chi_E(x)$ is a two-component wavefunction corresponding to an incident electron impinging on the proximitized region of the quantum Hall edge.  

Care must be taken to ensure proper boundary conditions due to the off-diagonal terms in $h$ that contain derivatives with discontinuous prefactors.  It is useful to work in the basis $\chi_{\pm,E}(x) = \chi_{1,E}(x) \pm \chi_{2,E}(x)$; defining $v_\pm(x) = v \pm \Delta(x)$, the matrix equation then reads
\begin{align}
    &-i\left[v_-(x)\partial_x + \frac{(\partial_x v_-(x))}{2}\right]\chi_{+,E} - \mu(x) \chi_{-,E} = E \chi_{+,E}
    \\
    &-i\left[v_+(x)\partial_x +  \frac{(\partial_x v_+(x))}{2}\right]\chi_{-,E} - \mu(x) \chi_{+,E} = E \chi_{-,E}.
\end{align}
The pieces involving $(\partial_x v_\pm(x))$, which produce delta functions, can be cancelled upon writing 
\begin{equation}
      \chi_{\pm,E}(x) = \left[\frac{v}{v_\mp(x)}\right]^{1/2} \bar \chi_{\pm,E}(x).
\end{equation}
In terms of $\bar \chi_{\pm,E}$ we obtain
\begin{align}
    &-iv \partial_x\bar\chi_{+,E} - \mu(x)\frac{v}{\sqrt{v_+(x)v_-(x)}} \bar \chi_{-,E} = E\frac{v}{v_-(x)} \bar\chi_{+,E}
    \\
    &-iv \partial_x\bar\chi_{-,E} - \mu(x)\frac{v}{\sqrt{v_+(x)v_-(x)}} \bar \chi_{+,E} = E\frac{v}{v_+(x)} \bar\chi_{-,E}.
\end{align}
The rescaled wavefunction components $\bar\chi_{\pm,E}$ are continuous at $x_i$ and $x_f$, which can be verified by integrating the above equations across an infinitesimal window around these points.  Correspondingly, the original $\chi_{\pm,E}$ functions exhibit nontrivial boundary conditions given by
\begin{subequations}\label{eq:boundary}
        \begin{align} 
        \chi_+(x_i^+) = \sqrt{\frac{v}{v - \Delta_{\rm sc}}} \chi_+(x_i^-) \\
       \chi_-(x_i^+) = \sqrt{\frac{v}{v + \Delta_{\rm sc}}} \chi_-(x_i^-) \\ 
       \chi_+(x_f^-) = \sqrt{\frac{v}{v - \Delta_{\rm sc}}} \chi_+(x_f^+) \\
       \chi_-(x_f^-) = \sqrt{\frac{v}{v + \Delta_{\rm sc}}} \chi_-(x_f^+).
        \end{align}
\end{subequations}
Discontinuity of the wavefunctions at $x_i$ and $x_f$ reflects the abrupt change in velocities of Majorana fermions, obtained by writing $\psi = \gamma_1 + i \gamma_2$, within the superconducting region.

The incoming piece of the wavefunction takes a simple plane wave form,
\begin{subequations}\label{eq:ansatz_x_leq_0}
    \begin{align} 
    \chi_{1,E}(x<x_i) = A^{\rm in}_
    {\rm e} e^{iEx/v}  \\
    \chi_{2,E}(x<x_i) = A^{\rm in}_{\rm h} e^{iEx/v},
    \end{align}
\end{subequations}
with $A^{\rm in}_{\rm h} = 0$ as appropriate for an incident electron.  
The outgoing part of the wavefunction similarly reads
\begin{subequations}\label{eq:ansatz_geq_l}
    \begin{align} 
    \chi_{1,E}(x>x_f) = A^{\rm out}_
    {\rm e} e^{iE x/v} \\
    \chi_{2,E}(x>x_f) = A^{\rm out}_
    {\rm h} e^{iEx/v}.
    \end{align}
\end{subequations}
In the superconducting region, the wavefunction is a superposition of plane-waves carrying different momenta $k_\pm$---see Eq.~\eqref{kpm}---due to the induced pairing:  
\begin{subequations}\label{eq:ansatz_superconducting}
        \begin{align} 
        \chi_{1,E}(x_i<x<x_f) &= c_{+} e^{ik_+x} + c_{-} e^{ik_-x}\\
        \chi_{2,E}(x_i<x<x_f) &= a_+ c_{+} e^{ik_+x} + a_- c_{-} e^{ik_-x} .
        \end{align}
    \end{subequations}
The electron and hole parts are related by coefficients
\begin{equation}
    a_{\pm} = \frac{\frac{v\mu_{\rm sc}}{E\Delta_{\rm sc}}\mp f(E)}{\frac{\mu_{\rm sc}}{E}-1}.
    \label{apm}
\end{equation}
Finally, the coefficients $A^{\rm out}_{\rm e}$, $A^{\rm out}_{\rm h}$, and $c_{\pm}$ can be expressed in terms of the incoming wavefunction coefficients using the boundary conditions in Eq.~\eqref{eq:boundary}.

\section{Vortex-free scattering matrix}
\label{appendix:vortex_free_S}

Using the results from App.~\ref{appendix:diagonalize}, we can determine the scattering matrix for the vortex-free case as follows.  The coefficients $c_{\pm}$ for the proximitized region relate to the incoming coefficients $A^{\rm in}_{\rm e}$ and $A^{\rm in}_{\rm h}$ as
\begin{equation}\label{eq:Mi_vortex_free_relate}
        \begin{pmatrix}
        c_{+} \\
        c_{-} 
        \end{pmatrix} 
        = e^{i E x_i/v}\begin{pmatrix}
          e^{-ik_+x_i} & 0 \\
          0 &  e^{-ik_-x_i}
        \end{pmatrix}
        M_i 
        \begin{pmatrix}
        A^{\rm in}_{\rm e} \\
        A^{\rm in}_ {\rm h}
        \end{pmatrix},
\end{equation}
while the outgoing coefficients $A^{\rm out}_{\rm e}$ and $A^{\rm out}_{\rm h}$ relate to $c_{\pm}$ as
\begin{equation}\label{eq:Mf_vortex_free_relate}
        \begin{pmatrix}
         A^{\rm out}_{\rm e} \\
         A^{\rm out}_{\rm h}
        \end{pmatrix} = e^{-i E x_f/v}M_f
        \begin{pmatrix}
         e^{ik_+x_f} & 0 \\
         0 & e^{ik_-x_f}
        \end{pmatrix}
        \begin{pmatrix}
        c_{+}  \\
        c_{-}
        \end{pmatrix}.
\end{equation}
Here $M_i$ and $M_f$ are matrices that encode the boundary conditions.  Because the boundaries at $x_i$ and $x_f$ are identical, we have $M_i = M_f^{-1}$.  The full expression for the (real) matrix $M_f$ reads
\begin{align} \label{M_f_full_expression}
    M_f &= \frac{1}{2}\begin{pmatrix}
         1 & -1 \\
         1 & 1
        \end{pmatrix}
        \begin{pmatrix}
         \sqrt{\frac{v-\Delta_{\rm sc}}{v}} & 0 \\
         0 & \sqrt{\frac{v+\Delta_{\rm sc}}{v}}
        \end{pmatrix}
        \begin{pmatrix}
        1 & 1 \\
        -1 & 1
        \end{pmatrix} 
        \nonumber \\
        &\times\left( \begin{array}{cc} 1 & 1 \\ a_+ & a_- \end{array}\right)
\end{align}
where $a_\pm$ are given in Eq.~\eqref{apm}.  In the upper line of $M_f$, the matrix on the left rotates to basis of Majorana fermions $\gamma_{1,2}$, the middle matrix incorporates the velocity factors in the boundary condition from Eq.~\eqref{eq:boundary}, and the rightmost matrix rotates back to the original basis; the matrix on the lower line of $M_f$ encodes continuity of the resulting rescaled wavefunctions.
The scattering matrix can then be expressed in terms of the diagonal unitary matrix $D(x_f-x_i)$ from Eq.~\eqref{eq:transmission_matrix} via
\begin{equation}
    S_0(x_f-x_i) = e^{iE(x_i-x_f)/v}M_f D(x_f-x_i) M_i. \label{S_Mi_Mf}
\end{equation}

The expression for the scattering matrix in Eq.~\eqref{S_Mi_Mf} is not unique.  In particular, since $D(x_f-x_i)$ is diagonal, we can write
\begin{equation}
D(x_f-x_i) = P D(x_f-x_i) P^{-1} 
\end{equation}
for
\begin{equation}
   P = 
    \begin{pmatrix}
     \beta & 0 \\
     0 & \beta'
    \end{pmatrix}
\end{equation}
with arbitrary $\beta,\beta'$.   We can choose $\beta$ and $\beta'$ such that  $P^{-1}M_i = \mathcal{O}$ is an orthogonal matrix, in which case $M_f P = \mathcal{O}^T$ and the scattering matrix takes the form
\begin{align}\label{eq:S0-final}
    S_0(x_f-x_i) &= e^{i\omega} \mathcal{O}^T D(x_f-x_i)\mathcal{O}
\end{align}
quoted in Eq.~\eqref{eq:scattering_matrix_orthogonal_no_vortex}.  To see that such a choice is possible, we first write $M_f$ more succinctly as
\begin{equation}
    M_f = 
    \begin{pmatrix}
     a_{11} & a_{12} \\
     a_{21} & a_{22}
    \end{pmatrix}.
\end{equation}
Our explicit solution from Eq.~\eqref{M_f_full_expression} reveals that $a_{11}a_{12} + a_{21}a_{22} = 0$ [which underlies the `hidden' unitarity of $S_0$ as expressed in Eq.~\eqref{S_Mi_Mf}].  The conditions $\det(M_f P) = 1$ and $(M_f P)^T = (M_f P)^{-1}$ can then be satisfied by choosing
\begin{align}
    \beta = \frac{1}{\sqrt{a_{11}^2 + a_{12}^2}},~~~~\beta' = - \frac{a_{21}}{a_{12}} \beta. 
\end{align}
The resulting expression for $S_0$ in Eq.~\eqref{eq:S0-final} has the virtue of being manifestly unitary.  Additionally, its physical meaning is more transparent: $\mathcal{O}$ rotates from the original electron-hole basis to the basis of eigenstates in the proximitized region; $D(x_f-x_i)$ adds the corresponding phase factors accumulated by these eigenmodes on crossing the superconductor, after which $\mathcal{O}^T$ rotates back to the original basis.

\section{Scattering matrix in the single-vortex case}\label{appendix:general}

To model the coupling to a single vortex, we supplement the vortex-free Hamiltonian conveniently expressed in Eq.~\eqref{eq:vortex_free_H_bdg}
with the vortex terms in Eqs.~\eqref{eq:Hv} and \eqref{eq:Hint}, repeated here for clarity:
\begin{align}
    H_v &= \sum_{n = 0}^{n_{\rm max}} \epsilon\left(n+\frac{1}{2}\right) {a}_n^{\dagger}{a}_n \label{eq:appendix_Hv} \\
    H_{\rm int} &=  \sum_{n = 0}^{n_{\rm max}} [t{a}_n \psi(x_1) + t' a_n^{\dagger}{\psi}(x_1) + h.c.]. \label{eq:appendix_Hint}
\end{align}

(If desired, one can straightforwardly adapt the calculations below to solve the problem with $n$-dependent $t, t'$ couplings, but we choose not to do so here for simplicity.)
In the presence of vortex hybridation terms, edge excitations with energy $E$ are created by operators of the form 
\begin{equation}
    \Gamma_E^\dagger = \sum_{n = 0}^{n_{\rm max}} (\eta_{1,n} a_n^\dagger + \eta_{2,n} a_n) + \int_x [\chi_{1,E}(x)\psi^\dagger + \chi_{2,E}(x)\psi]. \label{eq:single_vortex_operator}
\end{equation}
The new $\eta_{1,n}$ and $\eta_{2,n}$ components encode probability weight on the $n^{\rm th}$ vortex level; these pieces also depend on energy but we suppress that dependence for notational brevity.  

We derive the wavefunctions from  the full Hamiltonian $H$ by evaluating $[H,\Gamma_E^\dagger] = E \Gamma_E^\dagger$ and equating parts with the same operators.  This procedure yields the following equations:
\begin{multline}
    [-iv\partial_x-\mu(x)]\chi_{1,E}(x) + \frac{i}{2}\{\Delta(x),\partial_x\}\chi_{2,E}(x) \\
    + \sum_{n = 0}^{n_{\rm max}}[t'^*\eta_{1,n} + t\eta_{2,n}]\delta(x-x_1) = E\chi_{1,E}(x) \label{eq:EOM1}
\end{multline}
\begin{multline}
    [-iv\partial_x+\mu(x)]\chi_{2,E}(x) + \frac{i}{2}\{\Delta(x),\partial_x\}\chi_{1,E}(x) \\
    - \sum_{n = 0}^{n_{\rm max}}[t\eta_{1,n} + t'\eta_{2,n}]\delta(x-x_1) = E\chi_{2,E}(x) \label{eq:EOM2}
\end{multline}
with
\begin{subequations}\label{eq:eta}
\begin{align}
    \eta_{1,n} &= \frac{t'\chi_{1,E}(x_1)-t\chi_{2,E}(x_1)}{E-\epsilon\left(n+\frac{1}{2}\right)} \\
    \eta_{2,n} &= \frac{t\chi_{1,E}(x_1)-t'^*\chi_{2,E}(x_1)}{E-\epsilon\left(n+\frac{1}{2}\right)}.
\end{align}
\end{subequations}
Away from the vortex position $x_1$, Eqs.~\eqref{eq:EOM1} and \eqref{eq:EOM2} map onto the vortex-free problem solved in App.~\ref{appendix:diagonalize}.  We can thus read off the form of the wavefunctions in those regions from our previous solution, except that the amplitudes in the proximitized region [Eq.~\eqref{eq:ansatz_superconducting}] will differ on the two sides of the vortex.  More precisely, we now have
\begin{subequations}\label{before}        \begin{align} 
        \chi_{1,E}(x_i<x<x_1) &= c_{+} e^{ik_+x} + c_{-} e^{ik_-x}\\
        \chi_{2,E}(x_i<x<x_1) &= a_+ c_{+} e^{ik_+x} + a_- c_{-} e^{ik_-x} 
        \end{align}
\end{subequations}
for the paired region before the vortex and 
\begin{subequations}\label{after}        \begin{align} 
        \chi_{1,E}(x_1<x<x_f) &= c'_{+} e^{ik_+x} + c'_{-} e^{ik_-x}\\
        \chi_{2,E}(x_1<x<x_f) &= a_+ c'_{+} e^{ik_+x} + a_- c'_{-} e^{ik_-x}
        \end{align}
\end{subequations}
after the vortex.  Our previous solution already relates $c_{\pm}$ to the incoming amplitudes, and $c'_{\pm}$ to the outgoing amplitudes.  Here we simply need to relate $c_{\pm}$ and $c'_{\pm}$.  

Integrating Eqs.~\eqref{eq:EOM1} and \eqref{eq:EOM2} over an infinitesimal window around $x_1$ yields
\begin{multline}\label{eq:EOM1_integrated}
    -iv[\chi_{1,E}(x_1^+) - \chi_{1,E}(x_1^-)] + i\Delta_{\rm sc}[\chi_{2,E}(x_1^+) - \chi_{2,E}(x_1^-)] \\
    + \sum_{n = 0}^{n_{\rm max}}[t'^{*}\eta_{1,n} + t\eta_{2,n}] = 0
\end{multline}
and
\begin{multline}\label{eq:EOM2_integrated}
    -iv[\chi_{2,E}(x_1^+) - \chi_{2,E}(x_1^-)] + i\Delta_{\rm sc}[\chi_{1,E}(x_1^+) - \chi_{1,E}(x_1^-)] \\
    + \sum_{n = 0}^{n_{\rm max}}[t\eta_{1,n} + t'\eta_{2,n}] = 0.
\end{multline}
By plugging in Eq.~\eqref{eq:eta} with
\begin{subequations}\label{eq:bc_x0}
    \begin{align}
        \chi_{j,E}(x_1) &= \frac{1}{2}[\chi_{j,E}(x_1^+) + \chi_{j,E}(x_1^-)] 
    \end{align}
\end{subequations}
and inserting the wavefunctions from  Eqs.~\eqref{before} and \eqref{after}, we obtain
\begin{multline}
     ia(c_{+}'-c_{+})e^{ik_+x_1} +  ib(c_{-}'-c_{-})e^{ik_-x_1} \\
     + f_1(c_{+}'+c_{+})e^{ik_+x_1} + g_1(c_{-}'+c_{-})e^{ik_-x_1}= 0 \label{eq:EOM1_final}
\end{multline}
and
\begin{multline}
     ic(c_{+}'-c_{+})e^{ik_+x_1} + i d(c_{-}'-c_{-})e^{ik_-x_1} \\
     - f_2(c_{+}'+c_{+})e^{ik_+x_1} - g_2(c_{-}'+c_{-})e^{ik_-x_1}= 0. \label{eq:EOM2_final}
\end{multline}
Above we defined shorthand notation
\begin{subequations}\label{eq:abcd}
    \begin{align}
        a &= -v + \Delta_{\rm sc} a_+ \\
        b &= -v + \Delta_{\rm sc}a_-  \\
        c &= -va_+ + \Delta_{\rm sc}  \\
        d &= -va_- + \Delta_{\rm sc} 
    \end{align}
\end{subequations}
and
\begin{subequations}
    \begin{align}
           f_1 = & \frac{1}{2}\left[(|t'|^2 + t^2 - 2a_+tt'^{*}) s(E)
           + (|t'|^2-t^2) h(E)\right] \\
           f_2 = & \frac{1}{2}\left[(2tt'-a_+|t'|^2 - a_+ t^2) s(E)  + a_+(|t'|^2-t^2)h(E)\right]  \\
           g_1 = & \frac{1}{2}\left[(|t'|^2 + t^2 - 2a_-tt'^{*}) s(E) 
           + (|t'|^2-t^2) h(E)\right] \\
           g_2 = & \frac{1}{2}\left[(2tt'-a_-|t'|^2 - a_- t^2) s(E)
        + a_-(|t'|^2-t^2) h(E)\right]
    \end{align}
    \label{fg}
\end{subequations}
with 
\begin{equation}\label{eq:f_g_converge}
    s(E) = \sum_{n = 0}^{n_{\rm max}}\frac{E}{E^2-\epsilon^2(n+\frac{1}{2})^2}
\end{equation}
and
\begin{equation}\label{eq:f_g_diverge}
    h(E) = \sum_{n = 0}^{n_{\rm max}}\frac{\epsilon(n+\frac{1}{2})}{E^2-\epsilon^2(n+\frac{1}{2})^2}.
\end{equation}
Notice that the sum in $s(E)$ diverges logarithmically with $n_{\rm max}$ whereas the sum in $h(E)$ is convergent.  We nevertheless keep both terms in our calculations.

Let us define a matrix
\begin{equation}
    \Tilde{M}_v = \begin{pmatrix}
        b_{11} & b_{12} \\ \label{eq:Mv_simplified}
        b_{21} & b_{22}
        \end{pmatrix}
\end{equation} that relates $c_{\pm}'$ to $c_{\pm}$ via
\begin{equation}\label{eq:relation_Mv}
        \begin{pmatrix}
         c_{+}' \\
        c_{-}'
        \end{pmatrix} = \begin{pmatrix}
          e^{-ik_+x_1} & 0 \\
          0 &  e^{-ik_-x_1}
        \end{pmatrix}\Tilde{M}_v\begin{pmatrix}
          e^{ik_+x_1} & 0 \\
          0 &  e^{ik_-x_1}
        \end{pmatrix}
        \begin{pmatrix}
        c_{+} \\
        c_{-}
        \end{pmatrix}.
\end{equation}
With the insertion of the diagonal matrices above, $\tilde M_v$ is independent of the vortex position $x_1$.

Solving Eqs.~\eqref{eq:EOM1_final} and \eqref{eq:EOM2_final} gives the matrix elements
\begin{subequations} \label{eq:Mv_elements}
    \begin{align}
    b_{11} &= \frac{(ic+f_2)(ib+g_1) - (ia-f_1)(id-g_2)}{(ic-f_2)(ib+g_1) - (ia+f_1)(id-g_2)}  \\
    b_{12} &= \frac{2i(dg_1+bg_2)}{(ic-f_2)(ib+g_1) - (ia+f_1)(id-g_2)} \\
    b_{21} &= \frac{-2i(cf_1+af_2)}{(ic-f_2)(ib+g_1) - (ia+f_1)(id-g_2)}  \\
    b_{22} &= \frac{(ic-f_2)(ib-g_1) - (ia+f_1)(id+g_2)}{(ic-f_2)(ib+g_1) - (ia+f_1)(id-g_2)} .
\end{align}
\end{subequations}
Upon incorporating our results from the vortex-free analysis, 
the single-vortex scattering matrix takes the form (up to an unimportant overall phase factor)
\begin{equation}
    S = M_f D(x_f-x_1) \Tilde{M}_v D(x_1-x_i) M_i. \label{eq:single_vortex_scattering_matrix_t0}
\end{equation}
Finally, we can trade in $M_{f,i}$ for orthogonal matrices as described in App.~\ref{appendix:vortex_free_S} to obtain the more illuminating alternate form
\begin{align}
    S &= [\mathcal{O}^T D(x_f-x_1) \mathcal{O}] M_v [\mathcal{O}^T D(x_1-x_i) \mathcal{O}] \label{eq:single_vortex_scattering_matrix_with_P} \\
    \nonumber
    &= S_0(x_f-x_1) M_v  S_0(x_1-x_i)
\end{align}
with 
\begin{equation}
    M_v = M_f \Tilde{M}_v M_i
\end{equation}
Note that both $S_0$ and $M_v$ are unitary (unitarity of the latter can be explicitly verified from the solution above).

Dramatic simplification arises in the toy limit discussed in Sec.~\ref{toylimit} where $\mu_{\rm sc} = 0$ and $t = t'$.  
Here the terms in Eq.~\eqref{eq:abcd} and \eqref{fg} simplify to 
\begin{subequations}\label{eq:fg_simplified}
    \begin{align}
    a &= c = -v + \Delta_{\rm sc} \\
    b &= -d = -v - \Delta_{\rm sc}  \\
    f_1 &= f_2 = 0  \\
    g_1 &= g_2 = t^2\sum_{n = 0}^{n_{\rm max}}\frac{2E}{E^2-\epsilon^2\left(n+\frac{1}{2}\right)}.
    \end{align}
\end{subequations}
The vortex matrix $\tilde M_v$ accordingly becomes
\begin{equation}
    \Tilde{M}_v = 
    \begin{pmatrix}
     1 & 0 \\
     0 & e^{i\theta_E} 
    \end{pmatrix},
\end{equation}
where
\begin{equation} \label{eq:thetaEp}
    e^{i\theta_E} = \frac{v+\Delta_{\rm sc} - it^2\sum_{n = 0}^{n_{\rm max}}\frac{2E}{E^2-\epsilon^2\left(n+\frac{1}{2}\right)^2}}{v+\Delta_{\rm sc} + it^2\sum_{n = 0}^{n_{\rm max}}\frac{2E}{E^2-\epsilon^2\left(n+\frac{1}{2}\right)^2}}  
\end{equation}
represents an additional phase acquired by the Majorana fermion $\gamma_2$ due to hybridization with the vortex.

\bibliography{citations.bib}

\begin{thebibliography}{34}%
\makeatletter
\providecommand \@ifxundefined [1]{%
 \@ifx{#1\undefined}
}%
\providecommand \@ifnum [1]{%
 \ifnum #1\expandafter \@firstoftwo
 \else \expandafter \@secondoftwo
 \fi
}%
\providecommand \@ifx [1]{%
 \ifx #1\expandafter \@firstoftwo
 \else \expandafter \@secondoftwo
 \fi
}%
\providecommand \natexlab [1]{#1}%
\providecommand \enquote  [1]{``#1''}%
\providecommand \bibnamefont  [1]{#1}%
\providecommand \bibfnamefont [1]{#1}%
\providecommand \citenamefont [1]{#1}%
\providecommand \href@noop [0]{\@secondoftwo}%
\providecommand \href [0]{\begingroup \@sanitize@url \@href}%
\providecommand \@href[1]{\@@startlink{#1}\@@href}%
\providecommand \@@href[1]{\endgroup#1\@@endlink}%
\providecommand \@sanitize@url [0]{\catcode `\\12\catcode `\$12\catcode
  `\&12\catcode `\#12\catcode `\^12\catcode `\_12\catcode `\%12\relax}%
\providecommand \@@startlink[1]{}%
\providecommand \@@endlink[0]{}%
\providecommand \url  [0]{\begingroup\@sanitize@url \@url }%
\providecommand \@url [1]{\endgroup\@href {#1}{\urlprefix }}%
\providecommand \urlprefix  [0]{URL }%
\providecommand \Eprint [0]{\href }%
\providecommand \doibase [0]{http://dx.doi.org/}%
\providecommand \selectlanguage [0]{\@gobble}%
\providecommand \bibinfo  [0]{\@secondoftwo}%
\providecommand \bibfield  [0]{\@secondoftwo}%
\providecommand \translation [1]{[#1]}%
\providecommand \BibitemOpen [0]{}%
\providecommand \bibitemStop [0]{}%
\providecommand \bibitemNoStop [0]{.\EOS\space}%
\providecommand \EOS [0]{\spacefactor3000\relax}%
\providecommand \BibitemShut  [1]{\csname bibitem#1\endcsname}%
\let\auto@bib@innerbib\@empty
\bibitem [{\citenamefont {Qi}\ \emph {et~al.}(2010)\citenamefont {Qi},
  \citenamefont {Hughes},\ and\ \citenamefont {Zhang}}]{Qi10}%
  \BibitemOpen
  \bibfield  {author} {\bibinfo {author} {\bibfnamefont {Xiao-Liang}\
  \bibnamefont {Qi}}, \bibinfo {author} {\bibfnamefont {Taylor~L.}\
  \bibnamefont {Hughes}}, \ and\ \bibinfo {author} {\bibfnamefont {Shou-Cheng}\
  \bibnamefont {Zhang}},\ }\bibfield  {title} {\enquote {\bibinfo {title}
  {Chiral topological superconductor from the quantum hall state},}\ }\href
  {\doibase 10.1103/PhysRevB.82.184516} {\bibfield  {journal} {\bibinfo
  {journal} {Phys. Rev. B}\ }\textbf {\bibinfo {volume} {82}},\ \bibinfo
  {pages} {184516} (\bibinfo {year} {2010})},\ \Eprint
  {http://arxiv.org/abs/arXiv:1003.5448} {arXiv:1003.5448} \BibitemShut
  {NoStop}%
\bibitem [{\citenamefont {Lindner}\ \emph {et~al.}(2012)\citenamefont
  {Lindner}, \citenamefont {Berg}, \citenamefont {Refael},\ and\ \citenamefont
  {Stern}}]{Lindner12}%
  \BibitemOpen
  \bibfield  {author} {\bibinfo {author} {\bibfnamefont {Netanel~H.}\
  \bibnamefont {Lindner}}, \bibinfo {author} {\bibfnamefont {Erez}\
  \bibnamefont {Berg}}, \bibinfo {author} {\bibfnamefont {Gil}\ \bibnamefont
  {Refael}}, \ and\ \bibinfo {author} {\bibfnamefont {Ady}\ \bibnamefont
  {Stern}},\ }\bibfield  {title} {\enquote {\bibinfo {title} {Fractionalizing
  majorana fermions: Non-abelian statistics on the edges of abelian quantum
  hall states},}\ }\href {\doibase 10.1103/PhysRevX.2.041002} {\bibfield
  {journal} {\bibinfo  {journal} {Phys. Rev. X}\ }\textbf {\bibinfo {volume}
  {2}},\ \bibinfo {pages} {041002} (\bibinfo {year} {2012})},\ \Eprint
  {http://arxiv.org/abs/arXiv:1204.5733} {arXiv:1204.5733} \BibitemShut
  {NoStop}%
\bibitem [{\citenamefont {Cheng}(2012)}]{Cheng12}%
  \BibitemOpen
  \bibfield  {author} {\bibinfo {author} {\bibfnamefont {Meng}\ \bibnamefont
  {Cheng}},\ }\bibfield  {title} {\enquote {\bibinfo {title} {Superconducting
  proximity effect on the edge of fractional topological insulators},}\ }\href
  {\doibase 10.1103/PhysRevB.86.195126} {\bibfield  {journal} {\bibinfo
  {journal} {Phys. Rev. B}\ }\textbf {\bibinfo {volume} {86}},\ \bibinfo
  {pages} {195126} (\bibinfo {year} {2012})},\ \Eprint
  {http://arxiv.org/abs/arXiv:1204.6084} {arXiv:1204.6084} \BibitemShut
  {NoStop}%
\bibitem [{\citenamefont {Clarke}\ \emph {et~al.}(2013)\citenamefont {Clarke},
  \citenamefont {Alicea},\ and\ \citenamefont {Shtengel}}]{Clarke13}%
  \BibitemOpen
  \bibfield  {author} {\bibinfo {author} {\bibfnamefont {David~J.}\
  \bibnamefont {Clarke}}, \bibinfo {author} {\bibfnamefont {Jason}\
  \bibnamefont {Alicea}}, \ and\ \bibinfo {author} {\bibfnamefont {Kirill}\
  \bibnamefont {Shtengel}},\ }\bibfield  {title} {\enquote {\bibinfo {title}
  {Exotic non-abelian anyons from conventional fractional quantum hall
  states},}\ }\href {\doibase 10.1038/ncomms2340} {\bibfield  {journal}
  {\bibinfo  {journal} {Nature Communications}\ }\textbf {\bibinfo {volume}
  {4}} (\bibinfo {year} {2013}),\ 10.1038/ncomms2340},\ \Eprint
  {http://arxiv.org/abs/arXiv:1204.5479} {arXiv:1204.5479} \BibitemShut
  {NoStop}%
\bibitem [{\citenamefont {Vaezi}(2013)}]{Vaezi13}%
  \BibitemOpen
  \bibfield  {author} {\bibinfo {author} {\bibfnamefont {Abolhassan}\
  \bibnamefont {Vaezi}},\ }\bibfield  {title} {\enquote {\bibinfo {title}
  {Fractional topological superconductor with fractionalized majorana
  fermions},}\ }\href {\doibase 10.1103/PhysRevB.87.035132} {\bibfield
  {journal} {\bibinfo  {journal} {Phys. Rev. B}\ }\textbf {\bibinfo {volume}
  {87}},\ \bibinfo {pages} {035132} (\bibinfo {year} {2013})},\ \Eprint
  {http://arxiv.org/abs/arXiv:1204.6245} {arXiv:1204.6245} \BibitemShut
  {NoStop}%
\bibitem [{\citenamefont {Vaezi}(2014)}]{Vaezi14}%
  \BibitemOpen
  \bibfield  {author} {\bibinfo {author} {\bibfnamefont {Abolhassan}\
  \bibnamefont {Vaezi}},\ }\bibfield  {title} {\enquote {\bibinfo {title}
  {Superconducting analogue of the parafermion fractional quantum hall
  states},}\ }\href {\doibase 10.1103/PhysRevX.4.031009} {\bibfield  {journal}
  {\bibinfo  {journal} {Phys. Rev. X}\ }\textbf {\bibinfo {volume} {4}},\
  \bibinfo {pages} {031009} (\bibinfo {year} {2014})},\ \Eprint
  {http://arxiv.org/abs/arXiv:1307.8069} {arXiv:1307.8069} \BibitemShut
  {NoStop}%
\bibitem [{\citenamefont {Mong}\ \emph {et~al.}(2014)\citenamefont {Mong},
  \citenamefont {Clarke}, \citenamefont {Alicea}, \citenamefont {Lindner},
  \citenamefont {Fendley}, \citenamefont {Nayak}, \citenamefont {Oreg},
  \citenamefont {Stern}, \citenamefont {Berg}, \citenamefont {Shtengel},\ and\
  \citenamefont {Fisher}}]{Mong14}%
  \BibitemOpen
  \bibfield  {author} {\bibinfo {author} {\bibfnamefont {Roger S.~K.}\
  \bibnamefont {Mong}}, \bibinfo {author} {\bibfnamefont {David~J.}\
  \bibnamefont {Clarke}}, \bibinfo {author} {\bibfnamefont {Jason}\
  \bibnamefont {Alicea}}, \bibinfo {author} {\bibfnamefont {Netanel~H.}\
  \bibnamefont {Lindner}}, \bibinfo {author} {\bibfnamefont {Paul}\
  \bibnamefont {Fendley}}, \bibinfo {author} {\bibfnamefont {Chetan}\
  \bibnamefont {Nayak}}, \bibinfo {author} {\bibfnamefont {Yuval}\ \bibnamefont
  {Oreg}}, \bibinfo {author} {\bibfnamefont {Ady}\ \bibnamefont {Stern}},
  \bibinfo {author} {\bibfnamefont {Erez}\ \bibnamefont {Berg}}, \bibinfo
  {author} {\bibfnamefont {Kirill}\ \bibnamefont {Shtengel}}, \ and\ \bibinfo
  {author} {\bibfnamefont {Matthew P.~A.}\ \bibnamefont {Fisher}},\ }\bibfield
  {title} {\enquote {\bibinfo {title} {Universal topological quantum
  computation from a superconductor-abelian quantum hall heterostructure},}\
  }\href {\doibase 10.1103/PhysRevX.4.011036} {\bibfield  {journal} {\bibinfo
  {journal} {Phys. Rev. X}\ }\textbf {\bibinfo {volume} {4}},\ \bibinfo {pages}
  {011036} (\bibinfo {year} {2014})},\ \Eprint
  {http://arxiv.org/abs/arXiv:1307.4403} {arXiv:1307.4403} \BibitemShut
  {NoStop}%
\bibitem [{\citenamefont {Clarke}\ \emph {et~al.}(2014)\citenamefont {Clarke},
  \citenamefont {Alicea},\ and\ \citenamefont {Shtengel}}]{Clarke14}%
  \BibitemOpen
  \bibfield  {author} {\bibinfo {author} {\bibfnamefont {David~J.}\
  \bibnamefont {Clarke}}, \bibinfo {author} {\bibfnamefont {Jason}\
  \bibnamefont {Alicea}}, \ and\ \bibinfo {author} {\bibfnamefont {Kirill}\
  \bibnamefont {Shtengel}},\ }\bibfield  {title} {\enquote {\bibinfo {title}
  {Exotic circuit elements from zero-modes in hybrid
  superconductor{\textendash}quantum-hall systems},}\ }\href {\doibase
  10.1038/nphys3114} {\bibfield  {journal} {\bibinfo  {journal} {Nature
  Physics}\ }\textbf {\bibinfo {volume} {10}},\ \bibinfo {pages} {877--882}
  (\bibinfo {year} {2014})},\ \Eprint {http://arxiv.org/abs/arXiv:1312.6123}
  {arXiv:1312.6123} \BibitemShut {NoStop}%
\bibitem [{\citenamefont {Alicea}\ and\ \citenamefont
  {Fendley}(2016)}]{Alicea16}%
  \BibitemOpen
  \bibfield  {author} {\bibinfo {author} {\bibfnamefont {Jason}\ \bibnamefont
  {Alicea}}\ and\ \bibinfo {author} {\bibfnamefont {Paul}\ \bibnamefont
  {Fendley}},\ }\bibfield  {title} {\enquote {\bibinfo {title} {Topological
  phases with parafermions: Theory and blueprints},}\ }\href {\doibase
  10.1146/annurev-conmatphys-031115-011336} {\bibfield  {journal} {\bibinfo
  {journal} {Annual Review of Condensed Matter Physics}\ }\textbf {\bibinfo
  {volume} {7}},\ \bibinfo {pages} {119--139} (\bibinfo {year} {2016})},\
  \Eprint {http://arxiv.org/abs/arXiv:1504.02476} {arXiv:1504.02476}
  \BibitemShut {NoStop}%
\bibitem [{\citenamefont {Takayanagi}\ and\ \citenamefont
  {Akazaki}(1998)}]{Takayanagi98}%
  \BibitemOpen
  \bibfield  {author} {\bibinfo {author} {\bibfnamefont {Hideaki}\ \bibnamefont
  {Takayanagi}}\ and\ \bibinfo {author} {\bibfnamefont {Tatsushi}\ \bibnamefont
  {Akazaki}},\ }\bibfield  {title} {\enquote {\bibinfo {title}
  {Semiconductor-coupled superconducting junctions using nbn electrodes with
  high ${H}_{c2}$ and ${T}_c$},}\ }\href@noop {} {\bibfield  {journal}
  {\bibinfo  {journal} {Physica B: Condensed Matter}\ }\textbf {\bibinfo
  {volume} {249-251}},\ \bibinfo {pages} {462--466} (\bibinfo {year}
  {1998})}\BibitemShut {NoStop}%
\bibitem [{\citenamefont {Komatsu}\ \emph {et~al.}(2012)\citenamefont
  {Komatsu}, \citenamefont {Li}, \citenamefont {Autier-Laurent}, \citenamefont
  {Bouchiat},\ and\ \citenamefont {Gu\'eron}}]{Komatsu12}%
  \BibitemOpen
  \bibfield  {author} {\bibinfo {author} {\bibfnamefont {Katsuyoshi}\
  \bibnamefont {Komatsu}}, \bibinfo {author} {\bibfnamefont {Chuan}\
  \bibnamefont {Li}}, \bibinfo {author} {\bibfnamefont {S.}~\bibnamefont
  {Autier-Laurent}}, \bibinfo {author} {\bibfnamefont {H.}~\bibnamefont
  {Bouchiat}}, \ and\ \bibinfo {author} {\bibfnamefont {S.}~\bibnamefont
  {Gu\'eron}},\ }\bibfield  {title} {\enquote {\bibinfo {title}
  {Superconducting proximity effect in long
  superconductor/graphene/superconductor junctions: From specular andreev
  reflection at zero field to the quantum hall regime},}\ }\href {\doibase
  10.1103/PhysRevB.86.115412} {\bibfield  {journal} {\bibinfo  {journal} {Phys.
  Rev. B}\ }\textbf {\bibinfo {volume} {86}},\ \bibinfo {pages} {115412}
  (\bibinfo {year} {2012})}\BibitemShut {NoStop}%
\bibitem [{\citenamefont {Rickhaus}\ \emph {et~al.}(2012)\citenamefont
  {Rickhaus}, \citenamefont {Weiss}, \citenamefont {Marot},\ and\ \citenamefont
  {Schönenberger}}]{Rickhaus12}%
  \BibitemOpen
  \bibfield  {author} {\bibinfo {author} {\bibfnamefont {Peter}\ \bibnamefont
  {Rickhaus}}, \bibinfo {author} {\bibfnamefont {Markus}\ \bibnamefont
  {Weiss}}, \bibinfo {author} {\bibfnamefont {Laurent}\ \bibnamefont {Marot}},
  \ and\ \bibinfo {author} {\bibfnamefont {Christian}\ \bibnamefont
  {Schönenberger}},\ }\bibfield  {title} {\enquote {\bibinfo {title} {Quantum
  hall effect in graphene with superconducting electrodes},}\ }\href {\doibase
  10.1021/nl204415s} {\bibfield  {journal} {\bibinfo  {journal} {Nano Letters}\
  }\textbf {\bibinfo {volume} {12}},\ \bibinfo {pages} {1942–1945} (\bibinfo
  {year} {2012})},\ \Eprint {http://arxiv.org/abs/arXiv:1303.3394}
  {arXiv:1303.3394} \BibitemShut {NoStop}%
\bibitem [{\citenamefont {Wan}\ \emph {et~al.}(2015)\citenamefont {Wan},
  \citenamefont {Kazakov}, \citenamefont {Manfra}, \citenamefont {Pfeiffer},
  \citenamefont {West},\ and\ \citenamefont {Rokhinson}}]{Zhong15}%
  \BibitemOpen
  \bibfield  {author} {\bibinfo {author} {\bibfnamefont {Zhong}\ \bibnamefont
  {Wan}}, \bibinfo {author} {\bibfnamefont {Aleksandr}\ \bibnamefont
  {Kazakov}}, \bibinfo {author} {\bibfnamefont {Michael~J.}\ \bibnamefont
  {Manfra}}, \bibinfo {author} {\bibfnamefont {Loren~N.}\ \bibnamefont
  {Pfeiffer}}, \bibinfo {author} {\bibfnamefont {Ken~W.}\ \bibnamefont {West}},
  \ and\ \bibinfo {author} {\bibfnamefont {Leonid~P.}\ \bibnamefont
  {Rokhinson}},\ }\bibfield  {title} {\enquote {\bibinfo {title} {Induced
  superconductivity in high-mobility two-dimensional electron gas in gallium
  arsenide heterostructures},}\ }\href {\doibase 10.1038/ncomms8426} {\bibfield
   {journal} {\bibinfo  {journal} {Nature Communications}\ }\textbf {\bibinfo
  {volume} {6}} (\bibinfo {year} {2015}),\ 10.1038/ncomms8426},\ \Eprint
  {http://arxiv.org/abs/arXiv:1503.09138} {arXiv:1503.09138} \BibitemShut
  {NoStop}%
\bibitem [{\citenamefont {Amet}\ \emph {et~al.}(2016)\citenamefont {Amet},
  \citenamefont {Ke}, \citenamefont {Borzenets}, \citenamefont {Wang},
  \citenamefont {Watanabe}, \citenamefont {Taniguchi}, \citenamefont {Deacon},
  \citenamefont {Yamamoto}, \citenamefont {Bomze}, \citenamefont {Tarucha},\
  and\ \citenamefont {et~al.}}]{Amet16}%
  \BibitemOpen
  \bibfield  {author} {\bibinfo {author} {\bibfnamefont {F.}~\bibnamefont
  {Amet}}, \bibinfo {author} {\bibfnamefont {C.~T.}\ \bibnamefont {Ke}},
  \bibinfo {author} {\bibfnamefont {I.~V.}\ \bibnamefont {Borzenets}}, \bibinfo
  {author} {\bibfnamefont {J.}~\bibnamefont {Wang}}, \bibinfo {author}
  {\bibfnamefont {K.}~\bibnamefont {Watanabe}}, \bibinfo {author}
  {\bibfnamefont {T.}~\bibnamefont {Taniguchi}}, \bibinfo {author}
  {\bibfnamefont {R.~S.}\ \bibnamefont {Deacon}}, \bibinfo {author}
  {\bibfnamefont {M.}~\bibnamefont {Yamamoto}}, \bibinfo {author}
  {\bibfnamefont {Y.}~\bibnamefont {Bomze}}, \bibinfo {author} {\bibfnamefont
  {S.}~\bibnamefont {Tarucha}}, \ and\ \bibinfo {author} {\bibnamefont
  {et~al.}},\ }\bibfield  {title} {\enquote {\bibinfo {title} {Supercurrent in
  the quantum hall regime},}\ }\href {\doibase 10.1126/science.aad6203}
  {\bibfield  {journal} {\bibinfo  {journal} {Science}\ }\textbf {\bibinfo
  {volume} {352}},\ \bibinfo {pages} {966–969} (\bibinfo {year} {2016})},\
  \Eprint {http://arxiv.org/abs/arXiv:1512.09083} {arXiv:1512.09083}
  \BibitemShut {NoStop}%
\bibitem [{\citenamefont {Lee}\ \emph {et~al.}(2017)\citenamefont {Lee},
  \citenamefont {Huang}, \citenamefont {Efetov}, \citenamefont {Wei},
  \citenamefont {Hart}, \citenamefont {Taniguchi}, \citenamefont {Watanabe},
  \citenamefont {Yacoby},\ and\ \citenamefont {Kim}}]{Lee17}%
  \BibitemOpen
  \bibfield  {author} {\bibinfo {author} {\bibfnamefont {Gil-Ho}\ \bibnamefont
  {Lee}}, \bibinfo {author} {\bibfnamefont {Ko-Fan}\ \bibnamefont {Huang}},
  \bibinfo {author} {\bibfnamefont {Dmitri~K.}\ \bibnamefont {Efetov}},
  \bibinfo {author} {\bibfnamefont {Di~S.}\ \bibnamefont {Wei}}, \bibinfo
  {author} {\bibfnamefont {Sean}\ \bibnamefont {Hart}}, \bibinfo {author}
  {\bibfnamefont {Takashi}\ \bibnamefont {Taniguchi}}, \bibinfo {author}
  {\bibfnamefont {Kenji}\ \bibnamefont {Watanabe}}, \bibinfo {author}
  {\bibfnamefont {Amir}\ \bibnamefont {Yacoby}}, \ and\ \bibinfo {author}
  {\bibfnamefont {Philip}\ \bibnamefont {Kim}},\ }\bibfield  {title} {\enquote
  {\bibinfo {title} {Inducing superconducting correlation in quantum hall edge
  states},}\ }\href {\doibase 10.1038/nphys4084} {\bibfield  {journal}
  {\bibinfo  {journal} {Nature Physics}\ }\textbf {\bibinfo {volume} {13}},\
  \bibinfo {pages} {693–698} (\bibinfo {year} {2017})},\ \Eprint
  {http://arxiv.org/abs/arXiv:1609.08104} {arXiv:1609.08104} \BibitemShut
  {NoStop}%
\bibitem [{\citenamefont {Park}\ \emph {et~al.}(2017)\citenamefont {Park},
  \citenamefont {Kim}, \citenamefont {Watanabe}, \citenamefont {Taniguchi},\
  and\ \citenamefont {Lee}}]{Park17}%
  \BibitemOpen
  \bibfield  {author} {\bibinfo {author} {\bibfnamefont {Geon-Hyoung}\
  \bibnamefont {Park}}, \bibinfo {author} {\bibfnamefont {Minsoo}\ \bibnamefont
  {Kim}}, \bibinfo {author} {\bibfnamefont {Kenji}\ \bibnamefont {Watanabe}},
  \bibinfo {author} {\bibfnamefont {Takashi}\ \bibnamefont {Taniguchi}}, \ and\
  \bibinfo {author} {\bibfnamefont {Hu-Jong}\ \bibnamefont {Lee}},\ }\bibfield
  {title} {\enquote {\bibinfo {title} {Propagation of superconducting coherence
  via chiral quantum-hall edge channels},}\ }\href {\doibase
  10.1038/s41598-017-11209-w} {\bibfield  {journal} {\bibinfo  {journal}
  {Scientific Reports}\ }\textbf {\bibinfo {volume} {7}} (\bibinfo {year}
  {2017}),\ 10.1038/s41598-017-11209-w}\BibitemShut {NoStop}%
\bibitem [{\citenamefont {Guiducci}\ \emph {et~al.}(2018)\citenamefont
  {Guiducci}, \citenamefont {Carrega}, \citenamefont {Biasiol}, \citenamefont
  {Sorba}, \citenamefont {Beltram},\ and\ \citenamefont {Heun}}]{Guiducci18}%
  \BibitemOpen
  \bibfield  {author} {\bibinfo {author} {\bibfnamefont {Stefano}\ \bibnamefont
  {Guiducci}}, \bibinfo {author} {\bibfnamefont {Matteo}\ \bibnamefont
  {Carrega}}, \bibinfo {author} {\bibfnamefont {Giorgio}\ \bibnamefont
  {Biasiol}}, \bibinfo {author} {\bibfnamefont {Lucia}\ \bibnamefont {Sorba}},
  \bibinfo {author} {\bibfnamefont {Fabio}\ \bibnamefont {Beltram}}, \ and\
  \bibinfo {author} {\bibfnamefont {Stefan}\ \bibnamefont {Heun}},\ }\bibfield
  {title} {\enquote {\bibinfo {title} {Toward quantum hall effect in a
  josephson junction},}\ }\href {\doibase 10.1002/pssr.201800222} {\bibfield
  {journal} {\bibinfo  {journal} {Phys. Status solidi ({RRL}) - Rapid Research
  Letters}\ }\textbf {\bibinfo {volume} {13}},\ \bibinfo {pages} {1800222}
  (\bibinfo {year} {2018})},\ \Eprint {http://arxiv.org/abs/arXiv:1805.02862}
  {arXiv:1805.02862} \BibitemShut {NoStop}%
\bibitem [{\citenamefont {Draelos}\ \emph {et~al.}(2018)\citenamefont
  {Draelos}, \citenamefont {Wei}, \citenamefont {Seredinski}, \citenamefont
  {Ke}, \citenamefont {Mehta}, \citenamefont {Chamberlain}, \citenamefont
  {Watanabe}, \citenamefont {Taniguchi}, \citenamefont {Yamamoto},
  \citenamefont {Tarucha}, \citenamefont {Borzenets}, \citenamefont {Amet},\
  and\ \citenamefont {Finkelstein}}]{Draelos18}%
  \BibitemOpen
  \bibfield  {author} {\bibinfo {author} {\bibfnamefont {Anne~W.}\ \bibnamefont
  {Draelos}}, \bibinfo {author} {\bibfnamefont {Ming~Tso}\ \bibnamefont {Wei}},
  \bibinfo {author} {\bibfnamefont {Andrew}\ \bibnamefont {Seredinski}},
  \bibinfo {author} {\bibfnamefont {Chung~Ting}\ \bibnamefont {Ke}}, \bibinfo
  {author} {\bibfnamefont {Yash}\ \bibnamefont {Mehta}}, \bibinfo {author}
  {\bibfnamefont {Russell}\ \bibnamefont {Chamberlain}}, \bibinfo {author}
  {\bibfnamefont {Kenji}\ \bibnamefont {Watanabe}}, \bibinfo {author}
  {\bibfnamefont {Takashi}\ \bibnamefont {Taniguchi}}, \bibinfo {author}
  {\bibfnamefont {Michihisa}\ \bibnamefont {Yamamoto}}, \bibinfo {author}
  {\bibfnamefont {Seigo}\ \bibnamefont {Tarucha}}, \bibinfo {author}
  {\bibfnamefont {Ivan~V.}\ \bibnamefont {Borzenets}}, \bibinfo {author}
  {\bibfnamefont {Fran{\c{c}}ois}\ \bibnamefont {Amet}}, \ and\ \bibinfo
  {author} {\bibfnamefont {Gleb}\ \bibnamefont {Finkelstein}},\ }\bibfield
  {title} {\enquote {\bibinfo {title} {Investigation of supercurrent in the
  quantum hall regime in graphene josephson junctions},}\ }\href {\doibase
  10.1007/s10909-018-1872-9} {\bibfield  {journal} {\bibinfo  {journal}
  {Journal of Low Temperature Physics}\ }\textbf {\bibinfo {volume} {191}},\
  \bibinfo {pages} {288--300} (\bibinfo {year} {2018})},\ \Eprint
  {http://arxiv.org/abs/arXiv:1801.01447} {arXiv:1801.01447} \BibitemShut
  {NoStop}%
\bibitem [{\citenamefont {Seredinski}\ \emph {et~al.}(2019)\citenamefont
  {Seredinski}, \citenamefont {Draelos}, \citenamefont {Arnault}, \citenamefont
  {Wei}, \citenamefont {Li}, \citenamefont {Fleming}, \citenamefont {Watanabe},
  \citenamefont {Taniguchi}, \citenamefont {Amet},\ and\ \citenamefont
  {Finkelstein}}]{Seredinski19}%
  \BibitemOpen
  \bibfield  {author} {\bibinfo {author} {\bibfnamefont {Andrew}\ \bibnamefont
  {Seredinski}}, \bibinfo {author} {\bibfnamefont {Anne}\ \bibnamefont
  {Draelos}}, \bibinfo {author} {\bibfnamefont {Ethan}\ \bibnamefont
  {Arnault}}, \bibinfo {author} {\bibfnamefont {Ming-Tso}\ \bibnamefont {Wei}},
  \bibinfo {author} {\bibfnamefont {Hengming}\ \bibnamefont {Li}}, \bibinfo
  {author} {\bibfnamefont {Tate}\ \bibnamefont {Fleming}}, \bibinfo {author}
  {\bibfnamefont {Kenji}\ \bibnamefont {Watanabe}}, \bibinfo {author}
  {\bibfnamefont {Takashi}\ \bibnamefont {Taniguchi}}, \bibinfo {author}
  {\bibfnamefont {François}\ \bibnamefont {Amet}}, \ and\ \bibinfo {author}
  {\bibfnamefont {Gleb}\ \bibnamefont {Finkelstein}},\ }\bibfield  {title}
  {\enquote {\bibinfo {title} {Quantum hall–based superconducting
  interference device},}\ }\href {\doibase 10.1126/sciadv.aaw8693} {\bibfield
  {journal} {\bibinfo  {journal} {Science Advances}\ }\textbf {\bibinfo
  {volume} {5}},\ \bibinfo {pages} {eaaw8693} (\bibinfo {year}
  {2019})}\BibitemShut {NoStop}%
\bibitem [{\citenamefont {Zhi}\ \emph {et~al.}(2019)\citenamefont {Zhi},
  \citenamefont {Kang}, \citenamefont {Su}, \citenamefont {Fan}, \citenamefont
  {Li}, \citenamefont {Pan}, \citenamefont {Zhao}, \citenamefont {Zhao},\ and\
  \citenamefont {Xu}}]{Zhi19}%
  \BibitemOpen
  \bibfield  {author} {\bibinfo {author} {\bibfnamefont {Jinhua}\ \bibnamefont
  {Zhi}}, \bibinfo {author} {\bibfnamefont {Ning}\ \bibnamefont {Kang}},
  \bibinfo {author} {\bibfnamefont {Feifan}\ \bibnamefont {Su}}, \bibinfo
  {author} {\bibfnamefont {Dingxun}\ \bibnamefont {Fan}}, \bibinfo {author}
  {\bibfnamefont {Sen}\ \bibnamefont {Li}}, \bibinfo {author} {\bibfnamefont
  {Dong}\ \bibnamefont {Pan}}, \bibinfo {author} {\bibfnamefont {S.~P.}\
  \bibnamefont {Zhao}}, \bibinfo {author} {\bibfnamefont {Jianhua}\
  \bibnamefont {Zhao}}, \ and\ \bibinfo {author} {\bibfnamefont {H.~Q.}\
  \bibnamefont {Xu}},\ }\bibfield  {title} {\enquote {\bibinfo {title}
  {Coexistence of induced superconductivity and quantum hall states in insb
  nanosheets},}\ }\href {\doibase 10.1103/PhysRevB.99.245302} {\bibfield
  {journal} {\bibinfo  {journal} {Phys. Rev. B}\ }\textbf {\bibinfo {volume}
  {99}},\ \bibinfo {pages} {245302} (\bibinfo {year} {2019})}\BibitemShut
  {NoStop}%
\bibitem [{\citenamefont {{Zhao}}\ \emph {et~al.}(2020)\citenamefont {{Zhao}},
  \citenamefont {{Arnault}}, \citenamefont {{Bondarev}}, \citenamefont
  {{Seredinski}}, \citenamefont {{Larson}}, \citenamefont {{Draelos}},
  \citenamefont {{Li}}, \citenamefont {{Watanabe}}, \citenamefont
  {{Taniguchi}}, \citenamefont {{Amet}}, \citenamefont {{Baranger}},\ and\
  \citenamefont {{Finkelstein}}}]{Zhao20}%
  \BibitemOpen
  \bibfield  {author} {\bibinfo {author} {\bibfnamefont {Lingfei}\ \bibnamefont
  {{Zhao}}}, \bibinfo {author} {\bibfnamefont {Ethan~G.}\ \bibnamefont
  {{Arnault}}}, \bibinfo {author} {\bibfnamefont {Alexey}\ \bibnamefont
  {{Bondarev}}}, \bibinfo {author} {\bibfnamefont {Andrew}\ \bibnamefont
  {{Seredinski}}}, \bibinfo {author} {\bibfnamefont {Trevyn F.~Q.}\
  \bibnamefont {{Larson}}}, \bibinfo {author} {\bibfnamefont {Anne~W.}\
  \bibnamefont {{Draelos}}}, \bibinfo {author} {\bibfnamefont {Hengming}\
  \bibnamefont {{Li}}}, \bibinfo {author} {\bibfnamefont {Kenji}\ \bibnamefont
  {{Watanabe}}}, \bibinfo {author} {\bibfnamefont {Takashi}\ \bibnamefont
  {{Taniguchi}}}, \bibinfo {author} {\bibfnamefont {Fran{\c{c}}ois}\
  \bibnamefont {{Amet}}}, \bibinfo {author} {\bibfnamefont {Harold~U.}\
  \bibnamefont {{Baranger}}}, \ and\ \bibinfo {author} {\bibfnamefont {Gleb}\
  \bibnamefont {{Finkelstein}}},\ }\bibfield  {title} {\enquote {\bibinfo
  {title} {{Interference of chiral Andreev edge states}},}\ }\href {\doibase
  10.1038/s41567-020-0898-5} {\bibfield  {journal} {\bibinfo  {journal} {Nature
  Physics}\ }\textbf {\bibinfo {volume} {16}},\ \bibinfo {pages} {862--867}
  (\bibinfo {year} {2020})},\ \Eprint {http://arxiv.org/abs/arXiv:1907.01722}
  {arXiv:1907.01722} \BibitemShut {NoStop}%
\bibitem [{\citenamefont {Önder Gül}\ \emph {et~al.}(2022)\citenamefont
  {Önder Gül}, \citenamefont {Ronen}, \citenamefont {Lee}, \citenamefont
  {Shapourian}, \citenamefont {Zauberman}, \citenamefont {Lee}, \citenamefont
  {Watanabe}, \citenamefont {Taniguchi}, \citenamefont {Vishwanath},
  \citenamefont {Yacoby},\ and\ \citenamefont {Kim}}]{Gul20}%
  \BibitemOpen
  \bibfield  {author} {\bibinfo {author} {\bibnamefont {Önder Gül}}, \bibinfo
  {author} {\bibfnamefont {Yuval}\ \bibnamefont {Ronen}}, \bibinfo {author}
  {\bibfnamefont {Si~Young}\ \bibnamefont {Lee}}, \bibinfo {author}
  {\bibfnamefont {Hassan}\ \bibnamefont {Shapourian}}, \bibinfo {author}
  {\bibfnamefont {Jonathan}\ \bibnamefont {Zauberman}}, \bibinfo {author}
  {\bibfnamefont {Young~Hee}\ \bibnamefont {Lee}}, \bibinfo {author}
  {\bibfnamefont {Kenji}\ \bibnamefont {Watanabe}}, \bibinfo {author}
  {\bibfnamefont {Takashi}\ \bibnamefont {Taniguchi}}, \bibinfo {author}
  {\bibfnamefont {Ashvin}\ \bibnamefont {Vishwanath}}, \bibinfo {author}
  {\bibfnamefont {Amir}\ \bibnamefont {Yacoby}}, \ and\ \bibinfo {author}
  {\bibfnamefont {Philip}\ \bibnamefont {Kim}},\ }\bibfield  {title} {\enquote
  {\bibinfo {title} {Andreev reflection in the fractional quantum hall
  state},}\ }\href {\doibase 10.1103/physrevx.12.021057} {\bibfield  {journal}
  {\bibinfo  {journal} {Physical Review X}\ }\textbf {\bibinfo {volume} {12}}
  (\bibinfo {year} {2022}),\ 10.1103/physrevx.12.021057},\ \Eprint
  {http://arxiv.org/abs/arXiv:2009.07836} {arXiv:2009.07836} \BibitemShut
  {NoStop}%
\bibitem [{\citenamefont {Hatefipour}\ \emph {et~al.}(2021)\citenamefont
  {Hatefipour}, \citenamefont {Cuozzo}, \citenamefont {Kanter}, \citenamefont
  {Strickland}, \citenamefont {Lu}, \citenamefont {Rossi},\ and\ \citenamefont
  {Shabani}}]{Hatefipour21}%
  \BibitemOpen
  \bibfield  {author} {\bibinfo {author} {\bibfnamefont {Mehdi}\ \bibnamefont
  {Hatefipour}}, \bibinfo {author} {\bibfnamefont {Joseph~J.}\ \bibnamefont
  {Cuozzo}}, \bibinfo {author} {\bibfnamefont {Jesse}\ \bibnamefont {Kanter}},
  \bibinfo {author} {\bibfnamefont {William}\ \bibnamefont {Strickland}},
  \bibinfo {author} {\bibfnamefont {Tzu-Ming}\ \bibnamefont {Lu}}, \bibinfo
  {author} {\bibfnamefont {Enrico}\ \bibnamefont {Rossi}}, \ and\ \bibinfo
  {author} {\bibfnamefont {Javad}\ \bibnamefont {Shabani}},\ }\href@noop {}
  {\enquote {\bibinfo {title} {Induced superconducting pairing in integer
  quantum hall edge states},}\ } (\bibinfo {year} {2021}),\ \Eprint
  {http://arxiv.org/abs/arXiv:2108.08899} {arXiv:2108.08899} \BibitemShut
  {NoStop}%
\bibitem [{\citenamefont {van Ostaay}\ \emph {et~al.}(2011)\citenamefont {van
  Ostaay}, \citenamefont {Akhmerov},\ and\ \citenamefont
  {Beenakker}}]{Ostaay11}%
  \BibitemOpen
  \bibfield  {author} {\bibinfo {author} {\bibfnamefont {J.~A.~M.}\
  \bibnamefont {van Ostaay}}, \bibinfo {author} {\bibfnamefont {A.~R.}\
  \bibnamefont {Akhmerov}}, \ and\ \bibinfo {author} {\bibfnamefont {C.~W.~J.}\
  \bibnamefont {Beenakker}},\ }\bibfield  {title} {\enquote {\bibinfo {title}
  {Spin-triplet supercurrent carried by quantum hall edge states through a
  josephson junction},}\ }\href {\doibase 10.1103/PhysRevB.83.195441}
  {\bibfield  {journal} {\bibinfo  {journal} {Phys. Rev. B}\ }\textbf {\bibinfo
  {volume} {83}},\ \bibinfo {pages} {195441} (\bibinfo {year} {2011})},\
  \Eprint {http://arxiv.org/abs/arXiv:1103.0887} {arXiv:1103.0887} \BibitemShut
  {NoStop}%
\bibitem [{\citenamefont {Zocher}\ and\ \citenamefont
  {Rosenow}(2016)}]{Zocher16}%
  \BibitemOpen
  \bibfield  {author} {\bibinfo {author} {\bibfnamefont {Bj\"orn}\ \bibnamefont
  {Zocher}}\ and\ \bibinfo {author} {\bibfnamefont {Bernd}\ \bibnamefont
  {Rosenow}},\ }\bibfield  {title} {\enquote {\bibinfo {title} {Topological
  superconductivity in quantum hall--superconductor hybrid systems},}\ }\href
  {\doibase 10.1103/PhysRevB.93.214504} {\bibfield  {journal} {\bibinfo
  {journal} {Phys. Rev. B}\ }\textbf {\bibinfo {volume} {93}},\ \bibinfo
  {pages} {214504} (\bibinfo {year} {2016})},\ \Eprint
  {http://arxiv.org/abs/arXiv:1408.6148} {arXiv:1408.6148} \BibitemShut
  {NoStop}%
\bibitem [{\citenamefont {Huang}\ and\ \citenamefont
  {Nazarov}(2017)}]{Huang17}%
  \BibitemOpen
  \bibfield  {author} {\bibinfo {author} {\bibfnamefont {Xiao-Li}\ \bibnamefont
  {Huang}}\ and\ \bibinfo {author} {\bibfnamefont {Yuli~V.}\ \bibnamefont
  {Nazarov}},\ }\bibfield  {title} {\enquote {\bibinfo {title} {Supercurrents
  in unidirectional channels originate from information transfer in the
  opposite direction: A theoretical prediction},}\ }\href {\doibase
  10.1103/PhysRevLett.118.177001} {\bibfield  {journal} {\bibinfo  {journal}
  {Phys. Rev. Lett.}\ }\textbf {\bibinfo {volume} {118}},\ \bibinfo {pages}
  {177001} (\bibinfo {year} {2017})},\ \Eprint
  {http://arxiv.org/abs/arXiv:1607.05940} {arXiv:1607.05940} \BibitemShut
  {NoStop}%
\bibitem [{\citenamefont {Gamayun}\ \emph {et~al.}(2017)\citenamefont
  {Gamayun}, \citenamefont {Hutasoit},\ and\ \citenamefont
  {Cheianov}}]{Gamayun17}%
  \BibitemOpen
  \bibfield  {author} {\bibinfo {author} {\bibfnamefont {Oleksandr}\
  \bibnamefont {Gamayun}}, \bibinfo {author} {\bibfnamefont {Jimmy~A.}\
  \bibnamefont {Hutasoit}}, \ and\ \bibinfo {author} {\bibfnamefont {Vadim~V.}\
  \bibnamefont {Cheianov}},\ }\bibfield  {title} {\enquote {\bibinfo {title}
  {Two-terminal transport along a proximity-induced superconducting quantum
  hall edge},}\ }\href@noop {} {\bibfield  {journal} {\bibinfo  {journal}
  {Physical Review B}\ }\textbf {\bibinfo {volume} {96}} (\bibinfo {year}
  {2017})},\ \Eprint {http://arxiv.org/abs/arXiv:1705.01546} {arXiv:1705.01546}
  \BibitemShut {NoStop}%
\bibitem [{\citenamefont {Chaudhary}\ and\ \citenamefont
  {MacDonald}(2020)}]{Chaudhary20}%
  \BibitemOpen
  \bibfield  {author} {\bibinfo {author} {\bibfnamefont {Gaurav}\ \bibnamefont
  {Chaudhary}}\ and\ \bibinfo {author} {\bibfnamefont {Allan~H.}\ \bibnamefont
  {MacDonald}},\ }\bibfield  {title} {\enquote {\bibinfo {title}
  {Vortex-lattice structure and topological superconductivity in the quantum
  hall regime},}\ }\href {\doibase 10.1103/PhysRevB.101.024516} {\bibfield
  {journal} {\bibinfo  {journal} {Phys. Rev. B}\ }\textbf {\bibinfo {volume}
  {101}},\ \bibinfo {pages} {024516} (\bibinfo {year} {2020})},\ \Eprint
  {http://arxiv.org/abs/arXiv:1903.12249} {arXiv:1903.12249} \BibitemShut
  {NoStop}%
\bibitem [{\citenamefont {Manesco}\ \emph {et~al.}(2021)\citenamefont
  {Manesco}, \citenamefont {Flór}, \citenamefont {Liu},\ and\ \citenamefont
  {Akhmerov}}]{Manesco21}%
  \BibitemOpen
  \bibfield  {author} {\bibinfo {author} {\bibfnamefont {Antonio L.~R.}\
  \bibnamefont {Manesco}}, \bibinfo {author} {\bibfnamefont {Ian~Matthias}\
  \bibnamefont {Flór}}, \bibinfo {author} {\bibfnamefont {Chun-Xiao}\
  \bibnamefont {Liu}}, \ and\ \bibinfo {author} {\bibfnamefont {Anton~R.}\
  \bibnamefont {Akhmerov}},\ }\href {\doibase 10.48550/ARXIV.2103.06722}
  {\enquote {\bibinfo {title} {Mechanisms of andreev reflection in quantum hall
  graphene},}\ } (\bibinfo {year} {2021}),\ \Eprint
  {http://arxiv.org/abs/arXiv:2103.06722} {arXiv:2103.06722} \BibitemShut
  {NoStop}%
\bibitem [{\citenamefont {Nikolaenko}\ and\ \citenamefont
  {Pientka}(2021)}]{Nikolaenko21}%
  \BibitemOpen
  \bibfield  {author} {\bibinfo {author} {\bibfnamefont {Alexander}\
  \bibnamefont {Nikolaenko}}\ and\ \bibinfo {author} {\bibfnamefont {Falko}\
  \bibnamefont {Pientka}},\ }\bibfield  {title} {\enquote {\bibinfo {title}
  {Topological superconductivity in proximity to type-{II} superconductors},}\
  }\href {\doibase 10.1103/physrevb.103.134503} {\bibfield  {journal} {\bibinfo
   {journal} {Physical Review B}\ }\textbf {\bibinfo {volume} {103}} (\bibinfo
  {year} {2021}),\ 10.1103/physrevb.103.134503},\ \Eprint
  {http://arxiv.org/abs/arXiv:2009.14221} {arXiv:2009.14221} \BibitemShut
  {NoStop}%
\bibitem [{\citenamefont {Kurilovich}\ \emph {et~al.}(2022)\citenamefont
  {Kurilovich}, \citenamefont {Raines},\ and\ \citenamefont
  {Glazman}}]{Kurilovich22}%
  \BibitemOpen
  \bibfield  {author} {\bibinfo {author} {\bibfnamefont {Vladislav~D.}\
  \bibnamefont {Kurilovich}}, \bibinfo {author} {\bibfnamefont {Zachary~M.}\
  \bibnamefont {Raines}}, \ and\ \bibinfo {author} {\bibfnamefont {Leonid~I.}\
  \bibnamefont {Glazman}},\ }\href@noop {} {\enquote {\bibinfo {title}
  {Disorder in andreev reflection of a quantum hall edge},}\ } (\bibinfo {year}
  {2022}),\ \Eprint {http://arxiv.org/abs/arXiv:2201.00273} {arXiv:2201.00273}
  \BibitemShut {NoStop}%
\bibitem [{\citenamefont {Schiller}\ \emph {et~al.}(2022)\citenamefont
  {Schiller}, \citenamefont {Katzir}, \citenamefont {Stern}, \citenamefont
  {Berg}, \citenamefont {Lindner},\ and\ \citenamefont {Oreg}}]{Schiller22}%
  \BibitemOpen
  \bibfield  {author} {\bibinfo {author} {\bibfnamefont {Noam}\ \bibnamefont
  {Schiller}}, \bibinfo {author} {\bibfnamefont {Barak~A.}\ \bibnamefont
  {Katzir}}, \bibinfo {author} {\bibfnamefont {Ady}\ \bibnamefont {Stern}},
  \bibinfo {author} {\bibfnamefont {Erez}\ \bibnamefont {Berg}}, \bibinfo
  {author} {\bibfnamefont {Netanel~H.}\ \bibnamefont {Lindner}}, \ and\
  \bibinfo {author} {\bibfnamefont {Yuval}\ \bibnamefont {Oreg}},\ }\href@noop
  {} {\enquote {\bibinfo {title} {Interplay of superconductivity and
  dissipation in quantum hall edges},}\ } (\bibinfo {year} {2022}),\ \Eprint
  {http://arxiv.org/abs/arXiv:2202.10475} {arXiv:2202.10475} \BibitemShut
  {NoStop}%
\bibitem [{\citenamefont {Caroli}\ \emph {et~al.}(1964)\citenamefont {Caroli},
  \citenamefont {{De Gennes}},\ and\ \citenamefont {Matricon}}]{Caroli64}%
  \BibitemOpen
  \bibfield  {author} {\bibinfo {author} {\bibfnamefont {C.}~\bibnamefont
  {Caroli}}, \bibinfo {author} {\bibfnamefont {P.G.}\ \bibnamefont {{De
  Gennes}}}, \ and\ \bibinfo {author} {\bibfnamefont {J.}~\bibnamefont
  {Matricon}},\ }\bibfield  {title} {\enquote {\bibinfo {title} {Bound fermion
  states on a vortex line in a type ii superconductor},}\ }\href {\doibase
  https://doi.org/10.1016/0031-9163(64)90375-0} {\bibfield  {journal} {\bibinfo
   {journal} {Physics Letters}\ }\textbf {\bibinfo {volume} {9}},\ \bibinfo
  {pages} {307--309} (\bibinfo {year} {1964})}\BibitemShut {NoStop}%
\bibitem [{\citenamefont {Dolcini}(2009)}]{Dolcini09}%
  \BibitemOpen
  \bibfield  {author} {\bibinfo {author} {\bibfnamefont {Fabrizio}\
  \bibnamefont {Dolcini}},\ }\href@noop {} {\enquote {\bibinfo {title}
  {Introduction to the scattering matrix formalism},}\ } (\bibinfo {year}
  {2009})\BibitemShut {NoStop}%
\end{thebibliography}%
\end{document}